\documentclass[12pt]{iopart}

\usepackage{iopams}

\expandafter\let\csname equation*\endcsname\relax
\expandafter\let\csname endequation*\endcsname\relax
\usepackage{graphicx}
\usepackage{amsmath,amssymb}
\usepackage{here}
\usepackage{bm}
\usepackage{framed}
\usepackage{ulem}
\usepackage[usenames]{color}

\begin{document}

\title[]{Thermal features of Heisenberg antiferromagnets on edge- versus corner-sharing
         triangular-based lattices: A message from spin waves}

\author{Shoji Yamamoto and Jun Ohara}

\address{Department of Physics, Hokkaido University,
             Sapporo 060-0810, Japan}
\ead{yamamoto@phys.sci.hokudai.ac.jp}
\begin{indented}
\item[]
\end{indented}
\vspace{-5mm}
\begin{abstract}
We propose a new scheme of modifying spin waves so as to describe the thermodynamic properties
of various noncollinear antiferromagnets with particular interest in a comparison between
edge- versus corner-sharing triangular-based lattices.
The well-known modified spin-wave theory for collinear antiferromagnets diagonalizes a bosonic
Hamiltonian subject to the constraint that the total staggered magnetization be zero.
Applying this scheme to frustrated noncollinear antiferromagnets ends in a poor
thermodynamics, missing the optimal ground state and breaking the local $\mathrm{U}(1)$
rotational symmetry.
We find such a plausible double-constraint condition for spin spirals as to spontaneously
go back to the traditional single-constraint condition at the onset of a collinear
N\'eel-ordered classical ground state.
We first diagonalize
only the bilinear terms in Holstein-Primakoff boson operators on the order of spin magnitude $S$
and then bring these linear spin waves into interaction in a perturbative rather than
variational manner.
We demonstrate specific-heat calculations in terms of thus-modified interacting spin waves
on various triangular-based lattices.
In zero dimension, modified-spin-wave findings in comparison with
finite-temperature Lanczos calculations turn out so successful as to reproduce
the monomodal and bimodal specific-heat temperature profiles of the triangular-based
edge-sharing Platonic and corner-sharing Archimedean polyhedral-lattice antiferromagnets,
respectively.
In two dimensions, high-temperature series expansions and tensor-network-based
renormalization-group calculations are still controversial especially at low temperatures,
and under such circumstances, modified spin waves interestingly predict that the specific heat
of the kagome-lattice antiferromagnet in the corner-sharing geometry remains having both
mid-temperature broad maximum and low-temperature narrow peak in the thermodynamic limit,
while the specific heat of the triangular-lattice antiferromagnet in the edge-sharing geometry
retains a low-temperature sharp peak followed by a mid-temperature weak anormaly
in the thermodynamic limit.
By further calculating one-magnon spectral functions in terms of our newly developed
double-constraint modified spin-wave theory, we reveal
that not only the elaborate modification scheme but also quantum corrections, especially those
caused by the $O(S^0)$ primary self-energies, are key ingredients in the successful description
of triangular-based-lattice noncollinear antiferromagnets  over the whole temperature range of
absolute zero to infinity.
\end{abstract}

%
%
%
%
%

\section{Introduction}
\label{S:I}
Triangular- and kagome-lattice antiferromagnets attract increasing interest in the context
of possible realization of spin-liquid (SL) states \cite{S016502,Z025003,K451}.
Among them, their quantum spin-$\frac{1}{2}$ nearest-neighbor Heisenberg models are
enthusiastically studied from various points of view such as
quantum dimer models \cite{M1881,M137202} intending to describe resonating valence bond (RVB)
phases \cite{A153,F23},
Gutzwiller-projected wave functions of fermionic spinons to yield a gapless $\mathrm{U(1)}$ SL
\cite{H014413,R117205,H224413},
the ``Amperean" \cite{L031017} spinon pairing to reduce the $\mathrm{U(1)}$ gauge group to
$\mathbb{Z}_2$ \cite{L067006},
bosonic spinons accompanied by topological vison excitations in a gapped $\mathbb{Z}_2$ SL
\cite{Q176401,P289}, and
projective-symmetry-group (PSG) classifications of Schwinger-boson \cite{W174423} and
Schwinger-fermion \cite{L224413} mean-field states.
In cooperation with various theoretical investigations, much effort is devoted to designing
candidate materials.
They say that
the hexagonal perovskite $\mathrm{Ba}_3\mathrm{CoSb}_2\mathrm{O}_9$ \cite{D8923} and
the zinc paratacamite $\mathrm{ZnCu}_3(\mathrm{OH})_6\mathrm{Cl}_2$ \cite{S13462}
known as herbertsmithite \cite{B527}
closely approximate the ideal spin-$\frac{1}{2}$ triangular- and kagome-lattice
Heisenberg antiferromagnets, respectively.

   While the nearest-neighbor pair-exchange-coupled spin-$\frac{1}{2}$ Heisenberg antiferromagnet
on the equilateral triangular lattice has an ordered ground state of
the $120^\circ$ N\'eel type \cite{H2531,J2727,B2590,B10048,A2483,C3899,W127004,S1766},
they readily become liquid with likely perturbations such as ring exchange \cite{M1064,S235115},
spatial anisotropy \cite{Y014408,S073025}, and/or next-nearest-neighbor exchange \cite{H207203}.
On the other hand, the nearest-neighbor pair-exchange-coupled spin-$\frac{1}{2}$ Heisenberg
antiferromagnet on the regular kagome lattice has no conventional N\'eel ordering
such as what they call $\bm{Q}=\bm{0}$ ($3$ spins per unit cell) and
$\sqrt{3}\times\sqrt{3}$ ($9$ spins per unit cell) \cite{H2899,S12377}
in its ground state,
with a possibility of the former and latter
being stabilized by possible Dzyaloshinskii-Moriya interaction
\cite{C140405,M064428,M267201,L224414,A027203} and
ferromagnetic next-nearest-neighbor exchange coupling
\cite{M184401,K104418,D097203}, respectively.
Besides conventional magnetically ordered states, the ferromagnetic superexchange favors
valence bond crystals (VBCs) with a rather large unit cell \cite{I100404,I115031}
such as a honeycomb pattern of resonating benzene-like arrangements of singlet bonds on
the hexagonal plaquettes with a $2\sqrt{3}\times 2\sqrt{3}$ ``supercell"
($36$ spins per unit cell) \cite{M5962,N214415,S180407,S144415}.
In any case the precise nature of the spin-disordered ground state of the simple nearest-neighbor
kagome-lattice antiferromagnetic Heisenberg model remains a subject of considerable debate.

   In response to stimulative experimental \cite{M077204,H107204,W144412,H406,M455} and/or
numerical \cite{Y1173,D067201} findings, various SL phases have been nominated for
a possible ground state, including
an algebraic or $\mathrm{U(1)}$ Dirac SL \cite{R117205,H224413,M027204,I020407,I060405},
whose low-energy physics is governed by four flavors of massless two-component Dirac fermions
coupled to a $\mathrm{U(1)}$ gauge filed,
a gapped $\mathbb{Z}_2$ SL \cite{W174423,L224413,P289},
which exhibits gapped vortex excitations of an emergent $\mathbb{Z}_2$ gauge field as well as
a flat spinon band at low energies, and
a gapped chiral SL \cite{M207204,M125127},
which simultaneously and spontaneously breaks space reflection and time reversal symmetries.
Since one or more gapped $\mathbb{Z}_2$ SLs may be emergent from their ``parent" gapless
$\mathrm{U(1)}$ SL by small pairings of fermionic spinons, such neighboring SLs are so close
in energy and thus hard to distinguish numerically.
Even state-of-the-art simulations based on tensor-network wavefunctions lead to a debate over
the $\mathrm{U(1)}$ Dirac SL \cite{L137202,J006}, $\mathbb{Z}_2$ gapped SL \cite{M235107}, and
VBC with a $36$-site unit cell \cite{E187203}.
Under such circumstances, a symmetry-based analysis of SL phases provides an otherwise unavailable
insight.
Wen \cite{W165113} employed PSGs to characterize hundreds of \textit{symmetric} SLs with
$\mathrm{SU(2)}$, $\mathrm{U(1)}$, or $\mathbb{Z}_2$ gauge structures in the context of
fermionic mean-field theories.
Wang and Vishwanath \cite{W174423} pushed his exploration further for Schwinger-boson mean-field
states, focusing on the Ising gauge group but accessing both SL and conventional magnetically
ordered states.
Besides classifying topological orders (gauge configurations) in SLs, the language of projective
symmetry can identify itinerant spinon excitation modes \cite{M041103} and further describe their
Raman response \cite{K214411,Y063701}.

   The triangular and kagome lattices can be viewed as edge- and corner-sharing triangles in
two dimensions, respectively, and their Heisenberg antiferromagnets turn out to have distinct
low-energy structures \cite{L2521,A224414}.
An antiferromagnetic order established in the thermodynamic limit manifests itself in
finite-size clusters, whose low-energy spectra can be described in terms of a ``quantum top"
\cite{B2590,B10048,A2483}.
In the case of the spin-$\frac{1}{2}$ antiferromagnetic Heisenberg Hamiltonian on the equilateral
triangular lattice of $L$ sites, the three identical ferromagnetic sublattices each have
a collective spin of length $L/6$ and they couple to reproduce rotationally invariant states,
which approximate the lowest-lying eigenstates in different subspaces labeled by total spin $S$,
$
   E_L(S)-E_L(S_{\mathrm{min}})
  =S(S+1)/(2I_L)
   \ (S>S_{\mathrm{min}}),
$
where $S_{\mathrm{min}}$ is either $0$ or $\frac{1}{2}$ according as $L$ is even or odd and
the inertia of the top $I_L$ is an extensive quantity, proportional to the perpendicular
susceptibility \cite{A2483} and therefore to $L$ \cite{B2590,B10048}.
Hence, plotting the lowest energy level as a function of $S(S+1)$ yields a ``Pisa tower"
\cite{M1064,L2521,A224414,B10048} with a slope $\propto 1/L$.
The Pisa tower is well separated from the softest magnons that converge to the ground state as
$1/\sqrt{L}$ \cite{A2483}.
Thus and thus, an extensive set of low-lying levels, including both magnetic and nonmagnetic ones,
collapse onto the ground state in the thermodynamic limit.
In the spin-$\frac{1}{2}$ regular kagome-lattice Heisenberg antiferromagnet, on the contrary,
there is absolutely no such low-energy structure.
The candidate levels to form a tower of states neither scale as $S(S+1)$ nor separate from
the above continuum of excitations.
The magnetic (triplet) excitations are separated from the ground state by a gap and this gap is
filled with a continuum of nonmagnetic (singlet) excitations adjacent to the ground state
\cite{W501,C104424,J117203,L155142}.
Such differences in level structure should manifest themselves in thermodynamics at low
temperatures.
Especially in the kagome-lattice antiferromagnet,
a large amount of low-temperature residual entropy may cause some additional structure,
whether a peak \cite{S2953} or shoulder \cite{S094423},
to the main Schottky maximum in the temperature profile of the specific heat.
While the spin susceptibility is thermally activated due to the singlet-triplet gap
\cite{Y1173,D067201}, the specific heat does not decrease exponentially in the gap
with a possibility of defining an extra energy scale set by low-lying singlets.

   We are thus motivated to perform a comparative study of Heisenberg antiferromagnets in
the triangular-based edge- versus corner-sharing geometry.
We calculate the specific-heat curves of Heisenberg antiferromagnets not only on
the two-dimensional regular triangular and kagome lattices but also on ``zero-dimensional"
analogs, i.e., Platonic and Archimedean polyhedra consisting of edge- and corner-sharing
triangles, respectively.
We aim to verify whether they have \textit{generic} thermal features, and if any, further
clarify why they behave so.
In order to explore finite-temperature properties of infinite systems, we modify the conventional
antiferromagnetic spin-wave (SW) thermodynamics \cite{K568} applicably to low-dimensional
spin spirals.
We make drastic and far-reaching reforms in the traditional modified SW (MSW) theory designed for
low-dimensional collinear antiferromagnets \cite{T2494,H4769,T5000,N034714,Y094412}.

   It is very hard to calculate thermodynamic quantities of quantum frustrated antiferromagnets
on an infinite lattice in two or more dimensions.
There are limited numerical approaches available, and what is worse, their findings are not
necessarily consistent with each other.
Among others, is the long-standing debate on the low-temperature profile of the magnetic
specific heat of the spin-$\frac{1}{2}$ kagome-lattice Heisenberg antiferromagnet
\cite{S2953,S094423}, originating from the milikelvin heat-capacity measurements of
${}^3\mathrm{He}$ adsorbed on graphite \cite{G1868}.
Early calculations of small clusters \cite{E2405,E6871} argued that the specific heat of
the nearest-neighbor pair-exchange-coupled spin-$\frac{1}{2}$ Heisenberg antiferromagnet on
the regular kagome lattice is very likely to have a low-temperature peak in addition to
the main maximum at $k_{\mathrm{B}}T/J\simeq 2/3$ with $J$ being the antiferromagnetic exchange.
The Lanczos diagonalization technique combined with random sampling \cite{J5065,S4512} and
the thermal-pure-quantum-states formulation of statistical mechanics \cite{S010401}
enabled us to calculate somewhat larger medium-sized clusters, but their findings are not yet
decisive of whether the low-temperature structure remains a peak \cite{S113702} or reduces to
a shoulder \cite{S010401,S094423} in the thermodynamic limit.
Ingeniously devised quantum Monte Carlo algorithms \cite{Z8436,N9174} revealed that
much larger clusters still exhibit a double-peak temperature profile in their specific-heat curves.
Recently tensor-network-based methods have come into use in exploring thermodynamics,
but their findings \cite{C1545} are quite otherwise in favor of the conclusion that
the low-temperature structure of the specific heat is a shoulderlike hump rather than
a true maximum.
While high-temperature series expansion may also be employed to calculate the specific heat
in the thermodynamic limit, its reliability has been under debate.
When analyzed through standard Pad\'{e} approximants, the specific-heat high-temperature series
expansion extrapolated down to absolute zero has a large missing entropy \cite{E6871}.
In an attempt to improve the convergence of standard Pad\'{e} approximations at low temperatures
much below $J/k_{\textrm{B}}$, Bernu and Misguich \cite{B134409} introduced such biased
approximants as to satisfy the energy and entropy sum rules obeyed by the specific heat.
In the same context, Rigol, Bryant, and Singh \cite{R187202,R061118} proposed a numerical
linked-cluster algorithm, intending to capture advantages of both high-temperature expansion and
exact diagonalization, together with sequence extrapolation techniques to accelerate
the convergence of linked clusters.
Both approaches \cite{R187202,R061118,M014417,M305} concluded that the missing entropy should
be compensated by the appearance of an additional low-temperature peak below the major maximum.

   Thus, even via the use of such a wide variety of tools, yet the issue remains to be settled.
That is the very reason why we develop a new ``language" of our own.
Our newly developed MSW theory is not exact, to be sure, but it is widely applicable to frustrated
antiferromagnets in all dimensions in the thermodynamic limit to capture their thermal features
very well, serving as a mirror of their ground states in the context of whether they are
classically ordered or quantum disordered.
We investigate temperature profiles of the specific heat for the nearest-neighbor
antiferromagnetic Heisenberg model
\begin{align}
   \mathcal{H}
  =J\sum_{\langle l,l'\rangle}
   \bm{S}_{\bm{r}_l}\cdot\bm{S}_{\bm{r}_{l'}}
  =\frac{J}{2}\sum_{l=1}^L\sum_{\kappa=1}^z
   \bm{S}_{\bm{r}_l}\cdot\bm{S}_{\bm{r}_l+\bm{\delta}_{l:\kappa}},
   \label{E:HeisenbergH}
\end{align}
where $J$ is assumed to be positive, $\bm{S}_{\bm{r}_l}$ are the vector spin operators of
magnitude $S$ attached to the site at $\bm{r}_l$, and $\bm{\delta}_{l:\kappa}$ are the vectors
connecting the site at $\bm{r}_l$ with its $z$ nearest neighbors.
We start with some bipartite lattices and then proceed to a variety of triangular-based
polyhedral and planar lattices.
We suppose that they each consist of $L$ sites with the unique coordination number $z$.

\section{Modified spin-wave theory of collinear antiferromagnets}
\label{S:MSWofCA}
Takahashi \cite{T2494} and Hirsch \textit{et al.} \cite{H4769,T5000} had an idea of so modifying
the conventional SW (CSW) theory as to describe collinear antiferromagnets in lower than three
dimensions at finite temperatures, where Holstein--Primakoff \cite{H1098} or
Dyson-Maleev \cite{D1217,D1230} bosons are constrained to keep the total staggered magnetization
zero via a Bogoliubov transformation dependent on temperature.
If we apply this pioneering but naive single-constraint (SC)-modification scheme to frustrated
noncollinear antiferromagnets as it is, what will happen?
Let us begin by calculating the prototypical antiferromagnetic MSW thermodynamics for the triangular
and kagome lattices in comparison with those for the bipartite square and honeycomb lattices,
intending to ascertain what is the problem in the prototypical MSW as well as CSW formalisms when
applied to spiral magnets.

   Considering that we treat antiferromagnetic spin spirals in various geometries,
we employ Holstein-Primakoff, rather than Dyson-Maleev, bosons.
Within the linear SW (LSW) treatment, they are no different from each other.
When we go beyond it to obtain interacting SWs (ISWs), the Holstein-Primakoff and Dyson-Maleev
representations of any spin Hamiltonian generally differ from each other,
the former remaining Hermitian but the latter becoming non-Hermitian.
As long as the spins align antiparallel among neighbors, however, the Wick or Hartree-Fock
decompositions of the Holstein-Primakoff and Dyson-Maleev bosonic Hamiltonians coincide
with each other up to $O(S^0)$ \cite{N034714,Y094412}.
This is no longer the case with any noncollinear alignment among the spins, whose bosonic
Hamiltonian reads a series in descending powers of $\sqrt{S}$ rather than $S$ and generally
contains terms consisting of an odd number of boson operators.
A usual treatment \cite{C144416,C144415} of such bosonic Hamiltonians consists of diagonalizing
the harmonic part exactly and then taking account of higher-order interactions as perturbations,
where Holstein-Primakoff bosons are much more tractable than Dyson-Maleev bosons.

\subsection{Bosonic Hamiltonian}
\label{SS:BH}

\begin{figure}[ht]
\begin{flushright}
\includegraphics[width=130mm]{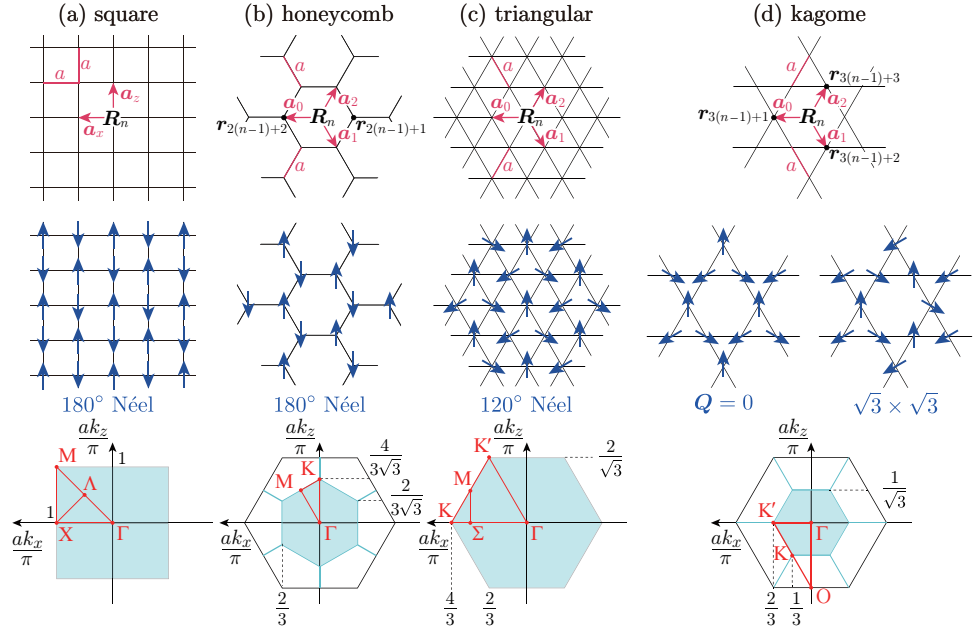}
\end{flushright}
\vspace{-7mm}
\caption{Various two-dimensional lattices
         [cf. \eqref{E:squareR}--\eqref{E:kagomeR} and \eqref{E:square_r}--\eqref{E:kagome_r}]
         together with their classical antiferromagnetic ground states
         [cf. \eqref{E:squareQ}--\eqref{E:kagomeQroot3}] and
         full paramagnetic---here in the sense of no interunit magnetic order---Brillouin zones
         being shaded.}
\label{F:VariousLattices}
\end{figure}

   In order to express the Heisenberg Hamiltonian \eqref{E:HeisenbergH} in terms of
Holstein-Primakoff bosons, we first transform it into the rotating frame with its $z$ axis
pointing along each local spin direction in the classical ground state.
We denote the local coordinate system by $(\tilde{x},\tilde{y},\tilde{z})$ distinguishably from
the laboratory frame $(x,y,z)$.
For coplanar antiferromagnets with their spins lying in the $z$-$x$ plane, for instance,
the spin components in the laboratory and rotating frames are related with each other as
\begin{align}
   \left[
    \begin{array}{c}
     S_{\bm{r}_{l}}^{z} \\
     S_{\bm{r}_{l}}^{x} \\
     S_{\bm{r}_{l}}^{y} \\
    \end{array}
   \right]
  =\left[
    \begin{array}{ccc}
     \cos\phi_{\bm{r}_{l}} &-\sin\phi_{\bm{r}_{l}} & 0 \\
     \sin\phi_{\bm{r}_{l}} & \cos\phi_{\bm{r}_{l}} & 0 \\
               0           &           0           & 1 \\
    \end{array}
   \right]
   \left[
    \begin{array}{c}
     S_{\bm{r}_{l}}^{\tilde{z}} \\
     S_{\bm{r}_{l}}^{\tilde{x}} \\
     S_{\bm{r}_{l}}^{\tilde{y}} \\
    \end{array}
   \right],
   \label{E:SinLandRframes2D}
\end{align}
where $\phi_{\bm{r}_l}$ is the angle formed by the axes $z$ and $\tilde{z}$ at $\bm{r}_l$.
The local coordinate notation is applicable to collinear antiferromagnets as well with substantial
advantage especially in our comparative study.
In the local coordinate system, \eqref{E:HeisenbergH} reads
\begin{align}
   &
   \mathcal{H}
  =J\sum_{\langle l,l'\rangle}
   \left[
    \left(
     S_{\bm{r}_l}^{\tilde{z}}S_{\bm{r}_{l'}}^{\tilde{z}}
    +S_{\bm{r}_l}^{\tilde{x}}S_{\bm{r}_{l'}}^{\tilde{x}}
    \right)
    \cos(\phi_{\bm{r}_l}-\phi_{\bm{r}_{l'}})
   +\left(
     S_{\bm{r}_l}^{\tilde{z}}S_{\bm{r}_{l'}}^{\tilde{x}}
    -S_{\bm{r}_l}^{\tilde{x}}S_{\bm{r}_{l'}}^{\tilde{z}}
    \right)
    \sin(\phi_{\bm{r}_l}-\phi_{\bm{r}_{l'}})
   \right.
   \allowdisplaybreaks
   \nonumber \\
   &\qquad\qquad
   \left.
   +S_{\bm{r}_l}^{\tilde{y}}S_{\bm{r}_{l'}}^{\tilde{y}}
   \right].
   \label{E:HeisenbergHinRframe}
\end{align}
When we employ the Holstein-Primakoff bosons
\begin{align}
   &
   S_{\bm{r}_l}^{\tilde{z}}
  =S-a_{\bm{r}_l}^\dagger a_{\bm{r}_l},\ 
   S_{\bm{r}_l}^{\tilde{x}}-iS_{\bm{r}_l}^{\tilde{y}}
  \equiv S_{\bm{r}_l}^{\tilde{+}\dagger}
  \equiv S_{\bm{r}_l}^{\tilde{-}}
  =a_{\bm{r}_l}^\dagger
   \sqrt{2S-a_{\bm{r}_l}^\dagger a_{\bm{r}_l}}
  \equiv \sqrt{2S}a_{\bm{r}_l}^\dagger\mathcal{R}_{\bm{r}_l}(S)
   \label{E:HPT}
\end{align}
and expand the square root $\mathcal{R}_{\bm{r}_l}(S)$ in descending powers of $S$,
\begin{equation}
   \mathcal{R}_{\bm{r}_l}(S)
  =1-\sum_{l=1}^\infty\frac{(2l-3)!!}{l!}
   \left(\frac{a_{\bm{r}_l}^\dagger a_{\bm{r}_l}}{4S}\right)^l,
   \label{E:Arl(S)}
\end{equation}
the Hamiltonian (\ref{E:HeisenbergHinRframe}) is expanded into the series
\begin{equation}
   \mathcal{H}
  =\sum_{m=0}^4
   \mathcal{H}^{\left(\frac{m}{2}\right)}
  +O\left(S^{-\frac{1}{2}}\right),
   \label{E:Hexpanded2D}
\end{equation}
where $\mathcal{H}^{\left(\frac{m}{2}\right)}$, on the order of $S^{\frac{m}{2}}$, read
\begin{align}
   &
   \mathcal{H}^{(2)}
  =JS^{2}
   \sum_{\langle l,l' \rangle}
   \cos
   \left(
    \phi_{\bm{r}_l}-\phi_{\bm{r}_{l'}}
   \right),
   \label{E:H(2)2D}
   \allowdisplaybreaks \\
   &
   \mathcal{H}^{\left(\frac{3}{2}\right)}
 =-J\sqrt{\frac{S^3}{2}}
   \sum_{\langle l,l' \rangle}
    \left(
     a_{\bm{r}_{l }}^\dagger-a_{\bm{r}_{l }}
    -a_{\bm{r}_{l'}}^\dagger+a_{\bm{r}_{l'}}
    \right)
   \sin
   \left(
    \phi_{\bm{r}_l}-\phi_{\bm{r}_{l'}}
   \right)
  =0,
   \label{E:H(3/2)2D}
   \allowdisplaybreaks \\
   &
   \mathcal{H}^{(1)}
 =-JS
   \sum_{\langle l,l' \rangle}
   \left[
    \vphantom{
    \frac{\cos
          \left(
           \phi_{\bm{r}_l}-\phi_{\bm{r}_{l'}}
          \right)
         -1}{2}
    }
    \left(
     a_{\bm{r}_{l }}^\dagger a_{\bm{r}_{l }}
    +a_{\bm{r}_{l'}}^\dagger a_{\bm{r}_{l'}}
    \right)
    \cos
    \left(
     \phi_{\bm{r}_l}-\phi_{\bm{r}_{l'}}
    \right)
   -\left(
     a_{\bm{r}_{l}}^\dagger a_{\bm{r}_{l'}}^\dagger
    +a_{\bm{r}_{l}}         a_{\bm{r}_{l'}}
    \right)
   \right.
   \allowdisplaybreaks
   \nonumber \\
   &\qquad\quad\times
    \frac{\cos
          \left(
           \phi_{\bm{r}_l}-\phi_{\bm{r}_{l'}}
          \right)
         -1}{2}
   \left.
   -\left(
     a_{\bm{r}_{l }}^\dagger a_{\bm{r}_{l'}}
    +a_{\bm{r}_{l'}}^\dagger a_{\bm{r}_{l}}         
    \right)
    \frac{\cos
          \left(
           \phi_{\bm{r}_l}-\phi_{\bm{r}_{l'}}
          \right)
         +1}{2}
   \right],
   \label{E:H(1)2D}
   \allowdisplaybreaks \\
   &
   \mathcal{H}^{\left(\frac{1}{2}\right)}
  =J\sqrt{\frac{S}{2}}
   \sum_{\langle l,l' \rangle}
   \left[
    \left(
     a_{\bm{r}_l}^\dagger+a_{\bm{r}_l}
    \right)
    a_{\bm{r}_{l'}}^\dagger a_{\bm{r}_{l'}}
   -a_{\bm{r}_{l }}^\dagger a_{\bm{r}_{l }}
    \left(
     a_{\bm{r}_{l'}}^\dagger+a_{\bm{r}_{l'}}
    \right)
     \vphantom{
     a_{\bm{r}_{l'}}^\dagger+a_{\bm{r}_{l'}}
     }
   \right]
   \sin
   \left(
    \phi_{\bm{r}_l}-\phi_{\bm{r}_{l'}}
   \right),
   \label{E:H(1/2)2D}
   \allowdisplaybreaks \\
   &
   \mathcal{H}^{(0)}
  =J
   \sum_{\langle l,l' \rangle}
   \left\{
    \vphantom{
    \frac{\cos
          \left(
           \phi_{\bm{r}_l}-\phi_{\bm{r}_{l'}}
          \right)
         +1}{8}}\!
    a_{\bm{r}_{l }}^\dagger a_{\bm{r}_{l }}
    a_{\bm{r}_{l'}}^\dagger a_{\bm{r}_{l'}}
    \cos
    \left(
     \phi_{\bm{r}_l}-\phi_{\bm{r}_{l'}}
    \right)
   -\left[
     a_{\bm{r}_l}^\dagger a_{\bm{r}_{l'}}^\dagger
     \left(
      a_{\bm{r}_{l }}^\dagger a_{\bm{r}_{l }}
     +a_{\bm{r}_{l'}}^\dagger a_{\bm{r}_{l'}}
     \right)
    \right.
   \right.
   \nonumber
   \allowdisplaybreaks \\
   &\qquad\quad
    \left.
    +\left(
      a_{\bm{r}_{l }}^\dagger a_{\bm{r}_{l }}
     +a_{\bm{r}_{l'}}^\dagger a_{\bm{r}_{l'}}
     \right)
     a_{\bm{r}_l}a_{\bm{r}_{l'}}
    \right]
    \frac{\cos
          \left(
           \phi_{\bm{r}_l}-\phi_{\bm{r}_{l'}}
          \right)
         -1}{8}
   -\left[
     a_{\bm{r}_{l }}^\dagger
     \left(
      a_{\bm{r}_{l }}^\dagger a_{\bm{r}_{l }}
     +a_{\bm{r}_{l'}}^\dagger a_{\bm{r}_{l'}}
     \right)
     a_{\bm{r}_{l'}}
    \right.
   \nonumber
   \allowdisplaybreaks \\
   &\qquad\quad
    \left.
    +a_{\bm{r}_{l'}}^\dagger
     \left(
      a_{\bm{r}_{l }}^\dagger a_{\bm{r}_{l }}
     +a_{\bm{r}_{l'}}^\dagger a_{\bm{r}_{l'}}
     \right)
     a_{\bm{r}_{l }}
    \right]
   \left.\!
    \frac{\cos
          \left(
           \phi_{\bm{r}_l}-\phi_{\bm{r}_{l'}}
          \right)
         +1}{8}
   \right\},
   \label{E:H(0)2D}
\end{align}
considering that
the sum of the relative rotation angles over the nearest neighbors of an arbitrary site
$
   \sum_{\kappa=1}^z
   \left(
    \phi_{\bm{r}_l+\bm{\delta}_{l:\kappa}}-\phi_{\bm{r}_l}
   \right)
$
equals zero or a multiple of $\pi$ according as $z$ is even or odd
in the classical ground states in question (Fig. \ref{F:VariousLattices}).

   The unit cells of the square and triangular lattices both can be reduced to a single site,
while those of the honeycomb and kagome lattices contain at least two and three sites,
respectively.
In any case, we label lattice sites ($l=1,\cdots,L$) with their belonging unit cells
($n=1,\cdots,N$) and further internal degrees of freedom ($\sigma=1,\cdots,L/N\equiv p$)
as $l\equiv p(n-1)+\sigma$.
We denote the primitive translation vectors of the square and triangular lattices by
\begin{align}
   \mathrm{square}:\ 
   &
   \bm{a}_z
  =a
   \left(
    \begin{array}{c}
     0 \\
     0 \\
     1 \\
    \end{array}
   \right),\ 
   \bm{a}_x
  =a
   \left(
    \begin{array}{c}
     1 \\
     0 \\
     0 \\
    \end{array}
   \right),
   \label{E:squarePTV}
   \allowdisplaybreaks \\
   \mathrm{triangular}:\ 
   &
   \bm{a}_1
  =a
   \left(
    \begin{array}{c}
    -\frac{1}{2} \\
     0 \\
    -\frac{\sqrt{3}}{2} \\
    \end{array}
   \right),\ 
   \bm{a}_2
  =a
   \left(
    \begin{array}{c}
    -\frac{1}{2} \\
     0 \\
     \frac{\sqrt{3}}{2} \\
    \end{array}
   \right),\ 
   \bm{a}_0
  \equiv
  -\bm{a}_1-\bm{a}_2
  =a
   \left(
    \begin{array}{c}
     1 \\
     0 \\
     0 \\
    \end{array}
   \right),
   \label{E:triangularPTV}
\end{align}
and then those of the honeycomb and kagome lattices are given by
$\bm{a}_0-\bm{a}_1$, $\bm{a}_1-\bm{a}_2$, $\bm{a}_2-\bm{a}_0$, and
$2\bm{a}_1$, $2\bm{a}_2$, $2\bm{a}_0$, respectively (cf. Fig. \ref{F:VariousLattices}).
The nearest-neighbor vectors of the $l$th site are given by
\begin{align}
   \mathrm{square}:\ 
   &
   \bm{\delta}_{l:\kappa}
  =\pm\bm{a}_z,\pm\bm{a}_x,
   \label{E:squaredelta}
   \allowdisplaybreaks \\
   \mathrm{honeycomb}:\ 
   &
   \bm{\delta}_{l:\kappa}
  =(-)^l\bm{a}_1,(-)^l\bm{a}_2,(-)^l\bm{a}_0,
   \label{E:honeycombdelta}
   \allowdisplaybreaks \\
   \mathrm{triangular}:\ 
   &
   \bm{\delta}_{l:\kappa}
  =\pm\bm{a}_1,\pm\bm{a}_2,\pm\bm{a}_0,
   \label{E:triangulardelta}
   \allowdisplaybreaks \\
   \mathrm{kagome}:\ 
   &
   \bm{\delta}_{l:\kappa}
  =\pm\bm{a}_{(l+1\ \mathrm{mod}\ 3)},
   \pm\bm{a}_{(l+2\ \mathrm{mod}\ 3)},
   \label{E:kagomedelta}
\end{align}
while the center of the $n$th unit cell is expressed as
\begin{align}
   \!\!\!\!\mathrm{square}:\ 
   &
   \bm{R}_n
  =n_1\bm{a}_z+n_2\bm{a}_x
  =a{\vphantom{n_2,0,n_1}}^{\mathrm{t}}\!\!
   \left(
    n_2,0,n_1
   \right),
   \label{E:squareR}
   \allowdisplaybreaks \\
   \!\!\!\!\mathrm{honeycomb}:\ 
   &
   \bm{R}_n
  =n_1(\bm{a}_2-\bm{a}_0)+n_2(\bm{a}_0-\bm{a}_1)
  =\sqrt{3}a{\vphantom{\frac{\sqrt{3}(n_2-n_1)}{2},0,\frac{n_2+n_1}{2}}}^{\mathrm{t}}\!\!
   \left[
    \frac{\sqrt{3}(n_2-n_1)}{2},0,\frac{n_2+n_1}{2}
   \right],
   \label{E:honeycombR}
   \allowdisplaybreaks \\
   \!\!\!\!\mathrm{triangular}:\ 
   &
   \bm{R}_n
  =n_1\bm{a}_2+n_2\bm{a}_0
  =a{\vphantom{n_2-\frac{n_1}{2},0,\frac{\sqrt{3}n_1}{2}}}^{\mathrm{t}}\!\!
   \left(
    n_2-\frac{n_1}{2},0,\frac{\sqrt{3}n_1}{2}
   \right),
   \label{E:triangularR}
   \allowdisplaybreaks \\
   \!\!\!\!\mathrm{kagome}:\ 
   &
   \bm{R}_n
  =2n_1\bm{a}_2+2n_2\bm{a}_0
  =2a{\vphantom{n_2-\frac{n_1}{2},0,\frac{\sqrt{3}n_1}{2}}}^{\mathrm{t}}\!\!
   \left(
    n_2-\frac{n_1}{2},0,\frac{\sqrt{3}n_1}{2}
   \right)
   \label{E:kagomeR}
\end{align}
(cf. Fig. \ref{F:VariousLattices})
with $n_1$ and $n_2$ being the unique set of integers to represent $\bm{R}_n$.
When we express each site position $\bm{r}_l\equiv\bm{r}_{p(n-1)+\sigma}$ as
\begin{align}
   \mathrm{square}:\ 
   &
   \bm{r}_{1(n-1)+\sigma}
  =\bm{R}_n,
   \label{E:square_r}
   \allowdisplaybreaks \\
   \mathrm{honeycomb}:\ 
   &
   \bm{r}_{2(n-1)+\sigma}
  =\bm{R}_n+(-)^\sigma\bm{a}_0,
   \label{E:honeycomb_r}
   \allowdisplaybreaks \\
   \mathrm{triangular}:\ 
   &
   \bm{r}_{1(n-1)+\sigma}
  =\bm{R}_n,
   \label{E:triangular_r}
   \allowdisplaybreaks \\
   \mathrm{kagome}:\ 
   &
   \bm{r}_{3(n-1)+\sigma}
  =\bm{R}_n+\bm{a}_{\sigma-1},
   \label{E:kagome_r}
\end{align}
and define the ``interunit" and/or ``intraunit" ordering wave vectors
\begin{align}
   \hspace*{-0.2cm}
   \mathrm{square}:\ 
   &
   \bm{Q}
  =\frac{1}{a}
   \left(
    \pi,0,\pi
   \right),\ 
   \bm{q}
  =\bm{0}
   \hspace*{-2.0cm}
   &
   (180^\circ\ \mathrm{N}\acute{\mathrm{e}}\mathrm{el}),
   &
   \label{E:squareQ}
   \allowdisplaybreaks \\
   \hspace*{-0.2cm}
   \mathrm{honeycomb:}\ 
   &
   \bm{Q}
  =\bm{0},\ 
   \bm{q}
  =\frac{1}{a}
   \left(
    \pi,0,0
   \right)
   \hspace*{-3.0cm}
   &
   (180^\circ\ \mathrm{N}\acute{\mathrm{e}}\mathrm{el}),
   &
   \label{E:honeycombQ}
   \allowdisplaybreaks \\
   \hspace*{-0.2cm}
   \mathrm{triangular:}\ 
   &
   \bm{Q}
  =\frac{1}{a}
   \left(
    \frac{4\pi}{3},0,0
   \right),\ 
   \bm{q}
  =\bm{0}
   \hspace*{-3.0cm}
   &
   (120^\circ\ \mathrm{N}\acute{\mathrm{e}}\mathrm{el}),
   &
   \label{E:triangularQ}
   \allowdisplaybreaks \\
   \hspace*{-0.2cm}
   \mathrm{kagome:}\ 
   &
   \bm{Q}
  =\bm{0},\ 
   \bm{q}
  =\frac{1}{\sqrt{3}a}
   \left(
    0,0,\frac{8\pi}{3}
   \right)
   \hspace*{-3.0cm}
   &
   (\bm{Q}=\bm{0}),
   &
   \label{E:kagomeQ=0}
   \allowdisplaybreaks \\
   \hspace*{-0.2cm}
   \mathrm{kagome:}\ 
   &
   \bm{Q}
  =\frac{1}{2a}
   \left(
    \frac{8\pi}{3},0,0
   \right),\ 
   \bm{q}
  =\bm{0}
   \hspace*{-3.0cm}
   &
   (\sqrt{3}\times\sqrt{3}),
   &
   \label{E:kagomeQroot3}
\end{align}
the rotation angle $\phi_{\bm{r}_l}\equiv\phi_{\bm{r}_{p(n-1)+\sigma}}$ can be given by
\begin{align}
   \mathrm{square}:\ 
   &
   \phi_{\bm{r}_{1(n-1)+\sigma}}
  =\bm{Q}\cdot\bm{R}_n,
   \label{E:squarephi}
   \allowdisplaybreaks \\
   \mathrm{honeycomb}:\ 
   &
   \phi_{\bm{r}_{2(n-1)+\sigma}}
  =\bm{Q}\cdot\bm{R}_n+\sigma\bm{q}\cdot\bm{a}_0,
   \label{E:honeycombphi}
   \allowdisplaybreaks \\
   \mathrm{triangular}:\ 
   &
   \phi_{\bm{r}_{1(n-1)+\sigma}}
  =\bm{Q}\cdot\bm{R}_n,
   \label{E:triangularphi}
   \allowdisplaybreaks \\
   \mathrm{kagome}:\ 
   &
   \phi_{\bm{r}_{3(n-1)+\sigma}}
  =\bm{Q}\cdot\bm{R}_n+\bm{q}\cdot\bm{a}_{\sigma-1}.
   \label{E:kagomephi}
\end{align}

\subsection{Traditional single-constraint condition}
\label{SS:SSCC}

   Intending to calculate thermal quantities of the high-temperature-superconductor-parent
material $\mathrm{La}_2\mathrm{CuO}_4$ in terms of SWs, Takahashi \cite{T1524,T487}
considered diagonalizing a bosonic Hamiltonian of the square-lattice antiferromagnet subject to
the constraint that the staggered-magnetization expectation value be zero at every temperature
$T$,
\begin{align}
   \langle
    \mathcal{M}^{\tilde{z}}
   \rangle_T
  \equiv
   \sum_{l=1}^L
   \langle
    S_{\bm{r}_{l}}^{\tilde{z}}
   \rangle_T
  =LS
  -\sum_{l=1}^L
   \langle
    a_{\bm{r}_l}^\dagger a_{\bm{r}_l}
   \rangle_T
  =0.
   \label{E:collinearMSWconstraintHPB}
\end{align}
Hirsch and Tang \cite{H4769} also initiated such an idea by truncating the bosonic Hamiltonian
at the harmonic part,
$
   \mathcal{H}
  \simeq
   \sum_{m=1}^2
   \mathcal{H}^{(m)}
  \equiv
   \mathcal{H}_{\mathrm{harm}},
$
and diagonalizing
\begin{align}
   \widetilde{\mathcal{H}}_{\mathrm{harm}}
  \equiv
   \mathcal{H}_{\mathrm{harm}}
  +\mu
   \sum_{l=1}^L
   \left(
    S-a_{\bm{r}_l}^\dagger a_{\bm{r}_l}
   \right)\ 
   \label{E:tildeHharmT}
\end{align}
with such $\mu$ as to satisfy the constraint condition \eqref{E:collinearMSWconstraintHPB}.
Several authors \cite{T2494,T5000,N034714,Y094412,T1524,T487,Y064426}
sophisticated the thus-modified LSWs (MLSWs) into modified ISWs (MISWs) taking account of
the quartic interaction \eqref{E:H(0)2D} within a mean-field approximation
[cf. \eqref{E:H(0)quad2D}].
As long as we try such a variational construction of MISWs for noncollinear antiferromagnets
\cite{X6177,H075017,H014415}, the cubic interaction \eqref{E:H(1/2)2D},
which conserves neither the total magnetization nor the number of magnons,
remains ineffective.
The interactions on the order of $S$ to a fractional power,
$\mathcal{H}^{\left(\frac{m}{2}\right)}$ ($m=1,-1,\cdots$),
are characteristic of noncollinear antiferromagnets.
Without them,
the $\bm{Q}=\bm{0}$ and $\sqrt{3}\times\sqrt{3}$ ground states of the kagome-lattice
antiferromagnet remain degenerate in energy with each other.

\subsection{Modified spin-wave interaction---Variational treatment}
\label{SS:MSWI---VT}

   Let us take a look at the variational MISW thermodynamics for noncollinear versus collinear
antiferromagnets before we take an alternative step forward.
In order to handle variational MISWs, we introduce the multivalued double-angle-bracket
notation applicable for various approximation schemes \cite{N034714,Y094412}
\begin{align}
   &
   \langle\!\langle\mathcal{A}\rangle\!\rangle
  \equiv
   \frac{1}{2}
   \langle\!\langle
    a_{\bm{r}_l}^\dagger a_{\bm{r}_l}
   +a_{\bm{r}_l+\bm{\delta}_{l:\kappa}}^\dagger a_{\bm{r}_l+\bm{\delta}_{l:\kappa}}
   \rangle\!\rangle
  =\langle\!\langle
    a_{\bm{r}_l}^\dagger a_{\bm{r}_l}
   \rangle\!\rangle,
   \label{E:<<A>>}
   \allowdisplaybreaks \\
   &
   \langle\!\langle\mathcal{B}\rangle\!\rangle
  \equiv
   \frac{1}{2}
   \langle\!\langle
    a_{\bm{r}_l}^\dagger a_{\bm{r}_l+\bm{\delta}_{l:\kappa}}^\dagger
   +\mathrm{H.c.}
   \rangle\!\rangle
  =\langle\!\langle
    a_{\bm{r}_l}^\dagger a_{\bm{r}_l+\bm{\delta}_{l:\kappa}}^\dagger
   \rangle\!\rangle,
   \label{E:<<B>>}
   \allowdisplaybreaks \\
   &
   \langle\!\langle\mathcal{C}\rangle\!\rangle
  \equiv
   \frac{1}{4}
   \langle\!\langle
    a_{\bm{r}_l                }^\dagger a_{\bm{r}_l                }^\dagger
   +a_{\bm{r}_l+\bm{\delta}_{l:\kappa}}^\dagger a_{\bm{r}_l+\bm{\delta}_{l:\kappa}}^\dagger
   +\mathrm{H.c.}
   \rangle\!\rangle
  =\frac{1}{2}
   \langle\!\langle
    a_{\bm{r}_l                }^\dagger a_{\bm{r}_l                }^\dagger
   +\mathrm{H.c.}
   \rangle\!\rangle
  =\langle\!\langle
    a_{\bm{r}_l                }^\dagger a_{\bm{r}_l                }^\dagger
   \rangle\!\rangle,
   \label{E:<<C>>}
   \allowdisplaybreaks \\
   &
   \langle\!\langle\mathcal{D}\rangle\!\rangle
  \equiv
   \frac{1}{2}
   \langle\!\langle
    a_{\bm{r}_l                }^\dagger a_{\bm{r}_l+\bm{\delta}_{l:\kappa}}
   +\mathrm{H.c.}
   \rangle\!\rangle
  =\langle\!\langle
    a_{\bm{r}_l                }^\dagger a_{\bm{r}_l+\bm{\delta}_{l:\kappa}}
   \rangle\!\rangle,
   \label{E:<<D>>}
\end{align}
which we shall read as
the quantum average in the Holstein-Primakoff-boson vacuum $\langle\ \,\rangle_0'$
for the LSW formalism,
the quantum average in the Bogoliubov-boson (magnon) vacuum $\langle\ \,\rangle_0$
for the Wick-decomposition-based ISW (WDISW) formalism,
or
the temperature-$T$ thermal average $\langle\ \,\rangle_T$
for the Hartree-Fock-decomposition-based ISW (HFISW) formalism.
Note that all the averages \eqref{E:<<A>>}--\eqref{E:<<D>>} are independent of the site indices
$\bm{r}_l$ and $\bm{\delta}_{l:\kappa}$ by virtue of translation and rotation symmetries.
We decompose the $O(S^0)$ quartic Hamiltonian \eqref{E:H(0)2D} into quadratic terms
\begin{align}
   &
   \mathcal{H}^{(0)}
  \simeq
  J\sum_{\langle l,l' \rangle}
   \left\{
    \left[
     \left(
      \langle\!\langle\mathcal{A}\rangle\!\rangle
     -\frac{\langle\!\langle\mathcal{B}\rangle\!\rangle
           +\langle\!\langle\mathcal{D}\rangle\!\rangle}
           {2}
     \right)
     \cos
     \left(
      \phi_{\bm{r}_l}-\phi_{\bm{r}_{l'}}
     \right)
    +\frac{\langle\!\langle\mathcal{B}\rangle\!\rangle
          -\langle\!\langle\mathcal{D}\rangle\!\rangle}
          {2}
    \right]
   \right.
   \nonumber
   \allowdisplaybreaks \\
   &\quad\times
    \left(
     a_{\bm{r}_{l }}^\dagger a_{\bm{r}_{l }}
    +a_{\bm{r}_{l'}}^\dagger a_{\bm{r}_{l'}}
    \right)
   +\left[
     \left(
      \langle\!\langle\mathcal{B}\rangle\!\rangle
     -\frac{2\langle\!\langle\mathcal{A}\rangle\!\rangle
           + \langle\!\langle\mathcal{C}\rangle\!\rangle}
           {4}
     \right)
     \cos
     \left(
      \phi_{\bm{r}_l}-\phi_{\bm{r}_{l'}}
     \right)
    +\frac{2\langle\!\langle\mathcal{A}\rangle\!\rangle
          - \langle\!\langle\mathcal{C}\rangle\!\rangle}
          {4}
    \right]
   \nonumber
   \allowdisplaybreaks \\
   &\quad\times
    \left(
     a_{\bm{r}_{l }}^\dagger a_{\bm{r}_{l'}}^\dagger
    +a_{\bm{r}_{l }}         a_{\bm{r}_{l'}}
    \right)
   +\left[
     \left(
      \langle\!\langle\mathcal{D}\rangle\!\rangle
     -\frac{2\langle\!\langle\mathcal{A}\rangle\!\rangle
           + \langle\!\langle\mathcal{C}\rangle\!\rangle}
           {4}
     \right)
     \cos
     \left(
      \phi_{\bm{r}_l}-\phi_{\bm{r}_{l'}}
     \right)
    -\frac{2\langle\!\langle\mathcal{A}\rangle\!\rangle
          - \langle\!\langle\mathcal{C}\rangle\!\rangle}
          {4}
    \right]
   \nonumber
   \allowdisplaybreaks \\
   &\quad\times
    \left(
     a_{\bm{r}_{l }}^\dagger a_{\bm{r}_{l'}}
    +a_{\bm{r}_{l'}}^\dagger a_{\bm{r}_{l }}
    \right)
   -\left[
     \frac{\langle\!\langle\mathcal{B}\rangle\!\rangle
          +\langle\!\langle\mathcal{D}\rangle\!\rangle}
          {8}
     \cos
     \left(
      \phi_{\bm{r}_l}-\phi_{\bm{r}_{l'}}
     \right)
    +\frac{\langle\!\langle\mathcal{B}\rangle\!\rangle
          -\langle\!\langle\mathcal{D}\rangle\!\rangle}
          {8}
    \right]
   \nonumber
   \allowdisplaybreaks \\
   &\quad\times
    \left(
     a_{\bm{r}_{l }}^\dagger a_{\bm{r}_{l }}^\dagger
    +a_{\bm{r}_{l'}}^\dagger a_{\bm{r}_{l'}}^\dagger
    +a_{\bm{r}_{l }}         a_{\bm{r}_{l }}
    +a_{\bm{r}_{l'}}         a_{\bm{r}_{l'}}
    \right)
   -\left[
     \vphantom{
     \frac{\langle\!\langle\mathcal{B}\rangle\!\rangle
          +\langle\!\langle\mathcal{D}\rangle\!\rangle}
          {2}
     }
     \langle\!\langle\mathcal{A}\rangle\!\rangle^2
    +\langle\!\langle\mathcal{B}\rangle\!\rangle^2
    +\langle\!\langle\mathcal{D}\rangle\!\rangle^2
    -\left(
      2\langle\!\langle\mathcal{A}\rangle\!\rangle
     + \langle\!\langle\mathcal{C}\rangle\!\rangle
     \right)
    \right.
   \nonumber
   \allowdisplaybreaks \\
   &\quad\times
    \left.
     \frac{\langle\!\langle\mathcal{B}\rangle\!\rangle
          +\langle\!\langle\mathcal{D}\rangle\!\rangle}
          {2}
    \right]
     \cos
     \left(
      \phi_{\bm{r}_l}-\phi_{\bm{r}_{l'}}
     \right)
   \left.
   -\left(
     2\langle\!\langle\mathcal{A}\rangle\!\rangle
    - \langle\!\langle\mathcal{C}\rangle\!\rangle
    \right)
    \frac{\langle\!\langle\mathcal{B}\rangle\!\rangle
         -\langle\!\langle\mathcal{D}\rangle\!\rangle}
         {2}
   \right\}
   \equiv
    \mathcal{H}^{(0)}_{\mathrm{quad}}
   \label{E:H(0)quad2D}
\end{align}
to have the tractable SW Hamiltonian
$
   \mathcal{H}
  \simeq\mathcal{H}^{(2)}+\mathcal{H}^{(1)}+\mathcal{H}^{(0)}_{\mathrm{quad}}
  \equiv\mathcal{H}_{\mathrm{quad}}
$
and define its MSW extension
\begin{align}
   \widetilde{\mathcal{H}}_{\mathrm{quad}}
  \equiv
   \mathcal{H}_{\mathrm{quad}}
  +\mu
   \sum_{l=1}^L
   \left(
    S-a_{\bm{r}_l}^\dagger a_{\bm{r}_l}
   \right)
   \label{E:tildeHquad}.
\end{align}
For collinear antiferromagnets,
every local rotation angle $\phi_{\bm{r}_l}$ reads either $0$ or $\pi$ and
every relative rotation angle $\phi_{\bm{r}_l+\bm{\delta}_{l:\kappa}}-\phi_{\bm{r}_l}$
becomes $\pi$ [cf. \eqref{E:squarephi} and \eqref{E:honeycombphi}],
resulting that their bosonic Hamiltonians $\mathcal{H}=\sum_{m=-2}^\infty\mathcal{H}^{-m}$
commute with the total uniform magnetization
\begin{align}
   \mathcal{M}^z
  \equiv
   \sum_{l=1}^L
   S_{\bm{r}_l}^z
  =\sum_{l=1}^L
   \left(
    S_{\bm{r}_l}^{\tilde{z}}\cos\phi_{\bm{r}_l}
   -S_{\bm{r}_l}^{\tilde{x}}\sin\phi_{\bm{r}_l}
   \right)
   \label{E:Mz}
\end{align}
in each order and \eqref{E:H(0)quad2D} reduces to
\begin{align}
   &
   \mathcal{H}_{\mathrm{quad}}^{(0)}
  =J
   \sum_{\langle l,l' \rangle}
   \left[
    \vphantom{a_{\bm{r}_{l }}^\dagger a_{\bm{r}_{l'}}^\dagger}
    \bigl(
     \langle\!\langle\mathcal{A}\rangle\!\rangle
    -\langle\!\langle\mathcal{B}\rangle\!\rangle
    \bigr)^2
   -\bigl(
     \langle\!\langle\mathcal{A}\rangle\!\rangle
    -\langle\!\langle\mathcal{B}\rangle\!\rangle
    \bigr)
    \left(
     a_{\bm{r}_{l }}^\dagger a_{\bm{r}_{l }}
    +a_{\bm{r}_{l'}}^\dagger a_{\bm{r}_{l'}}
    -a_{\bm{r}_{l }}^\dagger a_{\bm{r}_{l'}}^\dagger
    -a_{\bm{r}_{l }}         a_{\bm{r}_{l'}}
    \right)
   \right]
   \nonumber
   \allowdisplaybreaks \\
   &\qquad\quad\!\!\!\!
  \equiv
   \mathcal{H}_{\mathrm{BL}}^{(0)}
   \label{E:H(0)BL2D}
\end{align}
without any contribution from $\mathcal{C}$ and $\mathcal{D}$ to cause a net change in
the magnetization \eqref{E:Mz}.
If we define an effective Hamiltonian of the bilinear version,
\begin{align}
   \widetilde{\mathcal{H}}_{\mathrm{BL}}
  \equiv
   \mathcal{H}_{\mathrm{BL}}
  +\mu
   \sum_{l=1}^L
   \left(
    S-a_{\bm{r}_l}^\dagger a_{\bm{r}_l}
   \right);\ 
   \mathcal{H}_{\mathrm{BL}}
  \equiv
   \mathcal{H}^{(2)}+\mathcal{H}^{(1)}+\mathcal{H}^{(0)}_{\mathrm{BL}},
   \label{E:tildeHBL}
\end{align}
and demand that
$\langle\!\langle\mathcal{A}\rangle\!\rangle$ and $\langle\!\langle\mathcal{B}\rangle\!\rangle$
be the self-consistent Hartree-Fock fields to diagonalize \eqref{E:tildeHBL} subject to
the constraint condition \eqref{E:collinearMSWconstraintHPB},
we obtain Takahashi's MSW thermodynamics of a square-lattice antiferromagnet \cite{T2494}.
What will happen if we apply this Takahashi scheme and its some variations to noncollinear
antiferromagnets?

   Let us define the geometric functions
\begin{align}
   &
   \gamma_{\bm{k}_\nu:\sigma}
  \equiv
   \frac{1}{z}
   \sum_{\kappa=1}^z
   e^{i\bm{k}_\nu\cdot\bm{\delta}_{l:\kappa}}
  =\frac{1}{z}
   \sum_{\kappa=1}^z
   e^{i\bm{k}_\nu\cdot\bm{\delta}_{p(n-1)+\sigma:\kappa}}
   \ 
   (\sigma=1,\cdots,p)
   \label{E:square&honeycomb&trianbulargamma_k}
\end{align}
for the square ($z=4$, $p=1$), honeycomb ($z=3$, $p=2$), and triangular ($z=6$, $p=1$) lattices
and
\begin{align}
   &
   \gamma_{\bm{k}_\nu:\sigma}
  \equiv
   \sum_{\tau,\tau'=1}^{p}\sum_{\sigma'=1}^3
   \frac{|\varepsilon_{\tau\tau'\sigma'}|}{2}
   u_{\bm{k}_\nu:\tau \sigma}
   u_{\bm{k}_\nu:\tau'\sigma}
   \cos\bm{k}_\nu\cdot\bm{a}_{\sigma'-1}
   \ 
   (\sigma=1,\cdots,p)
   \label{E:kagomegamma_k}
\end{align}
with $u_{\bm{k}_\nu:\sigma'\sigma}$ satisfying the eigenvalue equations
\begin{align}
   &
   \left[
   \begin{array}{ccc}
    -2\gamma_{\bm{k}_\nu:\sigma} &  \cos\bm{k}_\nu\cdot\bm{a}_2 &  \cos\bm{k}_\nu\cdot\bm{a}_1 \\
     \cos\bm{k}_\nu\cdot\bm{a}_2 & -2\gamma_{\bm{k}_\nu:\sigma} &  \cos\bm{k}_\nu\cdot\bm{a}_0 \\
     \cos\bm{k}_\nu\cdot\bm{a}_1 &  \cos\bm{k}_\nu\cdot\bm{a}_0 & -2\gamma_{\bm{k}_\nu:\sigma} \\
   \end{array}
   \right]
   \left[
   \begin{array}{c}
    u_{\bm{k}_\nu:1\sigma} \\
    u_{\bm{k}_\nu:2\sigma} \\
    u_{\bm{k}_\nu:3\sigma} \\
   \end{array}
   \right]
  =0,
   \ 
   \sum_{\sigma'=1}^3
   u_{\bm{k}_\nu:\sigma'\sigma}^2
  \equiv
   1\ (\sigma=1,\cdots,p)
   \label{E:kagomeu_k}
\end{align}
for the kagome ($z=4$, $p=3$) lattice.
We number the three eigenmodes for the kagome lattice in ascending order,
\begin{align}
   &
   \gamma_{\bm{k}_\nu:1}
  \equiv
   \frac{1}{4}
   \left(
    1
   -\sqrt{8\prod_{\tau=1}^3\cos\bm{k}_\nu\cdot\bm{a}_{\tau-1}+1}
   \right),\ 
   \gamma_{\bm{k}_\nu:2}
  \equiv
  -\frac{1}{2},\ 
   \nonumber
   \allowdisplaybreaks \\
   &
   \gamma_{\bm{k}_\nu:3}
  \equiv
   \frac{1}{4}
   \left(
    1
   +\sqrt{8\prod_{\tau=1}^3\cos\bm{k}_\nu\cdot\bm{a}_{\tau-1}+1}
   \right).
   \label{E:kagomegamma_k:sigma}
\end{align}
Intending to diagonalize the effective Hamiltonian \eqref{E:tildeHquad}, we define
Fourier transforms of the Holstein-Primakoff boson operators as
\begin{align}
   &
   a_{\bm{k}_\nu:\sigma}
  =\frac{1}{\sqrt{N}}\sum_{n=1}^N
   e^{-i\bm{k}_\nu\cdot\bm{r}_{p(n-1)+\sigma}}a_{\bm{r}_{p(n-1)+\sigma}}
   \ 
   (\nu=1,\cdots,N;\ \sigma=1,\cdots,p),
   \label{E:FT}
\end{align}
where the wavevector $\bm{k}_\nu$ runs over the full paramagnetic Brillouin zone, as is shown in
Fig. \ref{F:VariousLattices}.
While their Bogoliubov transforms, i.e., the ideal MSW creation and annihilation operators,
are defined according to their belonging lattice, we eventually obtain a diagonal Hamiltonian
in the unified form
\begin{align}
   \!\!
   \widetilde{\mathcal{H}}_{\mathrm{quad}}
  =\sum_{l=0}^2 E^{(l)}
  +\sum_{\nu=1}^N
   \sum_{\sigma=1}^p
   \varepsilon_{\bm{k}_\nu:\sigma}
   \alpha_{\bm{k}_\nu:\sigma}^\dagger\alpha_{\bm{k}_\nu:\sigma}
  +\mu LS,
   \label{E:tildeHquaddiag}
\end{align}
where $E^{(2)}$ is the classical ground-state energy,
$E^{(1)}$ and $E^{(0)}$ are its $O(S^1)$ quantum and $O(S^0)$ variational \cite{X6177,H014415}
corrections, respectively,
and $\alpha_{\bm{k}_\nu:\sigma}^\dagger$
creates a magnon of the $\sigma$ species with wavevector $\bm{k}_\nu$ at a cost of energy
$\varepsilon_{\bm{k}_\nu:\sigma}$.
Further details are given by each lattice.
The square ($z=4$, $p=1$) and honeycomb ($z=3$, $p=2$) lattices have the expressions
\begin{align}
   &
   E^{(2)}
 =-\frac{z}{2}LJS^2,\ 
   E^{(1)}
 =-\frac{z}{2}L
   \left(
    JS-\frac{\mu}{z}
   \right)
  +\frac{1}{2}
   \sum_{\nu=1}^N\sum_{\sigma=1}^p
   \varepsilon_{\bm{k}_\nu:\sigma},\
   \nonumber
   \allowdisplaybreaks \\
   &
   E^{(0)}
  =\frac{z}{2}LJ
   \left[
    \bigl(
     \langle\!\langle\mathcal{A}\rangle\!\rangle
    -\langle\!\langle\mathcal{B}\rangle\!\rangle
    \bigr)^2
   +\langle\!\langle\mathcal{A}\rangle\!\rangle
   -\langle\!\langle\mathcal{B}\rangle\!\rangle
   \right],
   \label{E:square&honeycombE(l)}
   \allowdisplaybreaks \\
   &
   \omega_{\bm{k}_\nu:\sigma}
  \equiv
   \frac{\varepsilon_{\bm{k}_\nu:\sigma}}
        {zJ
         \bigl(
          S
         -\langle\!\langle\mathcal{A}\rangle\!\rangle
         +\langle\!\langle\mathcal{B}\rangle\!\rangle
         \bigr)}
   \equiv
    \sqrt{\left[
           1
          -\frac{\mu}
                {zJ
                 \bigl(
                  S
                 -\langle\!\langle\mathcal{A}\rangle\!\rangle
                 +\langle\!\langle\mathcal{B}\rangle\!\rangle
                 \bigr)}
          \right]^2
         -\left|
           \gamma_{\bm{k}_\nu:\sigma}
          \right|^2},
   \label{E:square&honeycombomega}
   \allowdisplaybreaks \\
   &
   \alpha_{ \bm{k}_\nu:\sigma}
  =a_{ \bm{k}_\nu:p+1-\sigma}
   \mathrm{cosh}\vartheta_{\bm{k}_\nu:\sigma}
  +a_{-\bm{k}_\nu:\sigma}^\dagger
   \frac{\gamma_{\bm{k}_\nu:\sigma}  }{|\gamma_{\bm{k}_\nu:\sigma}|}
   \mathrm{sinh}\vartheta_{\bm{k}_\nu:\sigma},\
   \nonumber
   \allowdisplaybreaks \\
   &
   \alpha_{-\bm{k}_\nu:\sigma}^\dagger
  =a_{ \bm{k}_\nu:\sigma}
   \frac{\gamma_{\bm{k}_\nu:\sigma}}{|\gamma_{\bm{k}_\nu:\sigma}|}
   \mathrm{sinh}\vartheta_{\bm{k}_\nu:\sigma}
  +a_{-\bm{k}_\nu:p+1-\sigma}^\dagger
   \mathrm{cosh}\vartheta_{\bm{k}_\nu:\sigma},
   \nonumber
   \allowdisplaybreaks \\
   &
   \mathrm{cosh}2\vartheta_{\bm{k}_\nu:\sigma}
  =\frac{1}{\omega_{\bm{k}_\nu:\sigma}}
   \left[
    1
   -\frac{\mu}
         {zJ
          \bigl(
           S
          -\langle\!\langle\mathcal{A}\rangle\!\rangle
          +\langle\!\langle\mathcal{B}\rangle\!\rangle
          \bigr)}
   \right],\ 
   \mathrm{sinh}2\vartheta_{\bm{k}_\nu:\sigma}
  =\frac{|\gamma_{\bm{k}_\nu:\sigma}|}{\omega_{\bm{k}_\nu:\sigma}}
   \label{E:square&honeycombBT}
\end{align}
in common.
The triangular ($z=6$, $p=1$) and kagome ($z=4$, $p=3$) lattices have the expressions
\begin{align}
   &
   E^{(2)}
 =-\frac{z}{2}L\frac{J}{2}S^2,\ 
   E^{(1)}
 =-\frac{z}{2}L
   \left(
    \frac{J}{2}S-\frac{\mu}{z}
   \right)
  +\frac{1}{2}
   \sum_{\nu=1}^N\sum_{\sigma=1}^p
   \varepsilon_{\bm{k}_\nu:\sigma},\ 
   \nonumber
   \allowdisplaybreaks \\
   &
   E^{(0)}
  =\frac{z}{2}L\frac{J}{2}
   \left[
    \left(
     \langle\!\langle\mathcal{A}\rangle\!\rangle
    -\frac{3\langle\!\langle\mathcal{B}\rangle\!\rangle}{2}
    +\frac{ \langle\!\langle\mathcal{D}\rangle\!\rangle}{2}
    \right)^2
   -\frac{3
          \bigl(
           \langle\!\langle\mathcal{B}\rangle\!\rangle
          -\langle\!\langle\mathcal{D}\rangle\!\rangle
          \bigr)^2}
         {4}
   \right.
   \nonumber
   \allowdisplaybreaks \\
   &\quad\quad\quad\!\!
   \left.
   \vphantom
   {\left(
     \langle\!\langle\mathcal{A}\rangle\!\rangle
    -\frac{3\langle\!\langle\mathcal{B}\rangle\!\rangle}{2}
    +\frac{ \langle\!\langle\mathcal{D}\rangle\!\rangle}{2}
    \right)^2}
   -\frac{\langle\!\langle\mathcal{B}\rangle\!\rangle
          \bigl(
           \langle\!\langle\mathcal{B}\rangle\!\rangle
          -\langle\!\langle\mathcal{C}\rangle\!\rangle
          \bigr)}
         {2}
   +\frac{3\langle\!\langle\mathcal{D}\rangle\!\rangle
          \bigl(
           \langle\!\langle\mathcal{D}\rangle\!\rangle
          -\langle\!\langle\mathcal{C}\rangle\!\rangle
          \bigr)}
         {2}
   +\langle\!\langle\mathcal{A}\rangle\!\rangle
   -\frac{3\langle\!\langle\mathcal{B}\rangle\!\rangle}{2}
   +\frac{ \langle\!\langle\mathcal{D}\rangle\!\rangle}{2}
   \right],
   \label{E:triangular&kagomeE(l)}
   \allowdisplaybreaks \\
   &
   \omega_{\bm{k}_\nu:\sigma}
  \equiv
   \frac{\varepsilon_{\bm{k}_\nu:\sigma}}{zJS}
  \equiv
   \sqrt{\left(
          \frac{1}{2}
         +\frac{\gamma_{\bm{k}_\nu:\sigma}}{4}
         +F
         +F'\gamma_{\bm{k}_\nu:\sigma}
         -\frac{\mu}{zJS}
         \right)^2
        -\left(
          \frac{3\gamma_{\bm{k}_\nu:\sigma}}{4}
         +G
         +G'\gamma_{\bm{k}_\nu:\sigma}
         \right)^2},
   \label{E:triangular&kagomeomega}
   \allowdisplaybreaks \\
   &
   \alpha_{ \bm{k}_\nu:\sigma}
  =\sum_{\tau=1}^3
   u_{ \bm{k}_\nu:\tau\sigma}
   \left(
    a_{ \bm{k}_\nu:\tau}
    \mathrm{cosh}\vartheta_{\bm{k}_\nu:\sigma}
   +a_{-\bm{k}_\nu:\tau}^\dagger
    \mathrm{sinh}\vartheta_{\bm{k}_\nu:\sigma}
   \right),\ 
   \nonumber
   \allowdisplaybreaks \\
   &
   \alpha_{-\bm{k}_\nu:\sigma}^\dagger
  =\sum_{\tau=1}^3
   u_{ \bm{k}_\nu:\tau\sigma}
   \left(
    a_{ \bm{k}_\nu:\tau}
    \mathrm{sinh}\vartheta_{\bm{k}_\nu:\sigma}
   +a_{-\bm{k}_\nu:\tau}^\dagger
    \mathrm{cosh}\vartheta_{\bm{k}_\nu:\sigma}
   \right),
   \nonumber
   \allowdisplaybreaks \\
   &
   \mathrm{cosh}2\vartheta_{\bm{k}_\nu:\sigma}
  =\frac{1}{\omega_{\bm{k}_\nu:\sigma}}
   \left(
    \frac{1}{2}
   +\frac{\gamma_{\bm{k}_\nu:\sigma}}{4}
   +F
   +F'\gamma_{\bm{k}_\nu:\sigma}
   -\frac{\mu}{zJS}
   \right),\ 
   \nonumber
   \allowdisplaybreaks \\
   &
   \mathrm{sinh}2\vartheta_{\bm{k}_\nu:\sigma}
  =\frac{1}{\omega_{\bm{k}_\nu:\sigma}}
   \left(
    \frac{3\gamma_{\bm{k}_\nu:\sigma}}{4}
   +G
   +G'\gamma_{\bm{k}_\nu:\sigma}
   \right)
   \label{E:triangular&kagomeBT}
\end{align}
in common with the abbreviations
\begin{align}
   &
   F
  \equiv
  -\frac{ \langle\!\langle\mathcal{A}\rangle\!\rangle}{2S}
  +\frac{3\langle\!\langle\mathcal{B}\rangle\!\rangle}{4S}
  -\frac{ \langle\!\langle\mathcal{D}\rangle\!\rangle}{4S},\ 
   F'
  \equiv
  -\frac{ \langle\!\langle\mathcal{A}\rangle\!\rangle}{4S}
  +\frac{3\langle\!\langle\mathcal{C}\rangle\!\rangle}{8S}
  -\frac{ \langle\!\langle\mathcal{D}\rangle\!\rangle}{2S},\ 
   \nonumber
   \allowdisplaybreaks \\
   &
   G
  \equiv
   \frac{ \langle\!\langle\mathcal{B}\rangle\!\rangle}{8S}
  -\frac{3\langle\!\langle\mathcal{D}\rangle\!\rangle}{8S},\ 
   G'
  \equiv
   \frac{3\langle\!\langle\mathcal{A}\rangle\!\rangle}{4S}
  -\frac{ \langle\!\langle\mathcal{B}\rangle\!\rangle}{2S}
  -\frac{ \langle\!\langle\mathcal{C}\rangle\!\rangle}{8S}
   \label{E:triangular&kagomeFF'GG'}
\end{align}
and the understanding that $u_{ \bm{k}_\nu:11}$ for the triangular lattice be unity.

\subsection{Variational single-constraint modified-spin-wave thermodynamics}
\label{SS:VSCMSWThD}

   Variational MISW---modified WDISW (MWDISW) and HFISW (MHFISW)---thermodynamics can be formulated
in terms of the self-consistent fields
$\langle\!\langle\mathcal{A}\rangle\!\rangle$ to $\langle\!\langle\mathcal{D}\rangle\!\rangle$,
which depend on how the SWs are interacting.
The MSW thermal distribution function reads
\begin{align}
   \langle
    \alpha_{\bm{k}_\nu:\sigma}^\dagger \alpha_{\bm{k}_\nu:\sigma}
   \rangle_T
   &
  =\frac
   {\mathrm{Tr}
    \bigl[
     e^{-\varepsilon_{\bm{k}_\nu:\sigma}
        \alpha_{\bm{k}_\nu:\sigma}^\dagger \alpha_{\bm{k}_\nu:\sigma}
       /k_{\mathrm{B}}T}
     \alpha_{\bm{k}_\nu:\sigma}^\dagger \alpha_{\bm{k}_\nu:\sigma}
    \bigr]}
   {\mathrm{Tr}
    \bigl[
     e^{-\varepsilon_{\bm{k}_\nu:\sigma}
         \alpha_{\bm{k}_\nu:\sigma}^\dagger \alpha_{\bm{k}_\nu:\sigma}
       /k_{\mathrm{B}}T}
    \bigr]}
  =\frac{1}
   {e^{\varepsilon_{\bm{k}_\nu:\sigma}/k_{\mathrm{B}}T}-1}
  \equiv
   \bar{n}_{\bm{k}_\nu:\sigma}
   \label{E:variationalMSWnk}
\end{align}
with $\varepsilon_{\bm{k}_\nu:\sigma}$ containing part or all of
$\langle\!\langle\mathcal{A}\rangle\!\rangle$ to $\langle\!\langle\mathcal{D}\rangle\!\rangle$.
Every time we encounter the double angle brackets
$\langle\!\langle\mathcal{A}\rangle\!\rangle$ to $\langle\!\langle\mathcal{D}\rangle\!\rangle$,
we read them according to the scheme of the time,
\begin{align}
   &
   \langle\mathcal{A}\rangle_0'
  =\langle\mathcal{B}\rangle_0'
  =0
   \hspace*{-2.0cm}
   &
   (\mathrm{MLSW});
   &
   \label{E:square&honeycomb<AandB>'0}
   \allowdisplaybreaks \\
   &
   \langle\mathcal{A}\rangle_0
  =\frac{1}{N}\sum_{\nu=1}^N
   \frac{1}{p}\sum_{\sigma=1}^p
   \frac{1}{2\omega_{\bm{k}_\nu:\sigma}}
   \left[
    1
   -\frac{\mu}
         {zJ
          \bigl(
           S
          -\langle\!\langle\mathcal{A}\rangle\!\rangle
          +\langle\!\langle\mathcal{B}\rangle\!\rangle
          \bigr)}
   \right]
  -\frac{1}{2},
   \hspace*{-2.0cm}
   &
   &
   \nonumber
   \allowdisplaybreaks \\
   &
   \langle\mathcal{B}\rangle_0
  =\frac{1}{N}\sum_{\nu=1}^N
   \frac{1}{p}\sum_{\sigma=1}^p
   \frac{|\gamma_{\bm{k}_\nu:\sigma}|^2}{2\omega_{\bm{k}_\nu:\sigma}}
   \hspace*{-5.0cm}
   &
   (\mathrm{MWDISW});
   &
   \label{E:square&honeycomb<AandB>0}
   \allowdisplaybreaks \\
   &
   \langle\mathcal{A}\rangle_T
  =\frac{1}{N}\sum_{\nu=1}^N
   \frac{1}{p}\sum_{\sigma=1}^p
   \frac{1}{\omega_{\bm{k}_\nu:\sigma}}
   \left[
    1
   -\frac{\mu}
         {zJ
          \bigl(
           S
          -\langle\!\langle\mathcal{A}\rangle\!\rangle
          +\langle\!\langle\mathcal{B}\rangle\!\rangle
          \bigr)}
   \right]
   \left(
    \bar{n}_{\bm{k}_\nu:\sigma}
   +\frac{1}{2}
   \right)
   -\frac{1}{2},
   \hspace*{-2.0cm}
   &
   &
   \nonumber
   \allowdisplaybreaks \\
   &
   \langle\mathcal{B}\rangle_T
  =\frac{1}{N}\sum_{\nu=1}^N
   \frac{1}{p}\sum_{\sigma=1}^p
   \frac{|\gamma_{\bm{k}_\nu:\sigma}|^2}{\omega_{\bm{k}_\nu:\sigma}}
   \left(
    \bar{n}_{\bm{k}_\nu:\sigma}
   +\frac{1}{2}
   \right)
   \hspace*{-2.0cm}
   &
   (\mathrm{MHFISW})
   &
   \label{E:square&honeycomb<AandB>T}
\end{align}
for the square ($z=4$, $p=1$) and honeycomb ($z=3$, $p=2$) lattices and
\begin{align}
   &\!\!\!
   \langle\mathcal{A}\rangle_0'
  =\langle\mathcal{B}\rangle_0'
  =\langle\mathcal{C}\rangle_0'
  =\langle\mathcal{D}\rangle_0'
  =0
   \hspace*{-4.0cm}
   &
   (\mathrm{MLSW});
   &
   \label{E:triangular&kagome<AtoD>'0}
   \allowdisplaybreaks \\
   &\!\!\!
   \langle\mathcal{A}\rangle_0
  =\frac{1}{N}\sum_{\nu=1}^N
   \frac{1}{p}\sum_{\sigma=1}^p
   \frac{1}{2\omega_{\bm{k}_\nu:\sigma}}
   \left(
    \frac{1}{2}
   +\frac{\gamma_{\bm{k}_\nu:\sigma}}{4}
   +F
   +F'\gamma_{\bm{k}_\nu:\sigma}
   -\frac{\mu}{zJS}
   \right)
  -\frac{1}{2},
   \hspace*{-4.0cm}
   &
   &
   \nonumber
   \allowdisplaybreaks \\
   &\!\!\!
   \langle\mathcal{B}\rangle_0
  =\frac{1}{N}\sum_{\nu=1}^N
   \frac{1}{p}\sum_{\sigma=1}^p
   \frac{\gamma_{\bm{k}_\nu:\sigma}}{2\omega_{\bm{k}_\nu:\sigma}}
   \left(
    \frac{3\gamma_{\bm{k}_\nu:\sigma}}{4}
   +G
   +G'\gamma_{\bm{k}_\nu:\sigma}
   \right),
   \hspace*{-4.0cm}
   &
   &
   \nonumber
   \allowdisplaybreaks \\
   &\!\!\!
   \langle\mathcal{C}\rangle_0
  =\frac{1}{N}\sum_{\nu=1}^N
   \frac{1}{p}\sum_{\sigma=1}^p
   \frac{1}{2\omega_{\bm{k}_\nu:\sigma}}
   \left(
    \frac{3\gamma_{\bm{k}_\nu:\sigma}}{4}
   +G
   +G'\gamma_{\bm{k}_\nu:\sigma}
   \right),
   \hspace*{-4.0cm}
   &
   &
   \nonumber
   \allowdisplaybreaks \\
   &\!\!\!
   \langle\mathcal{D}\rangle_0
  =\frac{1}{N}\sum_{\nu=1}^N
   \frac{1}{p}\sum_{\sigma=1}^p
   \left[
    \frac{\gamma_{\bm{k}_\nu:\sigma}}{2\omega_{\bm{k}_\nu:\sigma}}
    \left(
     \frac{1}{2}
    +\frac{\gamma_{\bm{k}_\nu:\sigma}}{4}
    +F
    +F'\gamma_{\bm{k}_\nu:\sigma}
    -\frac{\mu}{zJS}
    \right)
   -\frac{\gamma_{\bm{k}_\nu:\sigma}}{2}
   \right]
   \hspace*{-3.0cm}
   &
   &
   \nonumber
   \allowdisplaybreaks \\
   &\!\!\!
   \hspace*{-4.0cm}
   &
   (\mathrm{MWDISW});
   &
   \label{E:triangular&kagome<AtoD>0}
   \allowdisplaybreaks \\
   &\!\!\!
   \langle\mathcal{A}\rangle_T
  =\frac{1}{N}\sum_{\nu=1}^N
   \frac{1}{p}\sum_{\sigma=1}^p
   \frac{1}{\omega_{\bm{k}_\nu:\sigma}}
   \left(
    \frac{1}{2}
   +\frac{\gamma_{\bm{k}_\nu:\sigma}}{4}
   +F
   +F'\gamma_{\bm{k}_\nu:\sigma}
   -\frac{\mu}{zJS}
   \right)
   \left(
    \bar{n}_{\bm{k}_\nu:\sigma}
   +\frac{1}{2}
   \right)
  -\frac{1}{2},
   \hspace*{-4.0cm}
   &
   &
   \nonumber
   \allowdisplaybreaks \\
   &\!\!\!
   \langle\mathcal{B}\rangle_T
  =\frac{1}{N}\sum_{\nu=1}^N
   \frac{1}{p}\sum_{\sigma=1}^p
   \frac{\gamma_{\bm{k}_\nu:\sigma}}{\omega_{\bm{k}_\nu:\sigma}}
   \left(
    \frac{3\gamma_{\bm{k}_\nu:\sigma}}{4}
   +G
   +G'\gamma_{\bm{k}_\nu:\sigma}
   \right)
   \left(
    \bar{n}_{\bm{k}_\nu:\sigma}
   +\frac{1}{2}
   \right),
   \hspace*{-4.0cm}
   &
   &
   \nonumber
   \allowdisplaybreaks \\
   &\!\!\!
   \langle\mathcal{C}\rangle_T
  =\frac{1}{N}
              \sum_{\nu=1}^N
   \frac{1}{p}\sum_{\sigma=1}^p
   \frac{1}{\omega_{\bm{k}_\nu:\sigma}}
   \left(
    \frac{3\gamma_{\bm{k}_\nu:\sigma}}{4}
   +G
   +G'\gamma_{\bm{k}_\nu:\sigma}
   \right)
   \left(
    \bar{n}_{\bm{k}_\nu:\sigma}
   +\frac{1}{2}
   \right),
   \hspace*{-4.0cm}
   &
   &
   \nonumber
   \allowdisplaybreaks \\
   &\!\!\!
   \langle\mathcal{D}\rangle_T
  =\frac{1}{N}\sum_{\nu=1}^N
   \frac{1}{p}\sum_{\sigma=1}^p
   \left[
    \frac{\gamma_{\bm{k}_\nu:\sigma}}{\omega_{\bm{k}_\nu:\sigma}}
    \left(
     \frac{1}{2}
    +\frac{\gamma_{\bm{k}_\nu:\sigma}}{4}
    +F
    +F'\gamma_{\bm{k}_\nu:\sigma}
    -\frac{\mu}{zJS}
    \right)
    \left(
     \bar{n}_{\bm{k}_\nu:\sigma}
    +\frac{1}{2}
    \right)
   -\frac{\gamma_{\bm{k}_\nu:\sigma}}{2}
   \right]
   \hspace*{-4.0cm}
   &
   &
   \nonumber
   \allowdisplaybreaks \\
   &\!\!\!
   \hspace*{-4.0cm}
   &
   (\mathrm{MHFISW})
   &
   \label{E:triangular&kagome<AtoD>T}
\end{align}
for the triangular ($z=6$, $p=1$) and kagome ($z=4$, $p=3$) lattices, where they each still contain
$\langle\!\langle\mathcal{A}\rangle\!\rangle$ and $\langle\!\langle\mathcal{B}\rangle\!\rangle$
(collinear antiferromagnets) or
$\langle\!\langle\mathcal{A}\rangle\!\rangle$ to $\langle\!\langle\mathcal{D}\rangle\!\rangle$
(noncollinear antiferromagnets)
to be self-consistently determined.
The constraint condition is then written as
\begin{align}
   \langle
    \mathcal{M}^{\tilde{z}}
   \rangle_T
  =L
   \left(
    S-\langle\mathcal{A}\rangle_T
   \right)
  =0.
   \label{E:variationalMSWconstraint<<A>>}
\end{align}
In the MHFISW scheme, we solve the simultaneous equations
\eqref{E:square&honeycomb<AandB>T} plus \eqref{E:variationalMSWconstraint<<A>>}
for
$\langle\mathcal{A}\rangle_T$, $\langle\mathcal{B}\rangle_T$, and $\mu$
and
\eqref{E:triangular&kagome<AtoD>T} plus \eqref{E:variationalMSWconstraint<<A>>}
for
$\langle\mathcal{A}\rangle_T$ to $\langle\mathcal{D}\rangle_T$ and $\mu$.
In the MWDISW scheme, we solve the simultaneous equations
\eqref{E:square&honeycomb<AandB>0} plus \eqref{E:variationalMSWconstraint<<A>>}
for
$\langle\mathcal{A}\rangle_0$, $\langle\mathcal{B}\rangle_0$, and $\mu$
and
\eqref{E:triangular&kagome<AtoD>0} plus \eqref{E:variationalMSWconstraint<<A>>}
for
$\langle\mathcal{A}\rangle_0$ to $\langle\mathcal{D}\rangle_0$ and $\mu$.
In the MLSW scheme,
with \eqref{E:square&honeycomb<AandB>'0} or \eqref{E:triangular&kagome<AtoD>'0} in mind,
we have only to solve \eqref{E:variationalMSWconstraint<<A>>} for $\mu$.
We define the internal energy $E$ as
$
   \sum_{m=2}^4
   \langle
   \mathcal{H}^{\left(\frac{m}{2}\right)}
   \rangle_T
$
and
$
   \sum_{m=0}^4
   \langle
   \mathcal{H}^{\left(\frac{m}{2}\right)}
   \rangle_T
$
in the MLSW and MISW schemes, respectively, where the thermal average of the quartic Hamiltonian
is evaluated through the use of the Bloch-De Dominicis theorem \cite{B459}.
The thermal average of $\mathcal{H}^{\left(\frac{m}{2}\right)}$ with an odd $m$ generally vanishes,
while that of $\mathcal{H}^{\left(\frac{m}{2}\right)}$ with an even $m$ reads
\begin{align}
   &
   \langle\mathcal{H}^{(2)}\rangle_T
 =-\frac{z}{2}LJS^2,\ 
   \langle\mathcal{H}^{(1)}\rangle_T
  =zLJS
   \bigl(
    \langle\mathcal{A}\rangle_T
   -\langle\mathcal{B}\rangle_T
   \bigr),\ 
   \nonumber
   \allowdisplaybreaks \\
   &
   \langle\mathcal{H}^{(0)}\rangle_T
 =-\frac{z}{2}LJ
   \bigl(
    \langle\mathcal{A}\rangle_T
   -\langle\mathcal{B}\rangle_T
   \bigr)^2
   \label{E:square&honeycomb<H(l)>T}
\end{align}
for the square and honeycomb lattices and
\begin{align}
   &
   \langle\mathcal{H}^{(2)}\rangle_T
 =-\frac{z}{2}L\frac{J}{2}S^2,\ 
   \langle\mathcal{H}^{(1)}\rangle_T
  =zL\frac{J}{2}S
   \left(
    \langle\mathcal{A}\rangle_T
   -\frac{3\langle\mathcal{B}\rangle_T}{2}
   +\frac{ \langle\mathcal{D}\rangle_T}{2}
   \right),\ 
   \nonumber
   \allowdisplaybreaks \\
   &
   \langle\mathcal{H}^{(0)}\rangle_T
 =-\frac{z}{2}L\frac{J}{2}
   \left[
    \left(
     \langle\mathcal{A}\rangle_T
    -\frac{3\langle\mathcal{B}\rangle_T}{2}
    +\frac{ \langle\mathcal{D}\rangle_T}{2}
    \right)^2
   -\frac{3
          \bigl(
           \langle\mathcal{B}\rangle_T-\langle\mathcal{D}\rangle_T
          \bigr)^2}
         {4}
   \right.
   \nonumber
   \allowdisplaybreaks \\
   &\qquad\qquad
   \left.
   -\frac{\langle\!\langle\mathcal{B}\rangle\!\rangle
          \bigl(
           \langle\!\langle\mathcal{B}\rangle\!\rangle
          -\langle\!\langle\mathcal{C}\rangle\!\rangle
          \bigr)}
         {2}
   +\frac{3\langle\!\langle\mathcal{D}\rangle\!\rangle
          \bigl(
           \langle\!\langle\mathcal{D}\rangle\!\rangle
          -\langle\!\langle\mathcal{C}\rangle\!\rangle
          \bigr)}
         {2}
    \vphantom{\bigl(
               \langle\mathcal{A}\rangle_T
              -\frac{3\langle\mathcal{B}\rangle_T}{2}
              +\frac{ \langle\mathcal{D}\rangle_T}{2}
    \bigr)^2}
   \right]
   \label{E:triangular&kagome<H(l)>T}
\end{align}
for the triangular and kagome lattices.
\begin{figure}
\begin{flushright}
\includegraphics[width=130mm]{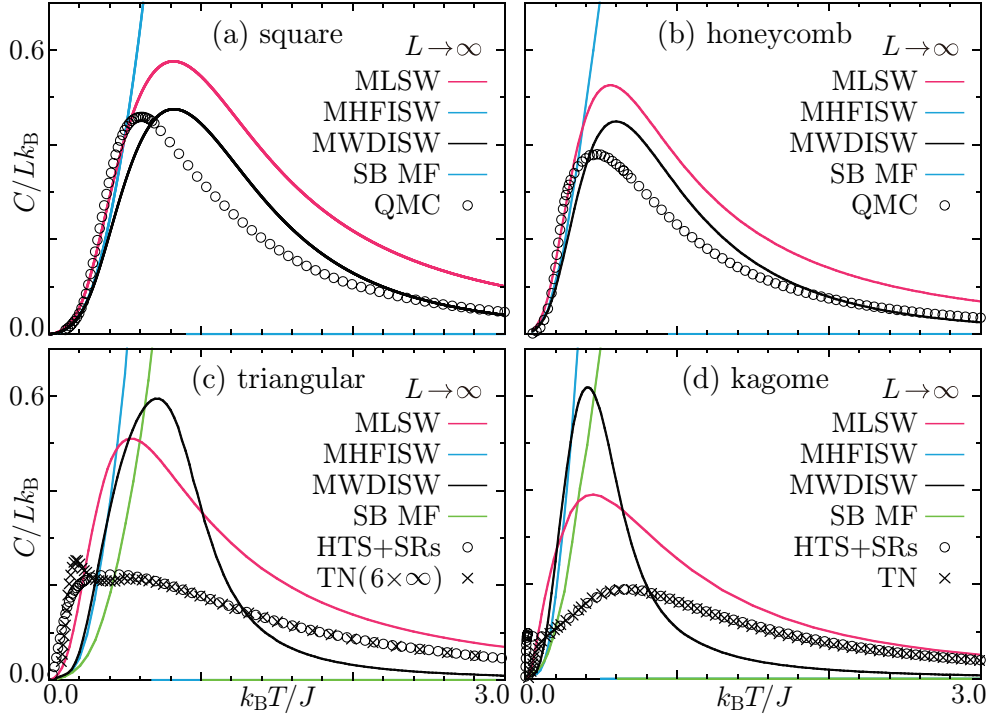}
\end{flushright}
\vspace{-7mm}
\caption{Traditional SC-MSW calculations of the specific heat $C\equiv\partial E/\partial T$
         [\eqref{E:square&honeycomb<H(l)>T}, \eqref{E:triangular&kagome<H(l)>T}]
         as a function of temperature for the $S=\frac{1}{2}$ Hamiltonian \eqref{E:HeisenbergH}
         on the $L\rightarrow\infty$ square (a), honeycomb (b), triangular (c), and kagome (d)
         lattices in comparison with quantum Monte Carlo calculations (QMC),
         optimal interpolations of Pad\'e approximants to high-temperature series subject to
         the energy and entropy sum rules ($\mathrm{HTS}+\mathrm{SRs}$)
         \cite{B134409,M014417,B140403},
         and tensor-network-based renormalization-group calculations (TN) \cite{C1545,C140404}.
         The findings referred to as TN in (d) \cite{C1545} are obtained on the infinite
         two-dimensional kagome lattice, while those as TN ($6\times\infty$) in (c) \cite{C140404}
         are extrapolations to the infinite cylindrical triangular lattice with hexagonal ends
         aslant.
         Schwinger-boson mean-field calculations (SB MF) are also shown for reference.}
\label{F:ConvSC-MSW&SB&QMC&TNofCin2D}
\end{figure}

   We show in Fig. \ref{F:ConvSC-MSW&SB&QMC&TNofCin2D} the thus-obtained MSW findings for
the specific heat $C\equiv\partial E/\partial T$, together with those obtained by
Auerbach-Arovas' Schwinger-boson mean-field theory \cite{Y064426,A617,A316,S5028},
in comparison with modern numerical approaches capable of touching bulk properties.
Without any frustration, we can calculate the internal energy for sufficiently large planes by
a quantum Monte Carlo method.
What Bernu and Misguich \cite{B134409,M014417} call the entropy method is a stable specific-heat
interpolation scheme between low and high temperatures intending to improve the convergence of
the high-temperature series expansion with the help of two sum rules on the energy and entropy.
This technique allows us to compute accurately the specific heat in the thermodynamic limit
possibly down to absolute zero, which is never the case with a direct Pad\'e analysis of
the series.
Various state-of-the-art renormalization group techniques based on tensor-network states
\cite{SpringerLNP964} are also a potentially powerful tool applicable to thermodynamics of
frustrated spin models without suffering from a negative-sign problem.
Linearized thermal tensor renormalization,
whether on the conventional Trotter-Suzuki-discretized linear quasicontinuous grid in
inverse-temperature $\beta\equiv 1/k_{\textrm{B}}T$ \cite{L127202},
on some interleaved $\beta$ grids \cite{C031082},
or in a Trotter-error-free manner based on the numerically exact Taylor series expansion of
the whole density operator $e^{-\beta\mathcal{H}}$ \cite{C161104},
indeed reveals the thermal quantities of quantum magnets on the linear-chain, square, honeycomb,
and triangular lattices of finite length and/or width
with pronounced precision and high efficiency.
However, elimination of the boundary effects is tricky in two dimensions \cite{C140404}
and any linearization of the kagome lattice in this context seems to be difficult.
Some algorithms address an infinite system directly without dealing with boundary effects or
finite-size corrections \cite{J250602}.
By mapping a two-dimensional quantum lattice model into a three-dimensional closed tensor network
and contracting the three-dimensional brick-wall tensor network with the imaginary time length
corresponding to temperature in question \cite{R134429}, finite-temperature properties of
even an infinite two-dimensional kagome antiferromagnet can be calculated \cite{C1545}.
However, the full update---contraction of the whole tensor network via the infinite time-evolving
block decimation algorithm \cite{O155117}---is awfully time-consuming and has to be reduced to
a cluster update \cite{C1545,L032114} in most cases, through the use of the Bethe approximation
\cite{C1545,L032114,L195137} for instance.
Selective discard of some or all of the loops in the original lattice
and inevitable truncation of the bond dimensions in contracting tensors
may yield significant errors near a critical point \cite{R134429} and/or at low temperatures
\cite{C140404}.

   With all these in mind, let us observe the traditional, here in the sense of constraining
the staggered magnetization to be zero, SC-MSW thermodynamics of collinear antiferromagnets,
Figs. \ref{F:ConvSC-MSW&SB&QMC&TNofCin2D}(a) and \ref{F:ConvSC-MSW&SB&QMC&TNofCin2D}(b), first.
While the lowest-order calculations, MLSW findings, succeed in designing antiferromagnetic peaks
of $C$, they are far from precise at low temperatures \cite{Y094412}.
The low-temperature quantitativity is significantly improved by taking account of the SW
interaction $\mathcal{H}^{(0)}$.
The MHFISW findings are highly precise at sufficiently low temperatures \cite{Y094412},
while they completely fail to reproduce the overall temperature dependences.
The worst of them is an artificial phase transition of the first order to
the trivial paramagnetic solution at a certain finite temperature.
The specific heat jumps down to zero when $k_{\mathrm{B}}T/J$ reaches $0.9108$ and $0.9568$
for the square and honeycomb lattices, respectively, where the bond order parameter
$\langle\mathcal{B}\rangle_T$ vanishes, satisfying
\begin{align}
   \left(
    1-\frac{\mu}{zJ\langle\mathcal{B}\rangle_T}
   \right)
   \langle\mathcal{B}\rangle_T
  =\frac{k_{\mathrm{B}}T}{zJ}
   \mathrm{ln}
   \left(
    1+\frac{1}{S}
   \right).
   \label{E:collinearPhTtoParamag}
\end{align}
The SB MF formulation \cite{A617,A316,S5028} coincides with the MHFISW thermodynamics
\cite{T2494,Y094412,T1524,Y064426} except for an overall numerical factor in each thermal
quantity.
By correcting what Arovas and Auerbach call ``the overcounting of the number of independent boson
degrees of freedom" \cite{A316,S5028}, the two approaches yield exactly the same specific heat
as a function of temperature.
The two schemes are no longer degenerate with each other in noncollinear antiferromagnets,
as is demonstrated in
Figs. \ref{F:ConvSC-MSW&SB&QMC&TNofCin2D}(c) and \ref{F:ConvSC-MSW&SB&QMC&TNofCin2D}(d).

   In order to retain the MHFISW precise low-temperature findings by all means and connect them
naturally with the correct high-temperature asymptotics,
we bring SWs into interaction in a different manner from the Hartree-Fock approximation.
A new treatment of the $O(S^0)$ quartic Hamiltonian $\mathcal{H}^{(0)}$ consists of applying
the Wick theorem \cite{W268} based on the magnon operators $\alpha_{\bm{k}_\nu:\sigma}^\dagger$
and $\alpha_{\bm{k}_\nu:\sigma}$ to it and neglecting the residual normal-ordered interaction
$:\!\mathcal{H}^{(0)}\!:$.
Then we have the bilinear Hamiltonian \eqref{E:H(0)BL2D} with
$\langle\!\langle\mathcal{A}\rangle\!\rangle$ and $\langle\!\langle\mathcal{B}\rangle\!\rangle$
read as the SW ground-state expectation values
$\langle\mathcal{A}\rangle_0$ and $\langle\mathcal{B}\rangle_0$
given by Eq. \eqref{E:square&honeycomb<AandB>0}
for the collinear antiferromagnets and
the quadratic Hamiltonian \eqref{E:H(0)quad2D} with
$\langle\!\langle\mathcal{A}\rangle\!\rangle$ to $\langle\!\langle\mathcal{D}\rangle\!\rangle$
read as $\langle\mathcal{A}\rangle_0$ to $\langle\mathcal{D}\rangle_0$
given by Eq. \eqref{E:triangular&kagome<AtoD>0}.
We continue to observe
Figs. \ref{F:ConvSC-MSW&SB&QMC&TNofCin2D}(a) and \ref{F:ConvSC-MSW&SB&QMC&TNofCin2D}(b).
Unlike MHFISWs, MWDISWs are free from thermal breakdown and succeed in reproducing
the Schottky-like peak of $C$ in a pretty good manner,
becoming degenerate with MHFISWs at sufficiently low temperatures and
giving correct high-temperature asymptotics much better than MLSWs.
The MWDISW thermodynamics is precise at both low and high temperatures and free from any
thermal breakdown.

   Next we observe the SC-MSW thermodynamics of noncollinear antiferromagnets,
Figs. \ref{F:ConvSC-MSW&SB&QMC&TNofCin2D}(c) and \ref{F:ConvSC-MSW&SB&QMC&TNofCin2D}(d).
While the numerical findings to compare are not necessarily conclusive at low temperatures,
the present MSW findings are all far from consistent with them on the whole.
In both cases, the MWDISW temperature profiles are neither reminiscent of the major broad maximum
consequent from the exchange coupling constant $J$ nor any reference for the high-temperature
asymptotics.
Especially for the kagome-lattice antiferromagnet, MWDISWs completely fail to guess
a low-temperature shoulder or additional peak below the main maximum.
As far as we bring SWs into interaction variationally, any scattering involving an odd number of
magnons does not play any role in designing thermodynamics.
$\mathcal{H}^{\left(\frac{m}{2}\right)}$,
or Hamiltonians on the order of $S$ to a fractional power in general,
must play a key role in creating thermal features peculiar to frustrated noncollinear
antiferromagnets, as will be demonstrated later, and therefore, we consider an alternative approach
to SW interactions intending to make them effective in addition.

\section{Modified spin-wave theory of noncollinear antiferromagnets}
\label{S:MSWofNCA}
What we newly construct is a double-constraint (DC)-MSW theory of antiferromagnetic spin spirals
in various geometries, where we perturbatively take account of up-to-$O(S^{-1})$ interactions
between DC MSWs.
In order to treat noncoplanar antiferromagnets as well,
we extend the local spin reference frame \eqref{E:SinLandRframes2D} to any rotation
in three dimensions,
rewrite the Hamiltonian \eqref{E:HeisenbergH} into the local coordinate system,
and then introduce Holstein-Primakoff bosons in the same way as \eqref{E:HPT}
to expand the Hamiltonian in descending powers of $\sqrt{S}$,
\begin{equation}
   \mathcal{H}
  =\sum_{m=-2}^4
   \mathcal{H}^{\left(\frac{m}{2}\right)}
  +O\left(S^{-\frac{3}{2}}\right),
   \label{E:Hexpanded3D}
\end{equation}
where $\mathcal{H}^{\left(\frac{m}{2}\right)}$, on the order of $S^{\frac{m}{2}}$, are explicitly
given in \ref{A:RFandCBH}.
We divide the bosonic Hamiltonian \eqref{E:Hexpanded3D} into the harmonic part
$\sum_{m=2}^4\mathcal{H}^{\left(\frac{m}{2}\right)}\equiv\mathcal{H}_{\mathrm{harm}}$, where
$\mathcal{H}^{(2)}$ is merely a constant and $\mathcal{H}^{\left(\frac{3}{2}\right)}$
always vanishes for symmetry reasons, and the interactions
$\sum_{m=-2}^1\mathcal{H}^{\left(\frac{m}{2}\right)}\equiv\mathcal{V}$
perturbing to the harmonic spin waves.
We first modify CLSWs into MLSWs by diagonalizing an effective harmonic Hamiltonian,
which we shall denote by $\widetilde{\mathcal{H}}_{\mathrm{harm}}$,
instead of the bare one $\mathcal{H}_{\mathrm{harm}}$,
and then perturb $\widetilde{\mathcal{H}}_{\mathrm{harm}}$ to fourth order at highest
in $\mathcal{V}$ within the order of $S^{-1}$ by calculating relevant one- and two-loop
corrections.
Such a perturbative treatment of SW interactions is rather orthodox than notable in itself,
especially in frustrated noncollinear antiferromagnets
\cite{C144416,C144415,C832,C207202,M094407,C237202}.
It is the elaborate constraint condition that distinguishes our MSW theory for frustrated
noncollinear antiferromagnets from any other previous related attempts.
We design a DC condition for SWs to satisfy, which spontaneously reduces to
the well-known traditional SC condition \eqref{E:collinearMSWconstraintHPB}
when the spins choose a collinear alignment as their classical ground state.

\subsection{Double-constraint condition}
\label{SS:DCC}

   \textit{The first requirement} originates from the traditional constraint condition
\eqref{E:collinearMSWconstraintHPB} to keep the number of Holstein-Primakoff bosons constant but
gives due consideration to general noncollinear antiferromagnets as well,
\begin{align}
   S
  -\langle\mathcal{A}\rangle_T
  =S
  -\langle
    a_{\bm{r}_l}^\dagger a_{\bm{r}_l}
   \rangle_T
  =\delta,
   \label{E:noncollinearMSWconstraintHPB1st}
\end{align}
where $\delta$ is a temperature-independent nonnegative parameter to regulate the number of
Holstein-Primakoff bosons and therefore quasiparticle antiferromagnons.
With vanishing $\delta$,
the generalized constraint condition \eqref{E:noncollinearMSWconstraintHPB1st} reduces to
the original one \eqref{E:collinearMSWconstraintHPB} designed for collinear antiferromagnets.
With this in mind, we require $\delta$ to minimize the ground-state energy.
When $\delta$ moves away from $0$ to $S$,
the ground-state energy, whether within or beyond the harmonic approximation,
monotonically increases in any collinear antiferromagnet but takes one and only minimum
in any noncollinear antiferromagnet.
Therefore, $\delta$ reasonably remains zero in collinear antiferromagnets, while tuning $\delta$ to
an optimal value further stabilizes SWs in noncollinear antiferromagnets.
Such determined $\delta$ dramatically improves the MSW description of the uniform magnetic
susceptibility as well as specific heat.

   Before detailing the specific-heat calculations, let us take a look at how hard it is for SWs
to describe magnetic susceptibilities of noncollinear antiferromagnets especially in lower than
three dimensions and how effectively our DC modification scheme works on them in this context.
In case the total magnetization in question
$\mathcal{M}^\lambda\equiv\sum_{l=1}^L S_{\bm{r}_l}^\lambda$
is not commutable with the Hamiltonian $\widetilde{\mathcal{H}}_{\mathrm{harm}}$,
we define the canonical correlation function \cite{K570} for general operators
$\mathcal{P}$ and $\mathcal{Q}$ as
\begin{align}
   \langle
    \mathcal{P};\mathcal{Q}
   \rangle_{\beta}
  \equiv
   \frac{1}{\beta\hbar}
   \int_0^{\beta\hbar}\!
   \frac
   {\mathrm{Tr}[e^{-\beta\widetilde{\mathcal{H}}_{\mathrm{harm}}}\mathcal{P}(\tau)\mathcal{Q}]}
   {\mathrm{Tr}[e^{-\beta\widetilde{\mathcal{H}}_{\mathrm{harm}}}]}
   d\tau,
   \label{E:CanonicalCorrelation}
\end{align}
where we denote (Euclidean) time evolution by
$
   \mathcal{P}(\tau)
  \equiv
   e^{ \tau\widetilde{\mathcal{H}}_{\mathrm{harm}}/\hbar}
   \mathcal{P}
   e^{-\tau\widetilde{\mathcal{H}}_{\mathrm{harm}}/\hbar}
$.
While any canonical correlation is symmetric,
$\langle\mathcal{P};\mathcal{Q}\rangle_\beta=\langle\mathcal{Q};\mathcal{P}\rangle_\beta$,
it is not necessarily the case with a simple canonical-ensemble average,
\begin{align}
   \langle
    \mathcal{P}\mathcal{Q}
   \rangle_{1/\beta k_{\mathrm{B}}}
  \equiv
   \frac
   {\mathrm{Tr}[e^{-\beta\widetilde{\mathcal{H}}_{\mathrm{harm}}}\mathcal{P}\mathcal{Q}]}
   {\mathrm{Tr}[e^{-\beta\widetilde{\mathcal{H}}_{\mathrm{harm}}}]}.
   \label{E:SimpleCorrelation}
\end{align}
We further introduce the spin fluctuation operator
$
   \delta S_{\bm{r}_l}^{\lambda}
  \equiv
   S_{\bm{r}_l}^\lambda
  -\langle
    S_{\bm{r}_l}^\lambda
   \rangle_{T}
$
and define its Fourier transform as
\begin{align}
   \delta S_{\bm{k}_\nu:\sigma}^\lambda
  =\frac{1}{\sqrt{N}}
   \sum_{n=1}^N
   e^{-i\bm{k}_\nu\cdot\bm{r}_{p(n-1)+\sigma}}
   \delta S_{\bm{r}_{p(n-1)+\sigma}}^\lambda,
   \label{E:FTforS}
\end{align}
just like \eqref{E:FT}.
Calculating canonical correlations between the zero-wave-vector Fourier components of
spin fluctuations with $\beta$ set equal to the inverse temperature $1/k_{\mathrm{B}}T$
gives the temperature-$T$ uniform magnetic susceptibility
\begin{align}
   \chi^{\lambda\lambda}
  =\frac{(g\mu_{\mathrm{B}})^{2}}{k_{\mathrm{B}}T}
   \frac{L}{p}
   \sum_{\sigma,\sigma'=1}^p
   \langle
    \delta S_{\bm{0}:\sigma }^{\lambda\,\dagger};
    \delta S_{\bm{0}:\sigma'}^{\lambda}
   \rangle_{1/k_{\mathrm{B}}T}.
   \label{E:chi}
\end{align}
When
$
   \sum_{\sigma=1}^p
   [S_{ \bm{k}_\nu:\sigma}^\lambda, 
   \widetilde{{\cal H}}_{\mathrm{harm}}]
  =0
$,
the relevant canonical correlation reduces to a static structure factor,
\begin{align}
   &
   \frac{1}{p}
   \sum_{\sigma,\sigma'=1}^{p}
   \langle
    \delta S_{ \bm{k}_{\nu}:\sigma }^{\lambda\,\dagger};
    \delta S_{ \bm{k}_{\nu}:\sigma'}^{\lambda}
   \rangle_{1/k_{\mathrm{B}}T}
   \nonumber
   \allowdisplaybreaks \\
   &
  =\frac{1}{L}
   \sum_{n,n'=1}^{N}
   \sum_{\sigma,\sigma'=1}^{p}
   e^{ i\bm{k}_{\nu}\cdot
   \left(
   {\bm{r}_{p(n-1)+\sigma}}-{\bm{r}_{p(n'-1)+\sigma'}}
   \right)}
   \langle
    \delta S_{\bm{r}_{p(n -1)+\sigma }}^{\lambda}
    \delta S_{\bm{r}_{p(n'-1)+\sigma'}}^{\lambda}
   \rangle_{T}
  \equiv
   S^{\lambda\lambda}(\bm{k}_{\nu}),
   \label{E:CanonicalCorrelationCollinear}
\end{align}
and therefore, the $\lambda\lambda$ diagonal element of the uniform susceptibility tensor times
temperature becomes the mean-square fluctuation of the uniform magnetization along
the $\lambda$ axis,
\begin{align}
   \frac{k_{\mathrm{B}}T}{(g\mu_{\mathrm{B}})^{2}}
   \chi^{\lambda\lambda}
  =\langle(\mathcal{M}^\lambda)^{2}\rangle_{T}
  -\langle \mathcal{M}^\lambda     \rangle_{T}^{2}
  =LS^{\lambda\lambda}(\bm{0}).
   \label{E:chiCollinear}
\end{align}
This is fundamentally the case with the $\mathrm{SU(2)}$ Heisenberg model, but its SW Hamiltonian
is no longer invariant under $\mathrm{SU(2)}$ rotations.
For the square- and honeycomb-lattice antiferromagnets,
$\widetilde{{\cal H}}_{\mathrm{harm}}$ \eqref{E:tildeHharmT} commutes with $\mathcal{M}^z$
and \eqref{E:chi} reads in the bosonic language
\begin{align}
   &
   \frac{\chi^{zz}}{(g\mu_{\mathrm{B}})^2}
  =\frac{1}{k_{\mathrm{B}}T}
   \sum_{\rho=1}^{p}\sum_{\nu=1}^{N}
   \bar{n}_{\bm{k}_{\nu}:\rho}
   \left(
    \bar{n}_{\bm{k}_{\nu}:\rho}+1
   \right),
   \label{E:chizzCollinear}
   \allowdisplaybreaks \\
   &
   \frac{\chi^{xx}}{(g\mu_{\mathrm{B}})^{2}}
  =\frac{\chi^{yy}}{(g\mu_{\mathrm{B}})^{2}}
  =\frac{N}{k_{\mathrm{B}}T}
   \sum_{\rho=1}^{p}
   \bigl(
    S-\langle {\cal A} \rangle_{T}
   \bigr)
   \left(
    \cosh 2\theta_{\bm{0}:\rho}
   -\sinh 2\theta_{\bm{0}:\rho}
   \right)
   \left(
    \bar{n}_{\bm{0}:\rho}+\frac{1}{2}
   \right).
   \label{E:chixx&yyCollinear}
\end{align}
Note that \eqref{E:chizzCollinear} satisfies \eqref{E:chiCollinear}.
The SC-MWDISW thermodynamics of the square-lattice antiferromagnet is highly successful
\cite{Y094412} for the susceptibility as well as the specific heat,
giving precise low-temperature analytics, avoiding the artificial discontinuous transition to
the paramagnetic phase at any finite temperature, and reproducing the correct high-temperature
asymptotics.
For the triangular- and kagome-lattice antiferromagnets,
$\widetilde{{\cal H}}_{\mathrm{harm}}$ [cf. \eqref{E:tildeHharmYO}]
no longer commutes with any of $\mathcal{M}^x$, $\mathcal{M}^y$, and $\mathcal{M}^z$,
and \eqref{E:chi} reads in the bosonic language
\begin{align}
   &\!
   \frac{\chi^{zz}}{(g\mu_{\mathrm{B}})^2}=\frac{\chi^{xx}}{(g\mu_{\mathrm{B}})^2}
  =\sum_{\rho,\rho'=1}^{p}\sum_{\kappa=\pm}
   \frac{e^{i\kappa(\bm{Q}+\bm{q})\cdot(\bm{a}_{\rho-1}-\bm{a}_{\rho'-1})}}{16}
   \nonumber
   \allowdisplaybreaks \\
   &\!\times
   \sum_{\sigma,\sigma'=1}^{p}
   \sum_{\nu=1}^{N}
   u_{\bm{k}_{\nu}             :\rho \sigma }
   u_{\bm{k}_{\nu}             :\rho'\sigma }
   u_{\bm{k}_{\nu}+\kappa\bm{Q}:\rho \sigma'}
   u_{\bm{k}_{\nu}+\kappa\bm{Q}:\rho'\sigma'}
   \left\{
    \bar{n}_{\bm{k}_{\nu}:\sigma}(\bar{n}_{\bm{k}_{\nu}+\kappa\bm{Q}:\sigma'}+1)
    \frac{e^{\beta(\varepsilon_{\bm{k}_{\nu}:\sigma}
                  -\varepsilon_{\bm{k}_{\nu}+\kappa\bm{Q}:\sigma'})}-1}
         {\varepsilon_{\bm{k}_{\nu}             :\sigma }
         -\varepsilon_{\bm{k}_{\nu}+\kappa\bm{Q}:\sigma'}}
   \right.
   \nonumber
   \allowdisplaybreaks \\
   &\!\quad\quad\times
    \bigl[
     (\cosh 2\vartheta_{\bm{k}_{\nu}             :\sigma }+1)
     (\cosh 2\vartheta_{\bm{k}_{\nu}+\kappa\bm{Q}:\sigma'}+1)
    + \sinh 2\vartheta_{\bm{k}_{\nu}             :\sigma }
      \sinh 2\vartheta_{\bm{k}_{\nu}+\kappa\bm{Q}:\sigma'}
    \bigr]
   \nonumber
   \allowdisplaybreaks \\
   &\!\ \ 
   -(\bar{n}_{\bm{k}_{\nu}:\sigma}+1)\bar{n}_{\bm{k}_{\nu}+\kappa\bm{Q}:\sigma'}
    \frac{e^{-\beta(\varepsilon_{\bm{k}_{\nu}             :\sigma }
                   -\varepsilon_{\bm{k}_{\nu}+\kappa\bm{Q}:\sigma'})}-1}
         {\varepsilon_{\bm{k}_{\nu}             :\sigma }
         -\varepsilon_{\bm{k}_{\nu}+\kappa\bm{Q}:\sigma'}}
   \nonumber
   \allowdisplaybreaks \\
   &\!\quad\quad\times
    \bigl[
     (\cosh 2\vartheta_{\bm{k}_{\nu}             :\sigma }-1)
     (\cosh 2\vartheta_{\bm{k}_{\nu}+\kappa\bm{Q}:\sigma'}-1)
    + \sinh 2\vartheta_{\bm{k}_{\nu}             :\sigma }
      \sinh 2\vartheta_{\bm{k}_{\nu}+\kappa\bm{Q}:\sigma'}
    \bigr]
   \nonumber
   \allowdisplaybreaks \\
   &\!\ \ 
   +\bar{n}_{\bm{k}_{\nu}:\sigma}\bar{n}_{\bm{k}_{\nu}+\kappa\bm{Q}:\sigma'}
    \frac{e^{\beta(\varepsilon_{\bm{k}_{\nu}             :\sigma }
                  +\varepsilon_{\bm{k}_{\nu}+\kappa\bm{Q}:\sigma'})}-1}
         {\varepsilon_{\bm{k}_{\nu}             :\sigma }
         +\varepsilon_{\bm{k}_{\nu}+\kappa\bm{Q}:\sigma'}}
   \nonumber
   \allowdisplaybreaks \\
   &\!\quad\quad\times
    \bigl[
     (\cosh 2\vartheta_{\bm{k}_{\nu}             :\sigma }+1)
     (\cosh 2\vartheta_{\bm{k}_{\nu}+\kappa\bm{Q}:\sigma'}-1)
    + \sinh 2\vartheta_{\bm{k}_{\nu}             :\sigma }
      \sinh 2\vartheta_{\bm{k}_{\nu}+\kappa\bm{Q}:\sigma'}
    \bigr]
   \nonumber
   \allowdisplaybreaks \\
   &\!\ \ 
   -(\bar{n}_{\bm{k}_{\nu}:\sigma}+1)(\bar{n}_{\bm{k}_{\nu}+\kappa\bm{Q}:\sigma'}+1)
    \frac{e^{-\beta(\varepsilon_{\bm{k}_{\nu}             :\sigma }
                   +\varepsilon_{\bm{k}_{\nu}+\kappa\bm{Q}:\sigma'})}-1}
         {\varepsilon_{\bm{k}_{\nu}             :\sigma }
         +\varepsilon_{\bm{k}_{\nu}+\kappa\bm{Q}:\sigma'}}
   \nonumber
   \allowdisplaybreaks \\
   &\!\quad\quad\times
    \bigl[
     (\cosh 2\vartheta_{\bm{k}_{\nu}             :\sigma }-1)
     (\cosh 2\vartheta_{\bm{k}_{\nu}+\kappa\bm{Q}:\sigma'}+1)
    + \sinh 2\vartheta_{\bm{k}_{\nu}             :\sigma }
      \sinh 2\vartheta_{\bm{k}_{\nu}+\kappa\bm{Q}:\sigma'}
    \bigr]
   \left.\!\!
    \vphantom{
    \frac{e^{-\beta(\varepsilon_{\bm{k}_{\nu}             :\sigma }
                   +\varepsilon_{\bm{k}_{\nu}+\kappa\bm{Q}:\sigma'})}-1}
         {\varepsilon_{\bm{k}_{\nu}             :\sigma }
         +\varepsilon_{\bm{k}_{\nu}+\kappa\bm{Q}:\sigma'}}}
   \right\}
   \nonumber
   \allowdisplaybreaks \\
   &\!
  +N
   \sum_{\rho,\rho'=1}^{p}\sum_{\kappa=\pm}
   \frac{e^{ i\kappa(\bm{Q}+\bm{q})\cdot(\bm{a}_{\rho-1}-\bm{a}_{\rho'-1})}}{4}
   \sum_{\sigma=1}^{p}
   u_{\kappa\bm{Q}:\rho \sigma}
   u_{\kappa\bm{Q}:\rho'\sigma}
   \left(
   \frac{S}{2}
  -\frac{\langle {\cal A} \rangle_{T}}{2}
  -\frac{\langle {\cal C} \rangle_{T}}{4}
   \right)
   \nonumber
   \allowdisplaybreaks \\
   &\!\quad\quad\times
   (\cosh 2\theta_{\kappa\bm{Q}:\sigma}
   +\sinh 2\theta_{\kappa\bm{Q}:\sigma})
   \left[
    \bar{n}_{\kappa\bm{Q}:\sigma}
    \frac{e^{\beta\varepsilon_{\kappa\bm{Q}:\sigma}}-1}
         {        \varepsilon_{\kappa\bm{Q}:\sigma}}
   -(\bar{n}_{\kappa\bm{Q}:\sigma}+1)
   \frac{e^{-\beta\varepsilon_{\kappa\bm{Q}:\sigma}}-1}
        {         \varepsilon_{\kappa\bm{Q}:\sigma}}
   \right]\!,\!
   \label{E:chizz&xxNoncollinear}
   \allowdisplaybreaks \\
   &\!
   \frac{\chi^{yy}}{(g\mu_{\mathrm{B}})^{2}}
  =N
   \sum_{\rho,\rho'=1}^{p}
   \sum_{\sigma=1}^{p}
   u_{\bm{0}:\rho \sigma}
   u_{\bm{0}:\rho'\sigma}
   \left(
    \frac{S}{2}
   -\frac{\langle {\cal A} \rangle_{T}}{2}
   +\frac{\langle {\cal C} \rangle_{T}}{4}
   \right)
   (\cosh 2\theta_{\bm{0}:\sigma}
   -\sinh 2\theta_{\bm{0}:\sigma})
   \nonumber
   \allowdisplaybreaks \\
   &\!\quad\quad\times
   \left[
    \bar{n}_{\bm{0}:\sigma}
    \frac{e^{\beta\varepsilon_{\bm{0}:\sigma}}-1}
         {        \varepsilon_{\bm{0}:\sigma}}
  -\left(
    \bar{n}_{\bm{0}:\sigma}+1
   \right)
   \frac{e^{-\beta\varepsilon_{\bm{0}:\sigma}}-1}
        {         \varepsilon_{\bm{0}:\sigma}}
   \right].
   \label{E:chiyyNoncollinear}
\end{align}
\begin{figure}
\begin{flushright}
\includegraphics[width=130mm]{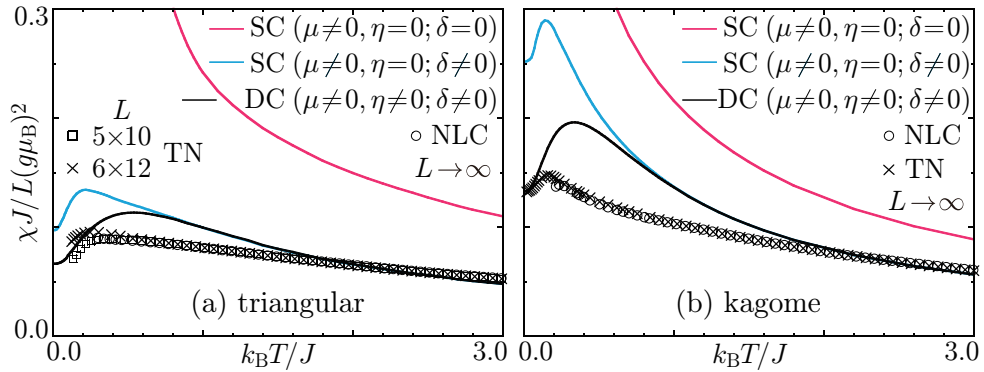}
\end{flushright}
\vspace{-7mm}
\caption{Traditional SC-MLSW, artificially tuned SC-MLSW [$\delta=0.2653$ (a), $\delta=0.1554$ (b)],
         and DC-MLSW calculations of the uniform magnetic susceptibility
         $\chi\equiv\sum_{\lambda=x,y,z}\chi^{\lambda\lambda}/3$
         [\eqref{E:chizz&xxNoncollinear}, \eqref{E:chiyyNoncollinear}]
         as a function of temperature for the $S=\frac{1}{2}$ Hamiltonian \eqref{E:HeisenbergH}
         on the $L\rightarrow\infty$ triangular (a) and kagome (b) lattices in comparison with
         high-temperature series expansions based on the numerical linked-cluster algorithm (NLC)
         \cite{R061118} and
         tensor-network-based renormalization-group calculations (TN) \cite{C1545,C140404}.
         Unlike the TN calculations on the kagome lattice \cite{C1545}, those on the triangular
         lattice \cite{C140404} are based on one-dimensional matrix product operators
         \cite{C031082} and $5\times 10\,$- and $6\times 12\,$-site cylindrical
         triangular lattices with pentagonal and hexagonal ends aslant, respectively,
         are thus calculated.}
\label{F:SC&tunedSC&DC-MLSWin2D}
\end{figure}

   We evaluate \eqref{E:chizz&xxNoncollinear} and \eqref{E:chiyyNoncollinear} modifying CLSWs under
various constraint conditions and compare them with numerical linked-cluster and
tensor-network-based renormalization-group calculations in Fig. \ref{F:SC&tunedSC&DC-MLSWin2D}.
Considering that SC MLSWs never fail to qualitatively reproduce
the overall temperature profile of the uniform magnetic susceptibility of any collinear
antiferromagnet \cite{Y094412}, their findings for the triangular- and kagome-lattice
antiferromagnets are far from successful especially at low temperatures.
They misread the per-site static uniform susceptibility as divergent and quasidivergent
with temperature going down to absolute zero in the triangular- and kagome-lattice
antiferromagnets, respectively.
The complete divergence of the zero-temperature per-site susceptibility of the triangular-lattice
antiferromagnet is because of SC MLSWs softening at $\bm{k}=\bm{Q}$
and artificially tuned SC MLSWs or more elaborate DC MLSWs solve this difficulty with their
$\bm{Q}$ modes hardening [cf. Eq. \eqref{E:chizz&xxNoncollinear} and
Figs. \ref{F:DispertionRelations}(b)--\ref{F:DispertionRelations}(d)].
Introducing the boson number tuning parameter $\delta$ into the first constraint
\eqref{E:noncollinearMSWconstraintHPB1st} dramatically improves the MSW thermodynamics
and imposing the second constraint
[cf. \eqref{E:noncollinearMSWconstraintHPB2nd} in the following] in addition on such MSWs
gives a still better description of the susceptibility at low temperatures.
In the high-temperature limit, on the other hand, SWs spontaneously meet the second requirement
\eqref{E:noncollinearMSWconstraintHPB2nd} inducing the relevant chemical potential $\eta$
to vanish.
Therefore, the superiority of DC MLSWs over SC MLSWs at high temperatures owes much to
tuning of the first constraint \eqref{E:noncollinearMSWconstraintHPB1st}.
In collinear antiferromagnets, SC MLSWs yield the correct high-temperature asymptotics
\cite{Y094412}
\begin{align}
   \lim_{T\rightarrow\infty}
   \frac{\chi}{L(g\mu_{\mathrm{B}})^2}
  \equiv
   \lim_{T\rightarrow\infty}
   \frac{\sum_{\lambda=x,y,z}\chi^{\lambda\lambda}}{3L(g\mu_{\mathrm{B}})^2}
  =\frac{S(S+1)}{3k_{\mathrm{B}}T},
   \label{E:chiTconvSC-MSW(t->oo)}
\end{align}
while in noncollinear antiferromagnets, SC MSWs constrained to
\eqref{E:noncollinearMSWconstraintHPB1st}, whether with or without $\delta$,
no longer hit the exact sum rule in general but say that
\begin{align}
   &\!\!\!\!
   \lim_{T\rightarrow\infty}
   \frac{\chi}{L(g\mu_{\mathrm{B}})^2}
  =\frac{S(S+1)}{3k_{\mathrm{B}}T}
   \left[
    \frac{(S-\delta)(S-\delta+1)}{S(S+1)}
   +\frac{1}{(S+1)\mathrm{ln}\left(1+\frac{1}{S-\delta}\right)}
   \right].
   \label{E:chiTimprSC-MSW(t->oo)}
\end{align}
In the case of $S=\frac{1}{2}$, SC MLSWs estimate
$
   \lim_{T\rightarrow\infty}
   \chi k_{\mathrm{B}}T/L(g\mu_{\mathrm{B}})^2
$
at $0.4017$, compared to the exact value $1/4$.
However, tuning $\delta$ optimally under the DC condition (cf. Table \ref{T:Optimaldelta})
yields much better estimates, $0.1963$ and $0.2769$ for the triangular and kagome-lattice
antiferromagnets, respectively.
We are thus convinced of the necessity of adjusting the traditional constraint condition
\eqref{E:collinearMSWconstraintHPB} to noncollinear antiferromagnets and
the validity of imposing the extended constraint condition
\eqref{E:noncollinearMSWconstraintHPB1st} on their Holstein-Primakoff bosons
so as to stabilize their ground states.

   We can have another good understanding of Eq. \eqref{E:noncollinearMSWconstraintHPB1st}
in a different aspect.
Let us recall the Schwinger boson representation
\begin{align}
   &
   S_{\bm{r}_l}^z
  =\frac{a_{\bm{r}_l:\uparrow  }^\dagger a_{\bm{r}_l:\uparrow  }
        -a_{\bm{r}_l:\downarrow}^\dagger a_{\bm{r}_l:\downarrow}}
        {2}
  \equiv
   \frac{n_{\bm{r}_l:\uparrow  }-n_{\bm{r}_l:\downarrow}}{2},\ 
   S_{\bm{r}_l}^x-iS_{\bm{r}_l}^y
  \equiv S_{\bm{r}_l}^{+\dagger}
  \equiv S_{\bm{r}_l}^{-}
  =a_{\bm{r}_l:\downarrow}^\dagger
   a_{\bm{r}_l:\uparrow  };
   \allowdisplaybreaks
   \nonumber
   \allowdisplaybreaks \\
   &
   \bm{S}_{\bm{r}_l}^2
  =\frac{n_{\bm{r}_l:\uparrow  }+n_{\bm{r}_l:\downarrow}}{2}
   \left(
    \frac{n_{\bm{r}_l:\uparrow  }+n_{\bm{r}_l:\downarrow}}{2}
   +1
   \right),
   \label{E:SB}
\end{align}
where physical states demand that the bosons should satisfy the constraint condition
\begin{align}
   \sum_{\sigma=\uparrow,\downarrow}
   a_{\bm{r}_l:\sigma}^\dagger a_{\bm{r}_l:\sigma}
  =2S
   \label{E:constraintSB}
\end{align}
at each lattice site.
In what are called Schwinger-boson mean-field theories
\cite{A617,A316,S5028,C12329,M8348,M47001,M123033},
the $L$ requirements \eqref{E:constraintSB} are relaxed and imposed only on average, i.e.,
merely a single Lagrange multiplier $\mu$ is introduced and the term
$
 \mu\sum_{l=1}^L
 \left(
  2S
 -\sum_{\sigma=\uparrow,\downarrow}a_{\bm{r}_l:\sigma}^\dagger a_{\bm{r}_l:\sigma}
 \right)
$
is added to the Hamiltonian.
The determination condition for $\mu$,
\begin{align}
   \frac{1}{L}\sum_{l=1}^L
   \sum_{\sigma=\uparrow,\downarrow}
   \langle
    a_{\bm{r}_l:\sigma}^\dagger a_{\bm{r}_l:\sigma}
   \rangle_T
  =2S,
   \label{E:constraintSBMF}
\end{align}
is reminiscent of the MSW constraint condition \eqref{E:collinearMSWconstraintHPB}.
In the Holstein-Primakoff representation, each spin is described by a single Bose oscillator
together with the nonholonomic constraint $0\leq a_{\bm{r}_l}^\dagger a_{\bm{r}_l}\leq 2S$,
whereas in the Schwinger representation, each spin is replaced by two bosons together with
the holonomic constraint \eqref{E:constraintSB}.
Then, there may be a Schwinger-representation analog of the generalized MSW constraint condition
\eqref{E:noncollinearMSWconstraintHPB1st}.
The idea of the number of bosons being tunable is practiced indeed in
an extended mean-field Schwinger boson framework developed by Messio \textit{et al.}
\cite{M064428,M267201,M207204,M125127} intending to explore chiral spin liquids
\cite{M267201,M207204,M125127} in the kagome-lattice antiferromagnet.
The Schwinger-boson mean-field solution with its bond order parameters being chosen as real
for the $S=\frac{1}{2}$ regular kagome-lattice Heisenberg antiferromagnet is a long-range
N\'eel-ordered phase of the $\sqrt{3}\times\sqrt{3}$ type with a gapless spinon spectrum.
However, if we allow the left-hand side of \eqref{E:constraintSBMF} to deviate from twice
the real spin quantum number, a SL phase characterized by a fully gapped spinon spectrum
stabilizes instead \cite{M064428,M267201,M207204} on the way of its value going away from unity
down to zero, the extreme quantum limit.
The analogy between the Holstein-Primakoff and Schwinger representations strongly motivates us
to tune $\delta$ in the generalized MSW constraint \eqref{E:noncollinearMSWconstraintHPB1st}.

   \textit{The second requirement} is related to spin rotational symmetry.
There is a crucial difference, for all similarities as have been noted above and observed in
Fig. \ref{F:ConvSC-MSW&SB&QMC&TNofCin2D}, between the Schwinger and Holstein-Primakoff bosons.
The Schwinger-boson mean-field formalism retains the $\mathrm{SU}(2)$ invariance of the original
Hamiltonian \eqref{E:HeisenbergH}, whereas every kind of SW theory, whether modified or not,
reduces this to $\mathrm{U}(1)$ or less.  
The Schwinger-boson representation of the $\mathrm{SU}(2)$ spin variables,
\eqref{E:SB} constrained to \eqref{E:constraintSB}, indeed gives an isotropic mean-field
expectation value of each spin component,
\begin{align}
   \langle
    S_{\bm{r}_{l}}^x S_{\bm{r}_{l}}^x
   \rangle_T
  =\langle
    S_{\bm{r}_{l}}^y S_{\bm{r}_{l}}^y
   \rangle_T
  =\langle
    S_{\bm{r}_{l}}^z S_{\bm{r}_{l}}^z
   \rangle_T
  =\frac{S(S+1)}{2},
   \label{E:<SxSx><SySy><SzSz>SBMF}
\end{align}
but misreads the spin magnitude as $3/2$ times as large as the correct value,
$
   \langle
    \bm{S}_{\bm{r}_l}^2
   \rangle_T
  =3S(S+1)/2
$
\cite{A617,A316}.
The Holstein-Primakoff-boson representation in the local spin reference frame \eqref{E:HPT}
gives the correct spin magnitude but breaks the original spin rotational symmetry,
\begin{align}
   &
   \langle
    S_{\bm{r}_{l}}^{\tilde{x}}S_{\bm{r}_{l}}^{\tilde{x}}
   \rangle_T
  =\left(
    S-\langle\mathcal{A}\rangle_T
   \right)
   \left(
    \frac{1}{2}+\langle\mathcal{A}\rangle_T
   \right)
  -\frac{\langle\mathcal{C}\rangle_T^2}{2}
  +\langle\mathcal{C}\rangle_T
   \left(
    S-\frac{1}{4}-\frac{3\langle\mathcal{A}\rangle_T}{2}
   \right)
  +O(S^{-1}),
   \nonumber
   \allowdisplaybreaks \\
   &
   \langle
    S_{\bm{r}_{l}}^{\tilde{y}}S_{\bm{r}_{l}}^{\tilde{y}}
   \rangle_T
  =\left(
    S-\langle\mathcal{A}\rangle_T
   \right)
   \left(
    \frac{1}{2}+\langle\mathcal{A}\rangle_T
   \right)
  -\frac{\langle\mathcal{C}\rangle_T^2}{2}
  -\langle\mathcal{C}\rangle_T
   \left(
    S-\frac{1}{4}-\frac{3\langle\mathcal{A}\rangle_T}{2}
   \right)
  +O(S^{-1}),
   \nonumber
   \allowdisplaybreaks \\
   &
   \langle
    S_{\bm{r}_{l}}^{\tilde{z}}S_{\bm{r}_{l}}^{\tilde{z}}
   \rangle_T
  =S^2
  +(1-2S)\langle\mathcal{A}\rangle_T
  +2\langle\mathcal{A}\rangle_T^2
  + \langle\mathcal{C}\rangle_T^2;\ \ 
   \sum_{\lambda=x,y,z}
   \langle
    S_{\bm{r}_{l}}^{\tilde{\lambda}}S_{\bm{r}_{l}}^{\tilde{\lambda}}
   \rangle_T
  =S(S+1).
   \label{E:<StildelambdaStildelambda>SW}
\end{align}
For collinear antiferromagnets, $\langle\mathcal{C}\rangle_T$ spontaneously vanishes to retain
a $\mathrm{U}(1)$ symmetry of each local spin operator in the rotating frame as well as
the global $\mathrm{U}(1)$ symmetry related to the conservation of $\mathcal{M}^z$ \eqref{E:Mz}.
For noncollinear antiferromagnets, the bosonic Hamiltonian ${\cal H}_{\mathrm{harm}}$ no longer
commutes with $\mathcal{M}^z$, yet we should expect each spin operator to possess
rotational symmetry about its local quantization axis $\tilde{z}$, which is met by demanding that
$
   \langle
    S_{\bm{r}_{l}}^{\tilde{x}}S_{\bm{r}_{l}}^{\tilde{x}}
   \rangle_T
  =\langle
    S_{\bm{r}_{l}}^{\tilde{y}}S_{\bm{r}_{l}}^{\tilde{y}}
   \rangle_T
$, i.e.,
\begin{align}
  \langle\mathcal{C}\rangle_T
  =\frac{\langle
          a_{\bm{r}_l}^\dagger a_{\bm{r}_l}^\dagger
         +a_{\bm{r}_l}         a_{\bm{r}_l}
         \rangle_T}{2}
  =0.
   \label{E:noncollinearMSWconstraintHPB2nd}
\end{align}

   Thus and thus, we introduce $2L$ Lagrange multipliers and diagonalize the effective harmonic
Hamiltonian
\begin{align}
   &
   \widetilde{{\cal H}}_{\mathrm{harm}}
  \equiv
   {\cal H}_{\mathrm{harm}}
  +\sum_{l=1}^L
   \mu_l
   \left(
    S-\delta-a_{\bm{r}_l}^\dagger a_{\bm{r}_l}
   \right)
  +\sum_{l=1}^L
   \frac{\eta_l}{2}
   \left(
    a_{\bm{r}_l}^\dagger a_{\bm{r}_l}^\dagger
   +a_{\bm{r}_l}         a_{\bm{r}_l}
   \right).
   \label{E:tildeHharmYO}
\end{align}
subject to \eqref{E:noncollinearMSWconstraintHPB1st} and \eqref{E:noncollinearMSWconstraintHPB2nd}
at each site.
Since the $2L$ Lagrange multipliers degenerate into merely two in practice as
$\mu_1=\cdots=\mu_L\equiv\mu$ and $\eta_1=\cdots=\eta_L\equiv\eta$, \eqref{E:tildeHharmYO} becomes
\begin{align}
   \widetilde{\mathcal{H}}_{\mathrm{harm}}
  =\sum_{l=1}^2 E^{(l)}
  +\sum_{\nu=1}^N
   \sum_{\sigma=1}^p
   \varepsilon_{\bm{k}_\nu:\sigma}
   \alpha_{\bm{k}_\nu:\sigma}^\dagger\alpha_{\bm{k}_\nu:\sigma}
  +\mu L(S-\delta)
   \label{E:tildeHharmdiag}
\end{align}
with $E^{(l)}$ and $\alpha_{\bm{k}_\nu:\sigma}^\dagger$ having the same meanings as those in
\eqref{E:tildeHquaddiag}.
For DC MLSWs in the triangular- and kagome-lattice antiferromagnets,
the creation energy \eqref{E:triangular&kagomeomega} and
the Bogoliubov transformation \eqref{E:triangular&kagomeBT} are rewritten to
\begin{align}
   &
   \omega_{\bm{k}_\nu:\sigma}
  \equiv
   \frac{\varepsilon_{\bm{k}_\nu:\sigma}}{zJS}
  \equiv
   \sqrt{\left(
          \frac{1}{2}
         +\frac{\gamma_{\bm{k}_\nu:\sigma}}{4}
         -\frac{\mu}{zJS}
         \right)^2
        -\left(
          \frac{3\gamma_{\bm{k}_\nu:\sigma}}{4}
         -\frac{\eta}{zJS}
         \right)^2},
   \label{E:triangular&kagomeomegaDCMLSW}
   \allowdisplaybreaks \\
   &
   \mathrm{cosh}2\vartheta_{\bm{k}_\nu:\sigma}
  =\frac{1}{\omega_{\bm{k}_\nu:\sigma}}
   \left(
    \frac{1}{2}
   +\frac{\gamma_{\bm{k}_\nu:\sigma}}{4}
   -\frac{\mu}{zJS}
   \right),
   \,\,\,
   \mathrm{sinh}2\vartheta_{\bm{k}_\nu:\sigma}
  =\frac{1}{\omega_{\bm{k}_\nu:\sigma}}
   \left(
    \frac{3\gamma_{\bm{k}_\nu:\sigma}}{4}
   -\frac{\eta}{zJS}
   \right).
   \label{E:triangular&kagomeBTDCMLSW}
\end{align}
For DC MLSWs in polyhedral-lattice antiferromagnets, we set $N$ and $p$ equal to $1$ and $L$,
respectively,
omitting their momentum indices as $\varepsilon_{\bm{k}_\nu:\sigma}\equiv\varepsilon_\sigma$ and
\begin{align}
   \alpha_{\bm{k}_\nu:\sigma}
  \equiv
   \alpha_\sigma
  =\sum_{l=1}^L
   \left(
    f_{\sigma l}a_{\bm{r}_l}+g_{\sigma l}^*a_{\bm{r}_l}^\dagger
   \right),
   \label{E:BT3D}
\end{align}
and make Bogoliubov bosons out of bare Holstein-Primakoff bosons by numerically obtaining
the coefficients $f_{\sigma l}$ and $g_{\sigma l}$.
Since spins are not necessarily coplanar in the classical ground states of
polyhedral-lattice antiferromagnets, we generalize the ground-state energy expressions into
three dimensions,
\begin{align}
   &
   E^{(1)}
  =JS
   \sum_{\langle l,l'\rangle}
   \left[
    \sin\phi_{{\bm r}_{l}}\sin\phi_{{\bm r}_{l'}}
    \cos
    \left(
     \theta_{{\bm r}_{l}}-\theta_{{\bm r}_{l'}}
    \right)
   +\cos\phi_{{\bm r}_{l}}\cos\phi_{{\bm r}_{l'}}
   \right]
  +\frac{L}{2}\mu
  +\frac{1}{2}
   \sum_{\nu=1}^N\sum_{\sigma=1}^p
   \varepsilon_{\bm{k}_\nu:\sigma},
   \nonumber
   \allowdisplaybreaks \\
   &
   E^{(2)}
  =JS^2
   \sum_{\langle l,l'\rangle}
   \left[
    \sin\phi_{{\bm r}_{l}}\sin\phi_{{\bm r}_{l'}}
    \cos
    \left(
     \theta_{{\bm r}_{l}}-\theta_{{\bm r}_{l'}}
    \right)
   +\cos\phi_{{\bm r}_{l}}\cos\phi_{{\bm r}_{l'}}
   \right].
   \label{E:E(l)3D}
\end{align}

\subsection{Modified-spin-wave interaction---Perturbative treatment}
\label{SS:MSWI---PT}

\begin{figure}
\begin{flushright}
\includegraphics[width=130mm]{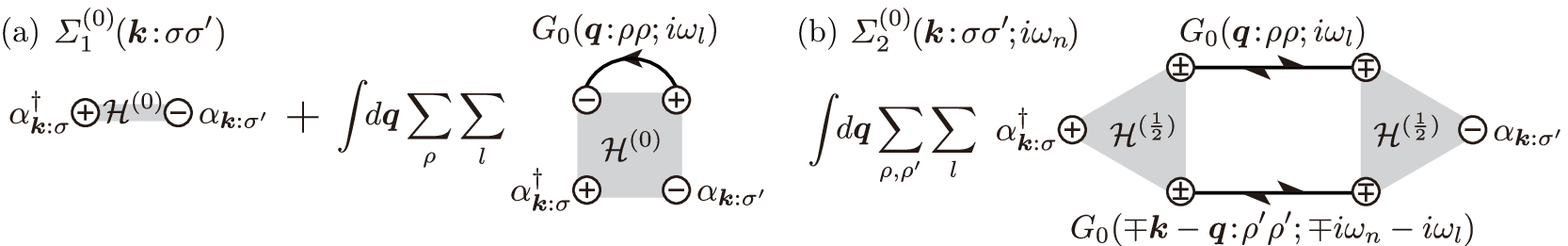}
\end{flushright}
\vspace{-7mm}
\caption{Diagrammatic representation of such self-energies as to give
         the leading $O(S^0)$ corrections
         $\Delta F_1^{(0)}$ \eqref{E:DeltaF_1^(0)} and $\Delta F_2^{(0)}$ \eqref{E:DeltaF_2^(0)}:
         (a) first-order and (b) second-order perturbations in
         $\mathcal{V}=\sum_{m=-2}^1\mathcal{H}^{\left(\frac{m}{2}\right)}$.
         The Holstein-Primakoff boson interactions $\mathcal{H}^{\left(\frac{m}{2}\right)}$
         each are rewritten in terms of Bogoliubov bosons and then normal-ordered.
         $\oplus$ and $\ominus$ denote creating and annihilating an MLSW (Bogoliubov boson),
         respectively.
         $\bm{q}$, $\rho$, $\rho'$, and $\omega_l$ are running indices.}
\label{F:PSE}
\end{figure}

   In order to investigate effects of the interactions $\mathcal{V}$ on temperature profiles of
the specific heat, we calculate their perturbative corrections to the DC-MLSW free energy.
The $l$th-order correction at temperature $T\equiv 1/\beta k_{\mathrm{B}}$ is calculated as
\begin{align}
   \Delta F_l
 =-\frac{(-1)^l}{\beta l!\hbar^l}
   \int\!\cdots\!\int_0^{\beta\hbar}\!
   d\tau_1 \cdots d\tau_l
   \langle
    \mathcal{T}
    \left[
     \mathcal{V}(\tau_1)\cdots\mathcal{V}(\tau_l)
    \right]
   \rangle_T,
   \label{E:DeltaFtolthOrder}
\end{align}
where we denote (Euclidean) time ordering operation by $\mathcal{T}$.
In the context of newly constructed perturbatively corrected (PC)-DC-MSW theory,
every thermal-bracket notation $\langle\ \,\rangle_T$ denotes the temperature-$T$ thermal average
with respect to MLSWs [cf. \eqref{E:SimpleCorrelation}] unless otherwise noted.
$\Delta F_l$ ($l=1,2,\cdots$) each make their own contribution on the order of $S^m$,
which we shall denote by $\Delta F_l^{(m)}$.
When we truncate any correction $\Delta F_l$ at the order of $S^{-1}$, all the remaining
corrections are given by
\begin{align}
   &
   \Delta F_1^{(0)}
  =\int_0^{\beta\hbar}\!\frac{d\tau_1}{\beta\hbar}
   \left\langle
    \mathcal{H}^{(0)}(\tau_1)
   \right\rangle_T,
   \label{E:DeltaF_1^(0)}
   \allowdisplaybreaks \\
   &
   \Delta F_1^{(-1)}
  =\int_0^{\beta\hbar}\!\frac{d\tau_1}{\beta\hbar}
   \left\langle
    \mathcal{H}^{(-1)}(\tau_1)
   \right\rangle_T,
   \label{E:DeltaF_1^(-1)}
   \allowdisplaybreaks \\
   &
   \Delta F_2^{(0)}
 =-\int\!\!\!\int_0^{\beta\hbar}\!\frac{d\tau_1 d\tau_2}{2\beta\hbar^2}
   \left\langle
    \mathcal{T}
    \left[
     \mathcal{H}^{\left(\frac{1}{2}\right)}(\tau_1)
     \mathcal{H}^{\left(\frac{1}{2}\right)}(\tau_2)
    \right]
   \right\rangle_T,
   \label{E:DeltaF_2^(0)}
   \allowdisplaybreaks \\
   &
   \Delta F_2^{(-1)}
 =-\int\!\!\!\int_0^{\beta\hbar}\!\frac{d\tau_1 d\tau_2}{2\beta\hbar^2}
   \left\{
    \left\langle
     \mathcal{T}
     \left[
      \mathcal{H}^{(0)}(\tau_1)
      \mathcal{H}^{(0)}(\tau_2)
     \right]
    \right\rangle_T
   +\left\langle
     \mathcal{T}
     \left[
      \mathcal{H}^{\left( \frac{1}{2}\right)}(\tau_1)
      \mathcal{H}^{\left(-\frac{1}{2}\right)}(\tau_2)
     \right]
    \right\rangle_T
    \vphantom{\mathcal{H}^{\left(\frac{1}{2}\right)}(\tau_1)}
   \right\},
   \label{E:DeltaF_2^(-1)}
   \allowdisplaybreaks \\
   &
   \Delta F_3^{(-1)}
  =\int\!\!\!\int\!\!\!\int_0^{\beta\hbar}\!\frac{d\tau_1 d\tau_2 d\tau_3}{6\beta\hbar^3}
    \left\langle
     \mathcal{T}
     \left[
      \mathcal{H}^{(0)}(\tau_1)
      \mathcal{H}^{\left(\frac{1}{2}\right)}(\tau_2)
      \mathcal{H}^{\left(\frac{1}{2}\right)}(\tau_3)
     \right]
    \right\rangle_T.
   \label{E:DeltaF_3^(-1)}
   \allowdisplaybreaks \\
   &
   \Delta F_4^{(-1)}
 =-\int\!\!\!\int\!\!\!\int\!\!\!\int_0^{\beta\hbar}\!
   \frac{d\tau_1 d\tau_2 d\tau_3 d\tau_4}{24\beta\hbar^4}
    \left\langle
     \mathcal{T}
     \left[
      \mathcal{H}^{\left(\frac{1}{2}\right)}(\tau_1)
      \mathcal{H}^{\left(\frac{1}{2}\right)}(\tau_2)
      \mathcal{H}^{\left(\frac{1}{2}\right)}(\tau_3)
      \mathcal{H}^{\left(\frac{1}{2}\right)}(\tau_4)
     \right]
    \right\rangle_T.
   \label{E:DeltaF_4^(-1)}
\end{align}
The $O(S^0)$ corrections $\Delta F_1^{(0)}$ and $\Delta F_2^{(0)}$ are both strictly calculated,
whereas the $O(S^{-1})$ corrections are approximately evaluated in an empirical manner.
It is interesting to observe temperature profiles of the specific heat for each order of $S$.
We define the up-to-$O(S^1)$ internal energy as
\begin{align}
   E
  =E_{\mathrm{harm}}
  \equiv
   \langle\mathcal{H}_{\mathrm{harm}}\rangle_T,
   \label{E:E(T)DC-MLSW}
\end{align}
the up-to-$O(S^0)$ internal energy as
\begin{align}
   E
  =E_{\mathrm{harm}}
  +\frac{\partial\beta\Delta F_1^{(0)}}{\partial\beta}
  +\frac{\partial\beta\Delta F_2^{(0)}}{\partial\beta},
   \label{E:E(T)PC-DC-MSWuptoS^0}
\end{align}
and the up-to-$O(S^{-1})$ internal energy as
\begin{align}
   &
   E
  \simeq
   E_{\mathrm{harm}}
  +\frac{\partial\beta\Delta F_1^{(0)}}{\partial\beta}
  +\frac{\partial\beta\Delta F_2^{(0)}}{\partial\beta}
  +\frac{\partial\beta\Delta F_2^{(-1)}}{\partial\beta}
  +\frac{\partial\beta\Delta F_3^{(-1)}}{\partial\beta}
  +\frac{\partial\beta\Delta F_4^{(-1)}}{\partial\beta}.
   \label{E:E(T)PC-DC-MSWuptoS^-1}
\end{align}
In terms of the unperturbed temperature Green functions
\begin{align}
   &
   G_0(\bm{k}_\nu:\sigma\sigma';\tau)
  \equiv
  -\left\langle
    \mathcal{T}
    \left[
     \alpha_{\bm{k}_\nu:\sigma }        (\tau)
     \alpha_{\bm{k}_\nu:\sigma'}^\dagger(0   )
    \right]
   \right\rangle_T,
   \nonumber
   \allowdisplaybreaks \\
   &
   G_0(\bm{k}_\nu:\sigma\sigma';i\omega_n)
  \equiv
   \int_0^{\beta\hbar}\!d\tau\,
   e^{i\omega_n\tau}
   G_0(\bm{k}_\nu:\sigma\sigma';\tau)
   \label{E:G0def}
\end{align}
with bosonic Matsubara frequencies $\omega_n\equiv 2n\pi/\beta\hbar$, the leading corrections
on the order of $S^0$, $\Delta F_1^{(0)}$ and $\Delta F_2^{(0)}$, can be written as
\begin{align}
   &
   \Delta F_1^{(0)}
  \equiv
   E^{(0)}
  +\frac{1}{\beta\hbar}
   \sum_{\nu=1}^N
   \sum_{\sigma,\sigma'=1}^p
   \sum_{n=-\infty}^\infty
   \varSigma_1^{(0)}(\bm{k}_\nu:\sigma\sigma')
   G_0(\bm{k}_\nu:\sigma'\sigma;i\omega_n),
   \label{E:DeltaF_1^(0)toSigma_1^(0)}
   \allowdisplaybreaks \\
   &
   \Delta F_2^{(0)}
  \equiv
   \frac{1}{\beta\hbar}
   \sum_{\nu=1}^N
   \sum_{\sigma,\sigma'=1}^p
   \sum_{n=-\infty}^\infty
   \varSigma_2^{(0)}(\bm{k}_\nu:\sigma\sigma';i\omega_n)
   G_0(\bm{k}_\nu:\sigma'\sigma;i\omega_n),
   \label{E:DeltaF_2^(0)toSigma_2^(0)}
\end{align}
where $E^{(0)}$ is the $O(S^0)$ perturbative correction to the DC-MLSW ground-state energy
$E^{(2)}+E^{(1)}$ \eqref{E:tildeHharmdiag},
\begin{align}
   &
   E^{(0)}
  =-J
   \sum_{\langle l,l'\rangle}
   \left\{
          \vphantom{\left[
                    \langle \mathcal{A} \rangle_{0} \langle \mathcal{D} \rangle_{0}
                   +\frac{\langle \mathcal{B} \rangle_{0} \langle \mathcal{C} \rangle_{0}
                   +\langle \mathcal{D} \rangle_{0}}{2}
                    \right]}
    \left[
     \langle \mathcal{A} \rangle_{0}^{2}
    +\langle \mathcal{B} \rangle_{0}^{2}
    +\langle \mathcal{D} \rangle_{0}^{2}
    +\langle \mathcal{A} \rangle_{0}
    \right]
    \left[
     \sin\phi_{{\bm r}_{l}}\sin\phi_{{\bm r}_{l'}}
     \cos
     \left(
      \theta_{{\bm r}_{l}}-\theta_{{\bm r}_{l'}}
     \right)
   \right.
    \right.
   \nonumber
   \allowdisplaybreaks \\
   &
    \left.
    +\cos\phi_{{\bm r}_{l}}\cos\phi_{{\bm r}_{l'}}
    \right]
   -\left[
          \vphantom{\frac{\langle \mathcal{D} \rangle_{0} \langle \mathcal{C} \rangle_{0}
                         +\langle \mathcal{B} \rangle_{0}}{2}}
     \langle \mathcal{A} \rangle_{0} \langle \mathcal{B} \rangle_{0}
    +\frac{\langle \mathcal{D} \rangle_{0} \langle \mathcal{C} \rangle_{0}
          +\langle \mathcal{B} \rangle_{0}}{2}
    \right]
    \left[
     \left(
      \cos\phi_{{\bm r}_{l}}\cos\phi_{{\bm r}_{l'}}-1
     \right)
     \cos
     \left(
      \theta_{{\bm r}_{l}}-\theta_{{\bm r}_{l'}}
     \right)
    \right.
   \nonumber
   \allowdisplaybreaks \\
   &
    \left.
   +\sin\phi_{{\bm r}_{l}}\sin\phi_{{\bm r}_{l'}}
    \right]
   -\left[
     \langle \mathcal{A} \rangle_{0} \langle \mathcal{D} \rangle_{0}
          \vphantom{\frac{\langle \mathcal{D} \rangle_{0} \langle \mathcal{C} \rangle_{0}
                         +\langle \mathcal{B} \rangle_{0}}{2}}
    +\frac{\langle \mathcal{B} \rangle_{0} \langle \mathcal{C} \rangle_{0}
          +\langle \mathcal{D} \rangle_{0}}{2}
    \right]
    \left[
     \left(
      \cos\phi_{{\bm r}_{l}}\cos\phi_{{\bm r}_{l'}}+1
     \right)
     \cos
     \left(
      \theta_{{\bm r}_{l}}-\theta_{{\bm r}_{l'}}
     \right)
    \right.
   \nonumber
   \allowdisplaybreaks \\
   &
    \left.
   +\sin\phi_{{\bm r}_{l}}\!\sin\phi_{{\bm r}_{l'}}
    \right]
   \left.\!\!
           \vphantom{\left[
                     \langle \mathcal{A} \rangle_{0} \langle \mathcal{D} \rangle_{0}
                    +\frac{\langle \mathcal{B} \rangle_{0} \langle \mathcal{C} \rangle_{0}
                    +\langle \mathcal{D} \rangle_{0}}{2}
                     \right]}
   \right\}
   \label{E:E(l)3D}
\end{align}
with $\langle\ \,\rangle_0$ denoting the quantum average in the DC-MLSW vacuum.
We refer to
$\varSigma_1^{(0)}(\bm{k}_\nu:\sigma\sigma')$
[Fig. \ref{F:PSE}(a) and Eq. \eqref{E:Sigma_1^(0)}] and
$\varSigma_2^{(0)}(\bm{k}_\nu:\sigma\sigma';i\omega_n)$
[Fig. \ref{F:PSE}(b) and Eq. \eqref{E:Sigma_2^(0)}]
as the primary self-energies \cite{McGrraw-Hill1971} and pick up only their iterations in
any higher-order corrections (cf. \ref{A:Akw} as well),
\begin{align}
   &
   \Delta F_2^{(-1)}
  \simeq
   \frac{1}{\beta\hbar}
   \sum_{\nu=1}^N
   \sum_{\sigma,\cdots,\sigma''''=1}^p
   \sum_{n=-\infty}^\infty
   \varSigma_1^{(0)}(\bm{k}_\nu:\sigma\sigma')
   G_0(\bm{k}_\nu:\sigma'\sigma'';i\omega_n)
   \varSigma_1^{(0)}(\bm{k}_\nu:\sigma''\sigma''')
   \nonumber
   \allowdisplaybreaks \\
   &\qquad\qquad\qquad\qquad\qquad\qquad\times
   G_0(\bm{k}_\nu:\sigma'''\sigma'''';i\omega_n),
   \label{E:approxDeltaF_2^(-1)}
   \allowdisplaybreaks \\
   &
   \Delta F_3^{(-1)}
  \simeq
   \frac{1}{\beta\hbar}
   \sum_{\nu=1}^N
   \sum_{\sigma,\cdots,\sigma''''=1}^p
   \sum_{n=-\infty}^\infty
   \varSigma_1^{(0)}(\bm{k}_\nu:\sigma\sigma')
   G_0(\bm{k}_\nu:\sigma'\sigma'';i\omega_n)
   \varSigma_2^{(0)}(\bm{k}_\nu:\sigma''\sigma''';i\omega_n)
   \nonumber
   \allowdisplaybreaks \\
   &\qquad\qquad\qquad\qquad\qquad\qquad\times
   G_0(\bm{k}_\nu:\sigma'''\sigma'''';i\omega_n),
   \label{E:approxDeltaF_3^(-1)}
   \allowdisplaybreaks \\
   &
   \Delta F_4^{(-1)}
  \simeq
   \frac{1}{\beta\hbar}
   \sum_{\nu=1}^N
   \sum_{\sigma,\cdots,\sigma''''=1}^p
   \sum_{n=-\infty}^\infty
   \varSigma_2^{(0)}(\bm{k}_\nu:\sigma\sigma';i\omega_n)
   G_0(\bm{k}_\nu:\sigma'\sigma'';i\omega_n)
   \varSigma_2^{(0)}(\bm{k}_\nu:\sigma''\sigma''';i\omega_n)
   \nonumber
   \allowdisplaybreaks \\
   &\qquad\qquad\qquad\qquad\qquad\qquad\times
   G_0(\bm{k}_\nu:\sigma'''\sigma'''';i\omega_n).
   \label{E:approxDeltaF_4^(-1)}
\end{align}
Such diagrams are emergent in
$\Delta F_2^{(-1)}$, $\Delta F_3^{(-1)}$, and $\Delta F_4^{(-1)}$
but absent in $\Delta F_1^{(-1)}$.
\begin{figure}
\begin{flushright}
\includegraphics[width=130mm]{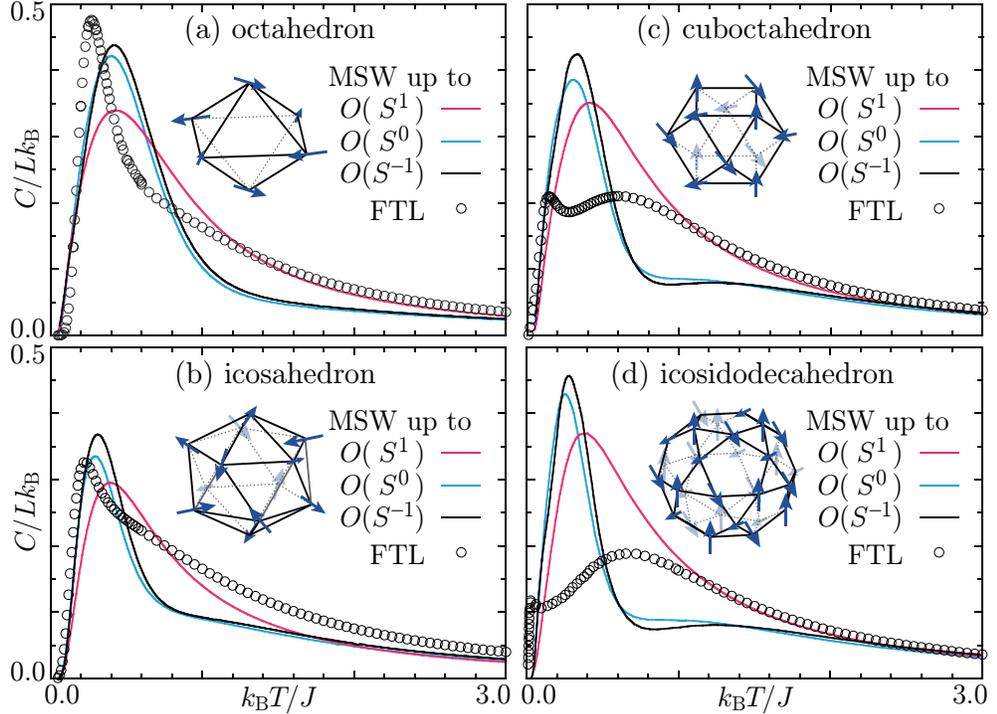}
\end{flushright}
\vspace{-7mm}
\caption{DC-MLSW and PC-DC-MSW calculations of the specific heat $C\equiv\partial E/\partial T$
         [\eqref{E:E(T)DC-MLSW}, \eqref{E:E(T)PC-DC-MSWuptoS^0}, \eqref{E:E(T)PC-DC-MSWuptoS^-1}]
         as a function of temperature for the $S=\frac{1}{2}$ Hamiltonian \eqref{E:HeisenbergH}
         on the octahedral (a), icosahedral (b), cuboctahedral (c), and icosidodecahedral (d)
         lattices in comparison with finite-temperature Lanczos calculations (FTL).}
\label{F:DC-MSW&FTLofCin0D}
\end{figure}

\subsection{Perturbative double-constraint modified-spin-wave thermodynamics}
\label{SS:PDCMSWThD}

\subsubsection{Triangular-based polyhedral-lattice antiferromagnets.}
\label{SSS:TBPAFM}

   Intending to verify to what extent our PC-DC-MSW scheme can depict the thermal features of
Heisenberg antiferromagnets in the triangular-based geometry, we first investigate various
polyhedral-lattice antiferromagnets whose thermal features can precisely be calculated by
the finite-temperature Lanczos method \cite{J5065,J1}.
In Fig. \ref{F:DC-MSW&FTLofCin0D}, we show the DC-MSW findings in comparison with
these numerically exact solutions.
As for their classical ground states from which Holstein-Primakoff bosons emerge,
the spins are not coplanar with a relative angle of $116.6^\circ$ between nearest neighbors
in the icosahedral antiferromagnet \cite{A100407}, whereas all spins are (assumed to be) coplanar
with a relative angle of $120^\circ$ between nearest neighbors in the rest three each
\cite{S6351,A100407}.
Similar to the kagome-lattice antiferromagnet, the cuboctahedral antiferromagnet has noncoplanar
singlet ground states degenerate in energy with the coplanar one \cite{S6351}.
These ground states of the four polyhedral-lattice antiferromagnets are schematically shown in
Fig. \ref{F:DC-MSW&FTLofCin0D}.
As for the boson number tuning parameter $\delta$,
we calculate the up-to-$O(S^m)$ ($m=1,0,-1$) DC-MSW ground-state energies $E|_{T=0}$
[\eqref{E:E(T)DC-MLSW}, \eqref{E:E(T)PC-DC-MSWuptoS^0}, \eqref{E:E(T)PC-DC-MSWuptoS^-1}]
as functions of $\delta$ given by hand and find the minimum point of each curve.
The thus-obtained optimal values of $\delta$ are listed in Table \ref{T:Optimaldelta}.
\begin{figure}
\begin{flushright}
\includegraphics[width=130mm]{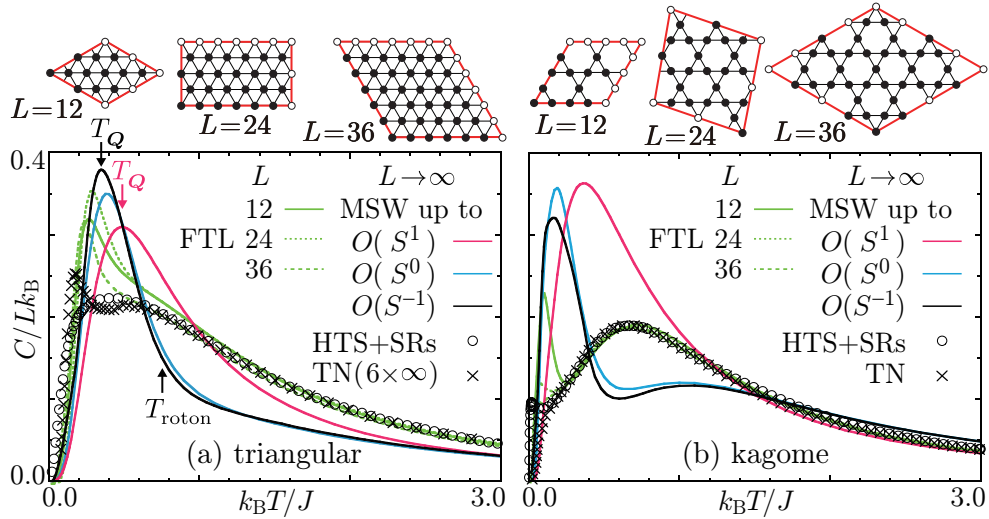}
\end{flushright}
\vspace*{-7mm}
\caption{DC-MLSW and PC-DC-MSW calculations of the specific heat $C\equiv\partial E/\partial T$
         [\eqref{E:E(T)DC-MLSW}, \eqref{E:E(T)PC-DC-MSWuptoS^0}, \eqref{E:E(T)PC-DC-MSWuptoS^-1}]
         as a function of temperature for the $S=\frac{1}{2}$ Hamiltonian \eqref{E:HeisenbergH}
         on the $L\rightarrow\infty$ triangular (a) and kagome (b) lattices in comparison with
         optimal interpolations of Pad\'e approximants to high-temperature series subject to
         the energy and entropy sum rules ($\mathrm{HTS}+\mathrm{SRs}$)
         \cite{B134409,M014417,B140403}
         and tensor-network-based renormalization-group calculations (TN) \cite{C1545,C140404}.
         The findings referred to as TN in (b) \cite{C1545} are obtained on the infinite
         two-dimensional kagome lattice, while those as TN ($6\times\infty$) in (a) \cite{C140404}
         are extrapolations to the infinite cylindrical triangular lattice with hexagonal ends
         aslant.
         Finite-temperature Lanczos calculations (FTL) for small periodic clusters ($L=12,24,36$)
         \cite{S235115,S094423} of the same Hamiltonian are also shown for reference.
         We denote the temperatures at which the up-to-$O(S^1)$ specific heat
         $C\equiv\partial E_{\mathrm{harm}}/\partial T$ \eqref{E:E(T)DC-MLSW} and
         up-to-$O(S^{-1})$ specific heat $C\equiv\partial E/\partial T$
         \eqref{E:E(T)PC-DC-MSWuptoS^-1} reach their maxima by $T_{\bm{Q}}$,
         which measure $0.4904J/k_{\mathrm{B}}$ and $0.3526J/k_{\mathrm{B}}$, respectively,
         and the temperature corresponding to the roton gap located at M in the Brillouin zone
         by $T_{\mathrm{roton}}$, which measures $0.7579J/k_{\mathrm{B}}$
         [cf. Fig. \ref{F:Akw}(b)].}
\label{F:DC-MSW&FTL&TNofCin2D}
\end{figure}
\begin{table}
\begin{minipage}{0.475\textwidth}
\centering
\begin{tabular}{lccc}
\hline \hline
\ DC MSW up to         & $ O(S^{ 1})$ & $ O(S^{ 0})$ & $ O(S^{-1})$ \\[0.5mm]
\hline
\ Octahedron           & $\ 0.1601\ $ & $\ 0.1624\ $ & $\ 0.1793\ $ \\
\ Icosahedron          & $\ 0.2702\ $ & $\ 0.2910\ $ & $\ 0.2685\ $ \\
\ Cuboctahedron        & $\ 0.1592\ $ & $\ 0.1623\ $ & $\ 0.1955\ $ \\
\ Icosidodecahedron\ \ & $\ 0.1325\ $ & $\ 0.1525\ $ & $\ 0.1580\ $ \\
\ Triangular           & $\ 0.2653\ $ & $\ 0.2851\ $ & $\ 0.2875\ $ \\
\ Kagome               & $\ 0.1554\ $ & $\ 0.1650\ $ & $\ 0.1633\ $ \\[0.5mm]
\hline \hline
\end{tabular}
\end{minipage}
\begin{minipage}{0.50\textwidth}
\caption{Such values of the boson number tuning parameter $\delta$ as to minimize
         the up-to-$O(S^m)$ ($m=1,0,-1$) DC-MSW ground-state energies $E|_{T=0}$
         [\eqref{E:E(T)DC-MLSW}, \eqref{E:E(T)PC-DC-MSWuptoS^0}, \eqref{E:E(T)PC-DC-MSWuptoS^-1}]
         for various noncollinear Heisenberg antiferromagnets in the triangular-based geometry.}
\label{T:Optimaldelta}
\end{minipage}
\end{table}

   The octahedron and icosahedron consist of edge-sharing triangles, while the cuboctahedron and
icosidodecahedron can be viewed as corner-sharing triangles.
The former and latter antiferromagnetic spin spirals yield such temperature profiles of
the specific heat as to bear remarkable resemblance to those of antiferromagnetic small clusters
of periodic triangular \cite{S235115,S073025,P035107} and kagome \cite{S094423,S010401,M305}
lattices (cf. Fig. \ref{F:DC-MSW&FTL&TNofCin2D}),
marked by
a quite sharp low-temperature maximum followed by a long gentle down slope from
intermediate to high temperatures \cite{K064453} and
a main round maximum at intermediate temperatures accompanied by an additional low-temperature
modest peak \cite{S535,N041101,S042110}, respectively.
The PC-DC-MSW findings for the former reproduce these temperature profiles quite well,
while those for the latter are indeed quantitatively less precise but intriguingly suggestive of
a distinct double-peak temperature profile.
These successful findings are not the case with DC MLSWs.
Even DC MLSWs may be better than SC MLSWs \cite{C280} at reproducing the overall
temperature profiles of thermal quantities, but they neither depict
the low-temperature steep peak of the antiferromagnetic specific heat characteristic of
the edge-sharing triangular geometry
nor hint at the double-peck structure of the antiferromagnetic specific heat peculiar to
the corner-sharing triangular geometry.
It is not until the fractional-power interactions $\mathcal{H}^{\left(\frac{m}{2}\right)}$ become
effective that DC MSWs give a fairly good description of the thermal features in various
triangular-based geometries.

\subsubsection{Triangular-based planar-lattice antiferromagnets.}
\label{SSS:TBPAFM}
\begin{table}[t]
\caption{DC-MSW estimates of the high-temperature asymptotic specific heat coefficient
         $A\equiv\lim_{T\rightarrow\infty}(k_{\mathrm{B}}T/J)^2C/k_{\mathrm{B}}$
         [cf. \eqref{E:C(T=oo)exact} and \eqref{E:C(T=oo)DC-MLSW}]
         in comparison with the exact values $z[S(S+1)]^2/6$
         for the $S=\frac{1}{2}$ Hamiltonian \eqref{E:HeisenbergH}
         on the $L\rightarrow\infty$ triangular and kagome lattices.}
\begin{center}
\begin{tabular}{lcccc}
\hline \hline
\ DC MSW up to\ & $ O(S^{ 1})$ & $ O(S^{ 0})$ & $ O(S^{-1})$ & exact        \\[0.5mm]
\hline
\ Triangular     & $\ 0.3674\ $ & $\ 0.6877\ $ & $\ 0.6751\ $ & $\ 0.5625\ $ \\
\ Kagome         & $\ 0.3350\ $ & $\ 0.4134\ $ & $\ 0.4104\ $ & $\ 0.3750\ $ \\
\hline \hline
\end{tabular}
\end{center}
\label{T:HighTcoefficientAofC}
\end{table}

   Now we proceed to the two-dimensional triangular- and kagome-lattice antiferromagnets
in the thermodynamic limit of our main interest.
In Fig. \ref{F:DC-MSW&FTL&TNofCin2D}, we present the PC-DC-MSW findings, whose optimal values of
$\delta$ are also listed in Table \ref{T:Optimaldelta}, in comparison with
numerical findings obtained by the high-temperature-series-expansion-based entropy method and
tensor-network-based renormalization-group approach.
Small clusters of periodic triangular lattice exhibit a quite sharp low-temperature maximum
in their temperature profiles of the specific heat in common,
while those of periodic kagome lattice exhibit a prominent low-temperature peak besides the main
round maximum in common.
The entropy method claims that the former and latter features do not and do survive in
the thermodynamic limit, respectively, whereas
the tensor-network approach claims that the former and latter features do and do not survive in
the thermodynamic limit, respectively.
The thermodynamic-limit properties of frustrated noncollinear antiferromagnets are thus hard to
evaluate even by such sophisticated methods.
Under such circumstances, our elaborate PC-DC-MSW theory predicts that
\textit{the former and latter features both survive in the thermodynamic limit}.
What a precious message from SWs!
This is their first clear report on thermodynamics of frustrated noncollinear quantum magnets.
PC DC MSWs succeed in reproducing such distinctive specific-heat curves of triangular-based
edge-sharing Platonic and corner-sharing Archimedean polyhedral-lattice antiferromagnets as well.
Note that MLSWs are ignorant of any individual fine structure of the specific heat.
Not only the elaborate modification scheme but also quantum corrections, especially those
caused by the $O(S^0)$ primary self-energies, are key ingredients in this context,
as will be discussed in more detail in Sect. \ref{S:SandD}.

   It is generally very hard to provide a precise thermodynamics of two- or higher-dimensional
noncollinear antiferromagnets over the whole temperature range of absolute zero to infinity.
Under such circumstances, our PC DC MISWs can give not only a qualitative description but also
semiquantitative information of thermal features as follows.
The exact high-temperature asymptotics of the specific heat is given by
\begin{align}
   \lim_{T\rightarrow\infty}
   \frac{C}{k_{\mathrm{B}}}
  =\frac{z}{6}
   \left[
    \frac{JS(S+1)}{k_{\mathrm{B}}T}
   \right]^2
  \equiv
   A\left(\frac{J}{k_{\mathrm{B}}T}\right)^2.
   \label{E:C(T=oo)exact}
\end{align}
DC MLSWs say that
\begin{align}
   \lim_{T\rightarrow\infty}
   \frac{C}{k_{\mathrm{B}}}
  =\frac{S-\delta+\frac{1}{2}}{\mathrm{ln}\left(1+\frac{1}{S-\delta}\right)}
   \left(\frac{JS}{k_{\mathrm{B}}T}\right)^2
   \sum_{\nu=1}^N\sum_{\sigma=1}^p
   \left(
    \frac{3}{4}z\gamma_{\bm{k}_\nu:\sigma}
   \right)^2
   \label{E:C(T=oo)DC-MLSW}
\end{align}
and the coefficient $A$ quantitatively ameliorates with perturbative corrections,
as is demonstrated in Table \ref{T:HighTcoefficientAofC}.
While PC DC MISWs overestimate and underestimate the specific heat at low and intermediate
temperatures, respectively, they satisfy fairly well a sum rule on the entropy,
\begin{align}
   \int_0^\infty\frac{C}{Lk_{\mathrm{B}}T}dT
  =\int_{T=0}^{T=\infty}\frac{d\mathcal{S}}{Lk_{\mathrm{B}}}
  =\mathrm{ln}(2S+1),
   \label{E:EntropySumRule}
\end{align}
as is demonstrated in Table \ref{T:EntropySumRule}.
Considering that a direct Pad\'e analysis of high-temperature expansion series of the specific heat
generally fails to satisfy this entropy sum rule \cite{B134409,M014417,S042110},
it is even surprising that our MSWs remain quantitative to this extent at high temperatures.
Although they are not necessarily successful in precisely reproducing a thermal quantity
at each temperature, yet they never fail to predict the overall thermal features.
This is because the DC-MLSW dispersion relations to construct the PC-DC-MSW thermodynamics
are equipped with the key ingredients in the low-lying energy spectra of
the triangular- and kagome-lattice antiferromagnets, which we shall further explain
in the next closing section.
\begin{table}
\begin{minipage}{0.48\textwidth}
\centering
\begin{tabular}{lccc}
\hline \hline
\ DC MSW up to & $ O(S^{ 1})$ & $ O(S^{ 0})$ & $ O(S^{-1})$ \\[0.5mm]
\hline
\ Triangular   & $\ \mathrm{ln}1.8230\ $ & $\ \mathrm{ln}1.7463\ $ & $\ \mathrm{ln}1.8267\ $ \\
\ Kagome       & $\ \mathrm{ln}2.1495\ $ & $\ \mathrm{ln}2.1030\ $ & $\ \mathrm{ln}2.1430\ $ \\
\hline \hline
\end{tabular}
\end{minipage}
\begin{minipage}{0.50\textwidth}
\caption{DC-MSW estimates of the entropy sum rule \eqref{E:EntropySumRule}
         for the $S=\frac{1}{2}$ Hamiltonian \eqref{E:HeisenbergH}
         on the $L\rightarrow\infty$ triangular and kagome lattices,
         compared to the exact value $\mathrm{ln}2$.}
\label{T:EntropySumRule}
\end{minipage}
\end{table}
\begin{figure}
\begin{flushright}
\includegraphics[width=130mm]{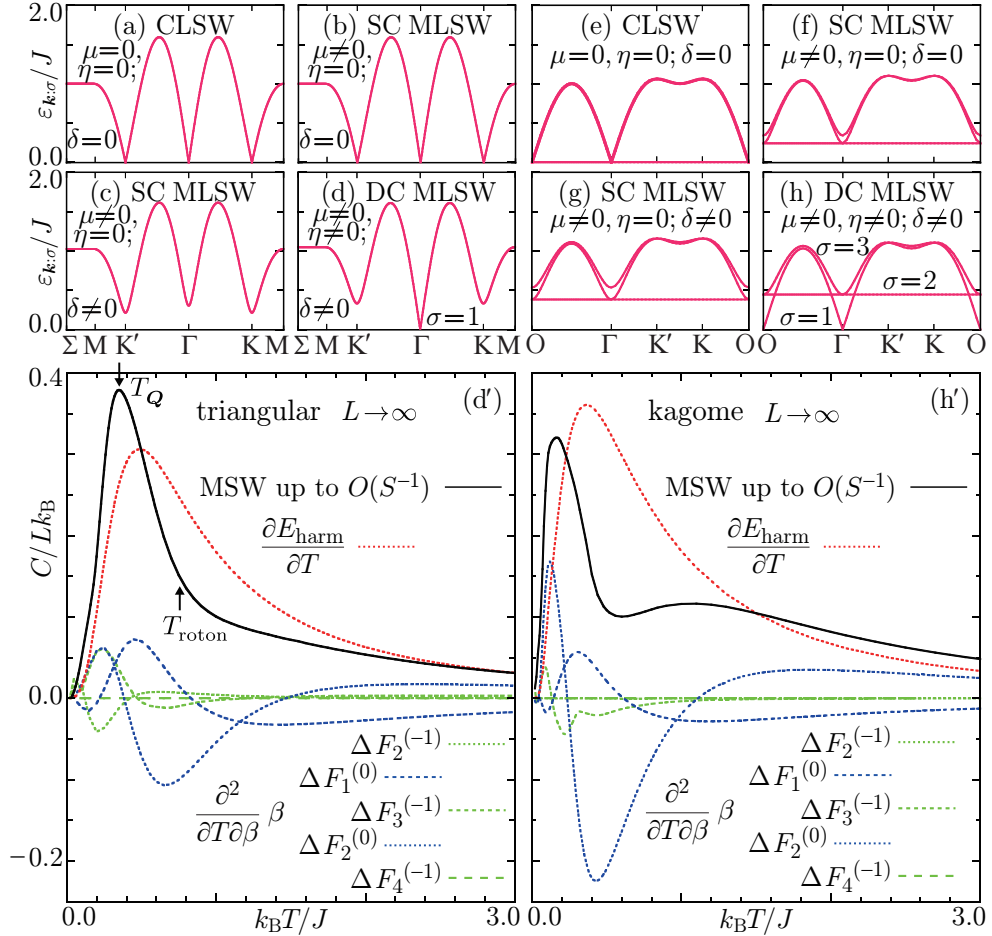}
\end{flushright}
\vspace*{-7mm}
\caption{CLSW [(a), (e)], traditional SC-MLSW [(b), (f)],
         artificially tuned SC-MLSW [$\delta=0.2653$ (c), $\delta=0.1554$ (g)], and
         DC-MLSW [(d), (h)] dispersion relations and
         up-to-$O(S^{-1})$ PC-DC-MSW calculations of the specific heat
         $C\equiv\partial E/\partial T$ \eqref{E:E(T)PC-DC-MSWuptoS^-1}
         as a function of temperature [($\mathrm{d}'$), ($\mathrm{h}'$)]
         obtainable from the DC-MLSW eigenspectra [(d), (h)]
         for the $S=\frac{1}{2}$ Hamiltonian \eqref{E:HeisenbergH} on the $L\rightarrow\infty$
         triangular [(a)--(d), ($\mathrm{d}'$)] and kagome [(e)--(h), ($\mathrm{h}'$)] lattices.
         The CLSW spectra are free from temperature dependence, while all the MLSW spectra,
         depending on temperature through the chemical potentials $\mu$ and $\eta$, are shown
         at $T=0$.
         The wavevectors move along the high-symmetry paths specified in
         Fig. \ref{F:VariousLattices}.
         On the triangular lattice, the ordering vectors
         $\bm{Q} \equiv\frac{1}{a}\left(\frac{4\pi}{3},0,0\right)$ and
         $\bm{Q}'\equiv\frac{1}{a}\left(\frac{2\pi}{3},0,\frac{2\pi}{\sqrt{3}}\right)$
         correspond to $\mathrm{K}$ and $\mathrm{K}'$, respectively,
         while on the kagome lattice, the ordering vectors
         $\bm{Q} \equiv\frac{1}{2a}\left(\frac{8\pi}{3},0,0\right)$ and
         $\bm{Q}'\equiv\frac{1}{2a}\left(\frac{4\pi}{3},0,\frac{4\pi}{\sqrt{3}}\right)$
         correspond to $\mathrm{K}$ and $\mathrm{K}'$, respectively.
         The specific-heat curves each are decomposed into
         the $O(S^1)$ MLSW contribution $\partial E_{\mathrm{harm}}/\partial T$ and
         the $O(S^m)$ $l$th-order perturbative corrections
         $\partial^2\beta\Delta F_l^{(m)}/\partial T\partial\beta$.
         In ($\mathrm{d}'$),
         we denote the temperature at which the specific heat
         $C\equiv\partial E/\partial T$ \eqref{E:E(T)PC-DC-MSWuptoS^-1}
         reaches its maximum by $T_{\bm{Q}}$, which measures $0.3526J/k_{\mathrm{B}}$,
         and the temperature corresponding to the roton gap located at M in the Brillouin zone
         by $T_{\mathrm{roton}}$, which measures $0.7579J/k_{\mathrm{B}}$
         [cf. Fig. \ref{F:Akw}(b)].}
\label{F:DispertionRelations}
\end{figure}

\section{Summary and discussion}
\label{S:SandD}
\subsection{Kagome-lattice Heisenberg antiferromagnet}
\label{SS:KLHAFM}
The low-energy Lanczos spectrum of the $S=\frac{1}{2}$ antiferromagnetic Heisenberg model
for a Kagome cluster of $L=48$ consists of a huge number of singlet states below the lowest-lying
triplet state \cite{L155142}.
With further increasing size $L$, the lowest triplet remains separated from the ground state by
a gap \cite{Y1173,D067201}, whereas the singlets in between presumably develop into a gapless
continuum adjacent to the ground state \cite{J117203,L155142,S094423}.
While the singlet-singlet gap strongly depends on which boundary condition is adopted,
toric or cylindric, the fact remains unchanged that it is always smaller than the singlet-triplet
gap \cite{Y1173}.
With all these in mind, let us observe MSW eigenspectra under various constraint conditions.
We show in Fig. \ref{F:DispertionRelations} the DC-MLSW dispersion relations and the consequent
up-to-$O(S^{-1})$ PC-DC-MSW specific-heat curves, together with the CLSW and SC-MLSW
dispersion relations, for the triangular- and kagome-lattice antiferromagnets, where
the DC-MLSW specific heat $\partial E_{\mathrm{harm}}/\partial T$ and
the perturbative corrections $\partial^2\beta\Delta F_l^{(m)}/\partial T\partial\beta$ to that
each are also shown separately.
CLSWs of the kagome-lattice Heisenberg antiferromagnets [Fig. \ref{F:DispertionRelations}(e)]
are well known to have a dispersionless zero-energy mode \cite{H2899,C832}, resulting that
we can neither calculate any kind of magnetic susceptibility nor evaluate any further perturbative
correction to the harmonic energy $E_{\mathrm{harm}}$ even at absolute zero \cite{C144415}.
Modifying these CLSWs under the traditional SC condition \eqref{E:collinearMSWconstraintHPB} pushes up
their whole eigenspectrum, lifting the degeneracy between the two dispersive bands
[Fig. \ref{F:DispertionRelations}(f)].
Further modifying them under the DC condition,
\eqref{E:noncollinearMSWconstraintHPB1st} plus \eqref{E:noncollinearMSWconstraintHPB2nd},
puts one of the dispersive modes back into its original gapless appearance,
keeping the flat band apart from the ground state [Fig. \ref{F:DispertionRelations}(h)].
Note that any artificially tuned SC condition, \eqref{E:noncollinearMSWconstraintHPB1st} without
\eqref{E:noncollinearMSWconstraintHPB2nd} but with a nonzero $\delta$, brings no qualitative change
into Fig. \ref{F:DispertionRelations}(f), i.e., induces no eigenstate to emerge below the flat band
[Fig. \ref{F:DispertionRelations}(g)].
The second constraint \eqref{E:noncollinearMSWconstraintHPB2nd} is necessary to recover a linear
Goldstone mode looking up at the floating flat band.
\begin{figure}[h]
\centering
\begin{minipage}{0.4\textwidth}
\centering
\includegraphics[width=85mm]{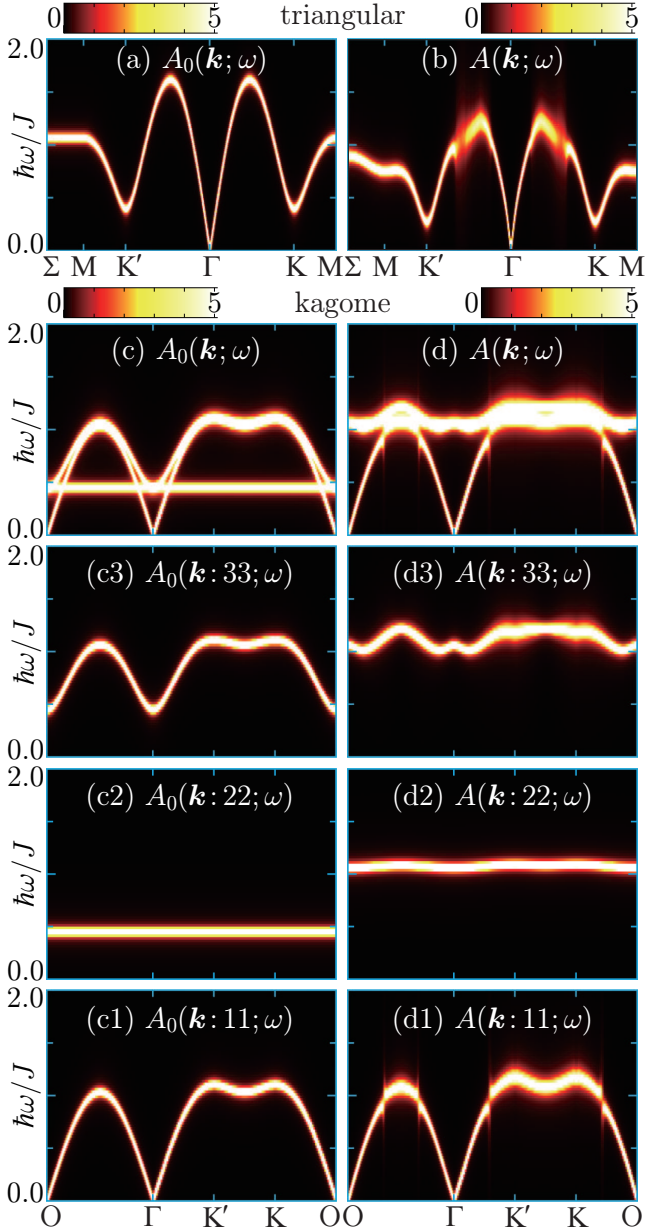}
\end{minipage}
\begin{minipage}{0.50\textwidth}
\caption{Intensity plots of the momentum and energy dependences of the DC-MLSW spectral function
         $A_0(\bm{k};\omega)\equiv\sum_{\sigma=1}^p A_0(\bm{k}:\sigma\sigma;\omega)$
         \eqref{E:A0kw} and its perturbative renormalization
         $A  (\bm{k};\omega)\equiv\sum_{\sigma=1}^p A  (\bm{k}:\sigma\sigma;\omega)$
         \eqref{E:Akw} evaluated within the on-shell approximation
         at $T=0$ for the $S=\frac{1}{2}$ Hamiltonian \eqref{E:HeisenbergH} on
         the $L\rightarrow\infty$ triangular [(a), (b)] and kagome
         [$(\mathrm{c})\equiv\sum_{\sigma=1}^3(\mathrm{c}\sigma)$,
          $(\mathrm{d})\equiv\sum_{\sigma=1}^3(\mathrm{d}\sigma)$] lattices.
         In the DC-MLSW calculations [(a), (c), (c1)--(c3)], every $\delta$-function peak
         [cf. Figs. \ref{F:DispertionRelations}(d), \ref{F:DispertionRelations}(h)]
         is Lorentzian-broadened with the use of an ``artificial" decay rate $\hbar\eta=0.01J$.
         Unlike bare magnons \eqref{E:tildeHharmdiag}, renormalized magnons \eqref{E:RenormEnergy}
         may have a finite decay rate of their own origin,
         $\varGamma_{\bm{k}:\sigma}$ \eqref{E:RenormDispRel&DecayRate}.
         Nevertheless we set their $\hbar\eta$ as well to $0.01J$;
         one is for the sake of comparison and another is for $\varGamma_{\bm{k}:\sigma}$ not being
         necessarily large enough to see over the whole Brillouin zone [(b), (d), (d1)--(d3)].
         The wavevectors move along the high-symmetry paths specified in
         Fig. \ref{F:VariousLattices}.}
\label{F:Akw}
\end{minipage}
\end{figure}

   Such an elaborate DC-MLSW spectrum bears some resemblance to the low-energy Lanczos spectrum
in that the dispersionless DC-MLSW mode may correspond to the low-lying triplet eigenstates
separated from the ground state by a gap and the linear DC-MLSW mode slipping under the flat band
can serve as the singlet eigenstates filling the singlet-triplet spin gap.
Although neither the total spin nor the total magnetization is a good quantum number
for the DC-MLSW Hamiltonian \eqref{E:tildeHharmYO}, yet its dispersionless and thus localized
excitation mode possibly has close relevance to the singlet-triplet spin gap.
The flat band of CLSWs in the kagome-lattice antiferromagnet consists of excitations localized
within an arbitrary hexagon of nearest-neighbor spins \cite{H2899}, and so is that of MLSWs,
no matter which modification scheme is imposed, SC or DC, and though the excitation energy is
no longer zero.
Interestingly enough, the VBC with a $36$-site unit cell \cite{M5962}, which is one of the most
promising candidate ground states for the kagome-lattice antiferromagnetic Heisenberg model,
has low-lying triplet excitations localized inside a hexagon of nearest-neighbor spins.
It actually has $18$ triplet modes in the reduced Brillouin zone and many of them are
dispersionless \cite{S180407,Y224424}.
In a bond operator mean-field theory \cite{Y224424}, the second-lowest flat mode, which lies
slightly higher than the lowest flat mode defining the spin gap, is completely localized within
either of the two ``perfect hexagons" in the $36$-site unit cell and therefore doubly degenerate.
Series expansions around the decoupled-dimer limit \cite{S180407} reveal doubly degenerate
flat modes of the same nature lying lowest in the spin-triplet channel.
They are indeed dispersionless in second-order perturbation theory \cite{S180407}, being
completely localized inside a perfect hexagon, but turn dispersive to the extent of about
one-hundredth of $J$ to third order in the expansion \cite{S144415}, weakly hopping from
one perfect hexagon to the next.
Our DC-MLSW flat mode [Fig. \ref{F:Akw}(c2)] also becomes weakly dispersive to the same extent
[Fig. \ref{F:Akw}(d2)] when it is renormalized with the primary self-energies
\eqref{E:Sigma_2^(0)}
[\eqref{E:Sigma_1^(0)} have no effect in this context, to be precise].
The primary self-energies \eqref{E:Sigma_1^(0)} push the gapped two modes upward
[Figs. \ref{F:Akw}(d2), \ref{F:Akw}(d3); \eqref{E:Sigma_2^(0)} have no effect in this context,
to be precise], retaining the Goldstone mode [Fig. \ref{F:Akw}(d1)],
so that a prominent low-temperature peak can emerge out of the main round maximum in
the temperature profile of the kagome-lattice-antiferromagnet specific heat.
It is likely that the four-boson scattering $\mathcal{H}^{(0)}$ is responsible for locating
the major broad maximum, while the three-boson scattering $\mathcal{H}^{\left(\frac{1}{2}\right)}$
designs the additional low-temperature modest peak.
For further details of spectral-function calculations, see \ref{A:Akw}.

   Once again, neither the well-pronounced low-temperature peak nor the mid-temperature broad
maximum is obtainable within any SC-MSW thermodynamics.
Just like the singlet states filling the spin gap in the Lanczos spectrum,
the sufficient density of antiferromagnon eigenstates lying below the flat band properly apart from
the ground state enables DC MSWs to reproduce the bimodal temperature profile of
the kagome-lattice-antiferromagnet specific heat.
It is not the case at all with DC MLSWs
[MSW up to $O(S^1)$ in Fig. \ref{F:DC-MSW&FTL&TNofCin2D}(b)]
but is indeed the case with up-to-$O(S^0)$ or higher DC MSWs
[MSW up to $O(S^0)$ and $O(S^{-1})$ in Fig. \ref{F:DC-MSW&FTL&TNofCin2D}(b)].
The bimodal temperature profile owes much to the second-order perturbative correction
$\Delta F_2^{(0)}$ originating from the three-boson scattering
$\mathcal{H}^{\left(\frac{1}{2}\right)}$ peculiar to noncollinear antiferromagnets.
The marked ups and downs of $\partial^2\beta\Delta F_2^{(0)}/\partial T\partial\beta$
[Fig. \ref{F:DispertionRelations}($\mathrm{h}'$)] are indeed thanks to the flat-band
[Fig. \ref{F:Akw}(c2)] contribution $\sigma=\sigma'=2$ in \eqref{E:DeltaF_2^(0)toSigma_2^(0)}.
We further find that $\partial\beta\Delta F_1^{(0)}/\partial\beta>0$,
while $\partial\beta\Delta F_2^{(0)}/\partial\beta<0$, i.e.,
the former and latter read repulsive and attractive interactions, respectively.
Considering that
at sufficiently low temperatures, the latter yields a remarkable amount of entropy,
whereas the former slightly cancel this
[See $k_{\mathrm{B}}T/J \lesssim 0.2$ in Fig. \ref{F:DispertionRelations}($\mathrm{h}'$)],
it is quite possible that bound states of antiferromagnons localized within a hexagon of
nearest-neighbor spins play a significant role in reproducing the low-temperature entropy
distinctive of the kagome-lattice antiferromagnet.
Indeed, Singh and Huse \cite{S144415} stated that low-lying triplet excitations against
the VBC with a $36$-site unit cell, including those localized inside a perfect hexagon,
attract one another and form many bound states in the spin-singlet channel.

\subsection{Triangular-lattice Heisenberg antiferromagnet}
\label{SS:TLHAFM}
Though CLSWs do not have a dispersionless zero-energy mode in the triangular-lattice
antiferromagnet \cite{J2727}, the fact remains that they cannot describe any magnetic properties
as functions of temperature at all.
Although SC MLSWs at $T=0$ have the same dispersion relation as CLSWs
[Figs. \ref{F:DispertionRelations}(a) and \ref{F:DispertionRelations}(b)],
yet SC MLSWs are no longer useless in thermodynamics with their chemical potential $\mu$ 
effective at every finite temperature, similar to those in the square-lattice antiferromagnet
\cite{Y094412}.
However, they misread the per-site static uniform susceptibility $\chi/L$ as divergent with
$T\rightarrow 0$ [Fig. \ref{F:SC&tunedSC&DC-MLSWin2D}(a)],
unlike those in the square-lattice antiferromagnet \cite{Y094412}.
For the square-lattice antiferromagnet, the MLSW Hamiltonian $\widetilde{{\cal H}}_{\mathrm{harm}}$
\eqref{E:tildeHharmT} commutes with the total magnetization $\mathcal{M}^z$,
so that
$
   \lim_{T\rightarrow 0}
   \left[
    \langle(\mathcal{M}^z)^{2}\rangle_{T}
   -\langle \mathcal{M}^z     \rangle_{T}^{2}
   \right]
$
becomes zero and $\lim_{T\rightarrow 0}\chi^{zz}/L$ stays finite [cf. \eqref{E:chiCollinear}],
while its constraint condition \eqref{E:collinearMSWconstraintHPB} makes both
$\chi^{xx}$ and $\chi^{yy}$ vanish [cf. \eqref{E:chixx&yyCollinear}].
SC MLSWs well sketch the square-lattice-antiferromagnet thermodynamics over the whole temperature
range of absolute zero to infinity, indeed \cite{Y094412}.
Although SC MLSWs soften at $\bm{k}=\pm\bm{Q}$ and $\bm{k}=\bm{0}$ with $T\rightarrow 0$ in
both the square- and triangular-lattice antiferromagnets, they yield convergent and divergent
$\chi/L$ with $T\rightarrow 0$ in the former and latter, respectively.
The difficulty in the latter lies in the expression \eqref{E:chizz&xxNoncollinear}.
In \eqref{E:chizz&xxNoncollinear},
$\lim_{T\rightarrow 0}\chi^{zz}/L$ and $\lim_{T\rightarrow 0}\chi^{xx}/L$ both diverge if
$\lim_{T\rightarrow 0}\varepsilon_{\pm\bm{Q}:1}=0$,
while in \eqref{E:chiyyNoncollinear},
$\lim_{T\rightarrow 0}\chi^{yy}/L$ vanishes even if
$\lim_{T\rightarrow 0}\varepsilon_{\bm{0}:1}=0$
by virtue of the vanishing factor
$\lim_{T\rightarrow 0}(\cosh 2\theta_{\bm{0}:1}-\sinh 2\theta_{\bm{0}:1})$.
Only the DC-MLSW dispersion relation Fig. \ref{F:DispertionRelations}(d)
reasonably---in the sense of minimizing the ground-state energy and keeping the excitation spectrum
gapless---suppresses the divergence of $\chi/L$ at $T=0$.
We can bring DC MLSWs into interaction with the use of the primary self-energies to give a still
better description of the susceptibility.
The tensor-network-based renormalization-group calculations \cite{C140404} are still limited to
particular geometries and subject to their boundaries at low temperatures of increasing interest.
Whether we use $d$-log Pad\'e and integrated differential approximants \cite{Z134422} or
employ particular sequence extrapolation techniques to accelerate the convergence \cite{R061118},
the extrapolations of high-temperature series work well down to the peak temperature but
begin to deviate from each other below the peak.
Under such circumstances, we may expect many useful pieces of information of
the DC-MSW thermodynamics reachable to both infinite-lattice and low-temperature limits.

   Developing an exponential tensor renormalization-group method on cylinder- and strip-shaped
triangular lattices, Chen \textit{et al.} \cite{C140404} calculated their thermal properties and
especially revealed two generic temperature scales,
$0.2\lesssim k_{\mathrm{B}}T_{\mathrm{low}}/J \lesssim 0.28$ and
$k_{\mathrm{B}}T_{\mathrm{high}}/J\simeq 0.55$,
at which the specific heat reaches its local maximum or exhibit a shoulder-like anormaly.
They claimed that $T_{\mathrm{low}}$ corresponds with the onset of the ``renormalized classical"
behavior \cite{C1057,C2344} that would be expected from a relevant semiclassical nonlinear sigma
model, whereas $T_{\mathrm{high}}$ originates in the quadratic roton-like excitation band
\cite{Z057201,Z224420}.

   Chakravarty, Halperin, and Nelson \cite{C1057,C2344} claimed that when a planar-lattice quantum
Heisenberg antiferromagnet has an ordered ground state, its long-wavelength behavior at certain
finite temperatures, which they designate as the renormalized classical regime, can be described
by an effective classical nonlinear sigma model in two spatial dimensions obtainable from
the pertinent quantum model in two spatial plus one temporal dimensions.
In this regime the correlation length reads
\begin{align}
   \frac{\xi}{a}
  =A
   \left(
    \frac{k_{\mathrm{B}}T}{B}
   \right)^x
   \mathrm{exp}
   \left(
    \frac{B}{k_{\mathrm{B}}T}
   \right)
   \left[
    1
   +O\left(
      \frac{k_{\mathrm{B}}T}{J}
     \right)
   \right],
   \label{E:RCxi}
\end{align}
where $A$ and $B$ depend on the spin-wave velocities and stiffness coefficients for twisting
the spins, both being renormalized by the quantum fluctuations at $T=0$.
The constant $A$ and power exponent $x$ are nonuniversal numbers in the sense that
they depend on our modeling, such as whether the Hamiltonian is a nearest-neighbor exchange model
or contains next-nearest-neighbor interactions in addition.
Still further, they depend on to which order we calculate, such as whether within
the one-loop approximation or up to two-loop corrections.
On the other hand, $B$ unambiguously sets the temperature scale for the correlations.
Vanishing $B$ signifies such strong quantum fluctuations as to destroy the long-range
magnetic order, but otherwise we can identify two distinct regions
separated by the crossover temperature $k_{\mathrm{B}}T=B$,
i.e., the renormalized classical behavior \cite{C1057,C2344} at low temperatures with
correlation length ultimately diverging exponentially as $T\rightarrow 0$ and quantum critical
behavior \cite{C1057,C2344,C11919} at intermediate temperatures with inverse correlation length
given by a certain power function of $T$.
Studying both quantum and classical---here in the sense of integrating out all quantum fluctuations
to obtain an effective classical model and neglecting any imaginary-time-derivative of
the three-component vector field from the beginning, respectively---nonlinear sigma models
suited to frustrated planar-lattice antiferromagnets assuming a noncollinearly ordered ground state,
Azaria, Delamotte, and Mouhanna \cite{A1762} calculated the two-loop
[in the context of calculating the (effective) classical model
(obtainable via the one-loop renormalization of the coupling constants)]
correlation length intending to describe quantum and classical Heisenberg antiferromagnets on
the triangular lattice at low temperatures.
They argued that the correlation length in the quantum model \cite{A1762} still diverges as
\eqref{E:RCxi}, where
\begin{align}
   &
   x
 =-\frac{1}{2},\ 
   B
  =\pi\tilde{\rho}_{\mathrm{s}}^\perp G(\tilde{\alpha})
  \equiv
   B_{\mathrm{quant}};\ 
   G(\tilde{\alpha})
  \equiv
   2+2(1+\tilde{\alpha})\frac{\mathrm{tan}^{-1}\sqrt{\tilde{\alpha}}}{\sqrt{\tilde{\alpha}}},\ 
   \tilde{\alpha}
  \equiv
   \frac{\tilde{\rho}_{\mathrm{s}}^\parallel}{\tilde{\rho}_{\mathrm{s}}^\perp}-1
   \label{E:RCxiBtriangular}
\end{align}
with $\tilde{\rho}_{\mathrm{s}}^\perp$ and $\tilde{\rho}_{\mathrm{s}}^\parallel$ being
the transverse (in-plane) and longitudinal spin stiffnesses renormalized by the quantum
fluctuations at $T=0$, respectively,
while the correlation length in the corresponding classical model \cite{A1762,A12612} also
diverges as \eqref{E:RCxi}, where
\begin{align}
   x
  =\frac{1}{2},\ 
   B
  =\pi\rho_{\mathrm{s}}^\perp G(\alpha)
  \equiv
   B_{\mathrm{class}};\ 
   \alpha
  \equiv
   \frac{\rho_{\mathrm{s}}^\parallel}{\rho_{\mathrm{s}}^\perp}-1
   \label{E:CxiBtriangular}
\end{align}
with $\rho_{\mathrm{s}}^\perp$ and $\rho_{\mathrm{s}}^\parallel$ being the bare stiffnesses at
$T=0$ instead.
Their prediction of the correlation length for the quantum model in the renormalized classical
regime \cite{A1762} is precisely available from a large-$N$ expansion based on symplectic symmetry
including fluctuations to order $1/N$ \cite{C601} as well, while that for the classical model
\cite{A1762,A12612} is well consistent with Monte Carlo simulations \cite{S13170}.
The bare spin stiffnesses are readily available by twisting N\'eel-ordered classical spins
by an infinitesimal angle per lattice constant along the relevant direction \cite{S7247}
or indirectly obtainable through \cite{D6797,C8891}
\begin{align}
   \rho_{\mathrm{s}}^\perp
  =\chi^\perp c_{\bm{Q}}^2,\ 
   \rho_{\mathrm{s}}^\parallel
  =\chi^\parallel c_{\bm{0}}^2,
   \label{E:Cchi&Cc=>Crho}
\end{align}
where $\chi^\perp$ and $\chi^\parallel$ are the bare transverse and longitudinal
susceptibilities at $T=0$, respectively,
\begin{align}
   \chi^\perp=\chi^\parallel
  =\frac{2\hbar^2}{9\sqrt{3}Ja^2},
   \allowdisplaybreaks
   \label{E:Cchi}
\end{align}
while $c_{\bm{Q}}$ and $c_{\bm{0}}$ are the LSW velocity at the ordering momenta
$\bm{k}=\pm{\bm{Q}}$ and that at $\bm{k}=\bm{0}$, respectively,
\begin{align}
   c_{\bm{Q}}
  =\frac{3\sqrt{3}}{2\sqrt{2}}\frac{JSa}{\hbar},\ 
   c_{\bm{0}}
  =\frac{3\sqrt{3}}{2}\frac{JSa}{\hbar}.
   \label{E:Cc}
\end{align}
In any case we have
\begin{align}
   \rho_{\mathrm{s}}^\perp
  \equiv
   \frac{\sqrt{3}}{4}JS^2,\ 
   \rho_{\mathrm{s}}^\parallel
  \equiv
   \frac{\sqrt{3}}{2}JS^2,
   \label{E:Crho}
\end{align}
and then \eqref{E:CxiBtriangular} reads
\begin{align}
   B_{\mathrm{class}}
  =\frac{\sqrt{3}}{4}\pi(2+\pi)JS^2
  =1.7486J,
   \label{E:B_RCClassical}
\end{align}
which was indeed demonstrated by the Monte Carlo calculations on the classical Heisenberg model
\cite{S13170}.

   In order to evaluate the actual quantum mechanical correlation length,
Elstner, Singh, and Young \cite{E1629} calculated series for
\begin{align}
   \xi^2
  =\frac{1}{L}
   \sum_{l,l'=1}^L
   \frac{(\bm{r}_l-\bm{r}_{l'})^2}{4S^{zz}(\bm{Q})}
   \langle
    S_{\bm{r}_l}^z S_{\bm{r}_{l'}}^z
   \rangle_T
   e^{i\bm{Q}\cdot(\bm{r}_l-\bm{r}_{l'})}
   \label{E:xi^2HTSE}
\end{align}
up to order $13$ in powers of the inverse temperature $1/k_{\mathrm{B}}T$ and found that
the quantity $(k_{\mathrm{B}}T/J)\mathrm{ln}(k_{\mathrm{B}}T\xi^2/Ja^2)$ extrapolate to about
$0.2\equiv 2B_{\mathrm{quant}}/J$ as $T\rightarrow 0$, which is apparently nonzero but merely
about $6\%$ of the corresponding classical value $2B_{\mathrm{class}}=3.4972J$.
It is this finding $2B_{\mathrm{quant}}/k_{\mathrm{B}}$ that Chen \textit{et al.} \cite{C140404}
identified their $T_{\mathrm{low}}$ with.
However, as the authors themselves \cite{E1629} pointed out,
there may be a crossover to the renormalized classical behavior
\eqref{E:RCxi} with \eqref{E:RCxiBtriangular} and thus an upturn \cite{Z224420,S13170,E1629,E5943}
of $(k_{\mathrm{B}}T/J)\mathrm{ln}(k_{\mathrm{B}}T\xi^2/Ja^2)$ with decreasing $T$ at some
temperature much lower than the high-temperature series expansion approach can reach,
resulting in a larger value of $B_{\mathrm{quant}}$.
Then, it is quite possible that the PC-DC-MSW specific-heat peak temperatures $T_{\bm{Q}}$ are
reasonable indeed.
In any case, the fact remains unchanged that the ratio of $B_{\mathrm{quant}}$ to
$B_{\mathrm{class}}$ on the triangular lattice \cite{Z224420,A1762,A12612,E1629,E5943,L9162}
is much smaller than that on the square lattice \cite{C1057,C2344,S7247,E5943} which is known
to be no smaller than $0.7$.
It must be the consequence of such strong quantum fluctuations in the triangular-lattice
antiferromagnet that the higher-order quantum corrections we take account of,
the lower peak temperature $T_{\bm{Q}}$ we have [Fig. \ref{F:DC-MSW&FTL&TNofCin2D}(a)].
The up-to-$O(S^m)$-MSW estimate of the peak temperature, $T_{\bm{Q}}^{(m)}$, monotonically and
significantly decreases with descending powers of $S$ as
$k_{\mathrm{B}}T_{\bm{Q}}^{(m)}/J=0.4904\,(m=1),\ 0.3894\,(m=0),\ 0.3526\,(m=-1)$.
On the other hand, it may be the consequence of much weaker quantum fluctuations in
the square-lattice antiferromagnet that the MLSW and MWDISW estimates of the peak temperature are almost
the same [Fig. \ref{F:ConvSC-MSW&SB&QMC&TNofCin2D}(a)].

   To end, we point out that the other temperature scale $T_{\mathrm{high}}$ \cite{C140404}
argued in relation to roton-like excitations \cite{C144416,Z057201,Z224420,S180403,G184403}
is also available by our PC DC MSWs.
The DC-MLSW and PC-DC-MSW temperature profiles of the triangular-lattice-antiferromagnet
specific heat look as though they have qualitatively different excitation mechanisms on
their down slopes from intermediate to high temperatures.
Quantum renormalization converts the saddle point at
$\mathrm{M}\equiv\frac{1}{a}\left(\pi,0,\frac{\pi}{\sqrt{3}}\right)$
in the DC-MLSW excitation energy surface
$\varepsilon_{\bm{k}:1}$ \eqref{E:triangular&kagomeomegaDCMLSW}
into a local minimum surrounded by flat parts [Figs. \ref{F:Akw}(a) and \ref{F:Akw}(b)].
Such local extrema, available on the ways from $\Gamma\equiv\frac{1}{a}\left(0,0,0\right)$ to
$\mathrm{K} \equiv\frac{1}{a}\left(\frac{4\pi}{3},0,0\right)$ and
$\mathrm{K}'\equiv\frac{1}{a}\left(\frac{2\pi}{3},0,\frac{2\pi}{\sqrt{3}}\right)$ as well
\cite{F031026,S144414},
may yield thermodynamic anomalies \cite{C144416,Z057201,Z224420}.
MSWs around the M point form in a quadratic band of roton-like excitations above the ground state
by $0.7579J\equiv k_{\mathrm{B}}T_{\mathrm{roton}}$ with stronger intensity than otherwise.
We indicate this temperature $T_{\mathrm{roton}}$ in
Figs. \ref{F:DC-MSW&FTL&TNofCin2D}(a) and \ref{F:DispertionRelations}($\mathrm{d}'$).
We cannot find any anormaly around $T_{\mathrm{roton}}$ in the DC-MLSW specific-heat curve,
while we find a steep-to-mild crossover as temperature increases across $T_{\mathrm{roton}}$
in the PC-DC-MSW specific-heat curves, where the $O(S^0)$ second-order perturbative correction
$\Delta F_2^{(0)}$ plays a prominent role, just like it does in yielding the bimodal temperature
profile of the kagome-lattice-antiferromagnet specific heat
[Figs. \ref{F:DispertionRelations}($\mathrm{d}'$) and \ref{F:DispertionRelations}($\mathrm{h}'$)].
There is a certain resemblance between the dispersionless DC-MLSW spectrum
$\varepsilon_{\bm{k}:1}$ around the M point in the triangular-lattice antiferromagnet and
the wholly flat DC-MLSW spectrum $\varepsilon_{\bm{k}:2}$ in the kagome-lattice antiferromagnet.
Without them, we do not have such marked ups and downs of
$\partial^2\beta\Delta F_2^{(0)}/\partial T\partial\beta$.
In the kagome-lattice antiferromagnet,
the wholly flat band gives such a prominent peak in the density of states to yield
the mid-temperature broad maximum,
whereas in the triangular-lattice antiferromagnet,
only a flat part of the whole spectrum is not sufficient to do this and does no more than
soften the down slope.

\subsection{Further possible applications}
\label{SS:FA}
The ground states of the triangular- and kagome-lattice antiferromagnets are so different from
each other as to be ordered and disordered, respectively, and their specific-heat curves are still
of different aspect such as a single-peak temperature profile containing two temperature scales
and an explicitly bimodal temperature profile, respectively.
Nevertheless, MSW excitations behind them have some similarities as well as differences.
This is understandable if their ground states sit close to each other sandwiching a quantum
critical point.
Additional ring-exchange interactions stabilize the nearest-neighbor pair-exchange-coupled
spin-$\frac{1}{2}$ Heisenberg antiferromagnet on the equilateral triangular lattice into
a disordered ground state \cite{M1064,S235115}, while
additional Dzyaloshinskii-Moriya interactions stabilize the nearest-neighbor pair-exchange-coupled
spin-$\frac{1}{2}$ Heisenberg antiferromagnet on the regular kagome lattice into
an ordered ground state \cite{C140405,M064428,M267201,L224414,A027203}.
If we monitor the triangular-lattice antiferromagnet with increasing ring-exchange interactions
and the kagome-lattice antiferromagnet with increasing Dzyaloshinskii-Moriya interactions
through the use of PC DC MSWs, we can indeed see single-to-double-peak and double-to-single-peak
crossovers of their specific-heat curves as an evidence of their locating in close vicinity of
a quantum critical point.

   In the framework of SW languages, frustrated noncollinear antiferromagnets are much less
tractable than collinear antiferromagnets, and still less are they when their quantum ground states
are disordered.
We have challenged this difficulty and just obtained a robust and eloquent PC-DC-MSW
thermodynamics, which is designed especially for noncollinear antiferromagnets but is not
inconsistent with the traditional SC-MSW thermodynamics for collinear antiferromagnets
\cite{T2494,T1524}.
Note that the present DC condition degenerates into the traditional SC condition for any collinear
ground state.

   Our PC-DC-MSW scheme is widely applicable to various frustrated noncollinear quantum magnets.
We may take further interest in describing giant molybdenum-oxide-based molecular spheres of
the Keplerate type with spin-higher-than-$\frac{1}{2}$ magnetic centers such as
$\{\mathrm{Mo}_{72}\mathrm{Cr}_{30}\}$ \cite{T6106,S224409} and
$\{\mathrm{Mo}_{72}\mathrm{Fe}_{30}\}$ \cite{M3238,M517,S224409,R094420}
containing $30$ $\mathrm{Cr}^{3+}$ ($S=\frac{3}{2}$) and $\mathrm{Fe}^{3+}$ ($S=\frac{5}{2}$)
ions, respectively, where neither exact diagonalization nor quantum Monte Carlo simulation is
feasible.
Considering that LSWs, whether modified or not, only give a poor description of the excitation
spectrum of the icosahedral keplerate cluster $\{\mathrm{Mo}_{72}\mathrm{Fe}_{30}\}$
\cite{W012415}, SC-MLSW thermodynamics \cite{C280} must be insufficient to capture its thermal
features and there may be something beyond Monte Carlo calculations of its classical analog
\cite{H2543}.


\ack
One of the authors (S. Y.) is grateful to B. Schmidt for useful pieces of information on his
finite-temperature Lanczos calculations.
This work is supported by JSPS KAKENHI Grant Number 22K03502.

\section*{Data availability statement}
Any data that support the findings of this study are included within the article.

\appendix
\section{Rotating frame and corresponding bosonic Hamiltonian}
\label{A:RFandCBH}
Suppose the spin components in the laboratory and rotating frames are related with each other as
\begin{align}
   \left[
    \begin{array}{c}
     S_{\bm{r}_{l}}^{z} \\
     S_{\bm{r}_{l}}^{x} \\
     S_{\bm{r}_{l}}^{y} \\
    \end{array}
   \right]
  =\left[
    \begin{array}{ccc}
     \cos\phi_{\bm{r}_{l}}                        &
    -\sin\phi_{\bm{r}_{l}}                        &
     0                                            \\
     \sin\phi_{\bm{r}_{l}}\cos\theta_{\bm{r}_{l}} &
     \cos\phi_{\bm{r}_{l}}\cos\theta_{\bm{r}_{l}} &
                         -\sin\theta_{\bm{r}_{l}} \\
     \sin\phi_{\bm{r}_{l}}\sin\theta_{\bm{r}_{l}} &
     \cos\phi_{\bm{r}_{l}}\sin\theta_{\bm{r}_{l}} &
                          \cos\theta_{\bm{r}_{l}} \\
    \end{array}
   \right]
   \left[
    \begin{array}{c}
     S_{\bm{r}_{l}}^{\tilde{z}} \\
     S_{\bm{r}_{l}}^{\tilde{x}} \\
     S_{\bm{r}_{l}}^{\tilde{y}} \\
    \end{array}
   \right],
   \label{E:SinLandRframes3D}
\end{align}
where $\theta_{\bm{r}_l}$ and $\phi_{\bm{r}_l}$ are the polar coordinates to specify
the rotating versus laboratory frame.
Substituting \eqref{E:SinLandRframes3D} into \eqref{E:HeisenbergH} and further applying
the Holstein-Primakoff transformation in the same way as \eqref{E:HPT} yields a bosonic
Hamiltonian in descending powers of $\sqrt{S}$, which we shall denote by
$
   \mathcal{H}
  =\sum_{m=-\infty}^4
   \mathcal{H}^{\left(\frac{m}{2}\right)}
$
with $\mathcal{H}^{\left(\frac{m}{2}\right)}$ being the Hamiltonian component on the order of
$S^{\left(\frac{m}{2}\right)}$.
We set free SWs, described by
$\sum_{m=2}^4\mathcal{H}^{\left(\frac{m}{2}\right)}\equiv\mathcal{H}_{\mathrm{harm}}$,
for the unperturbed Hamiltonian and regard all the rest
$\sum_{m=-\infty}^1\mathcal{H}^{\left(\frac{m}{2}\right)}\equiv\mathcal{V}$
as the perturbing interactions.
We perturb $\mathcal{H}_{\mathrm{harm}}$ to a certain order in $\mathcal{V}$.
We discard any correction beyond the order of $S^{-1}$ and then $\mathcal{V}$ may be truncated
at the order of $S^{-1}$ from the beginning, 
$\mathcal{V}=\sum_{m=-2}^1\mathcal{H}^{\left(\frac{m}{2}\right)}$.
Each Hamiltonian component reads
\begin{align}
   &
   \mathcal{H}^{(2)}
  =JS^{2}
   \sum_{\langle l,l' \rangle}
   \left[
    \sin\phi_{{\bm r}_{l}}\sin\phi_{{\bm r}_{l'}}
    \cos
    \left(
     \theta_{{\bm r}_{l}}-\theta_{{\bm r}_{l'}}
    \right)
   +\cos\phi_{{\bm r}_{l}}\cos\phi_{{\bm r}_{l'}}
   \right],
   \label{E:H(2)3D}
   \allowdisplaybreaks \\
   &
   \mathcal{H}^{\left(\frac{3}{2}\right)}
  =J\sqrt{\frac{S^3}{2}}
   \sum_{\langle l,l' \rangle}
   \left\{
    \vphantom{a_{{\bm r}_{l'}}^\dagger}
    a_{{\bm r}_{l}}^\dagger
    \left[
     \cos\phi_{{\bm r}_{l}}\sin\phi_{{\bm r}_{l'}}
     \cos
     \left(
      \theta_{{\bm r}_{l}}-\theta_{{\bm r}_{l'}}
     \right)
    -\sin\phi_{{\bm r}_{l }}\cos\phi_{{\bm r}_{l'}}
   \right.
    \right.
   \nonumber
   \allowdisplaybreaks \\
   &\quad
    \left.
   -i\sin\phi_{{\bm r}_{l'}}
     \sin
     \left(
      \theta_{{\bm r}_{l}}-\theta_{{\bm r}_{l'}}
     \right)
    \right]
   +a_{{\bm r}_{l}}
    \left[
     \cos\phi_{{\bm r}_{l}}\sin\phi_{{\bm r}_{l'}}
     \cos
     \left(
      \theta_{{\bm r}_{l}}-\theta_{{\bm r}_{l'}}
     \right)
    -\sin\phi_{{\bm r}_{l }}\cos\phi_{{\bm r}_{l'}}
    \right.
   \nonumber
   \allowdisplaybreaks \\
   &\quad
    \left.
   +i\sin\phi_{{\bm r}_{l'}}
     \sin
     \left(
      \theta_{{\bm r}_{l}}-\theta_{{\bm r}_{l'}}
     \right)
    \right]
   +a_{{\bm r}_{l'}}^\dagger 
    \left[
     \sin\phi_{{\bm r}_{l}}\cos\phi_{{\bm r}_{l'}}
     \cos
     \left(
      \theta_{{\bm r}_{l}}-\theta_{{\bm r}_{l'}}
     \right)
    -\cos\phi_{{\bm r}_{l}}\sin\phi_{{\bm r}_{l'}}
    \right.
   \nonumber
   \allowdisplaybreaks \\
   &\quad
    \left.
   +i\sin\phi_{{\bm r}_{l}}
     \sin
     \left(
      \theta_{{\bm r}_{l}}-\theta_{{\bm r}_{l'}}
     \right)
    \right]
   +a_{{\bm r}_{l'}}
    \left[
     \sin\phi_{{\bm r}_{l}}\cos\phi_{{\bm r}_{l'}}
     \cos
     \left(
      \theta_{{\bm r}_{l}}-\theta_{{\bm r}_{l'}}
     \right)
    -\cos\phi_{{\bm r}_{l}}\sin\phi_{{\bm r}_{l'}}
    \right.
   \nonumber
   \allowdisplaybreaks \\
   &\quad
    \left.
   -i\sin\phi_{{\bm r}_{l}}
     \sin
     \left(
      \theta_{{\bm r}_{l}}-\theta_{{\bm r}_{l'}}
     \right)
    \right]
   \left.\!\!
    \vphantom{a_{{\bm r}_{l'}}^\dagger}
   \right\}
   =0,
   \label{E:H(3/2)3D}
   \allowdisplaybreaks \\
   &
   \mathcal{H}^{(1)}
 =-\frac{JS}{2}
   \sum_{\langle l,l' \rangle}
   \left\{
   2\left(
    a_{{\bm r}_{l }}^\dagger a_{{\bm r}_{l }}
   +a_{{\bm r}_{l'}}^\dagger a_{{\bm r}_{l'}}
    \right)
    \left[
     \sin\phi_{{\bm r}_{l}}\sin\phi_{{\bm r}_{l'}}
     \cos
     \left(
      \theta_{{\bm r}_{l}}-\theta_{{\bm r}_{l'}}
     \right)
    +\cos\phi_{{\bm r}_{l}}\cos\phi_{{\bm r}_{l'}}
    \right]
   \right.
   \nonumber
   \allowdisplaybreaks \\
   &\quad
   -a_{{\bm r}_{l}}^\dagger a_{{\bm r}_{l'}}^\dagger 
    \left[
     \left(
      \cos\phi_{{\bm r}_{l}}\cos\phi_{{\bm r}_{l'}}-1
     \right)
     \cos
     \left(
      \theta_{{\bm r}_{l}}-\theta_{{\bm r}_{l'}}
     \right)
    +\sin\phi_{{\bm r}_{l}}\sin\phi_{{\bm r}_{l'}}
     \right.
   \nonumber
   \allowdisplaybreaks \\
   &\quad\quad\quad\quad\quad\!\!
     \left.
   +i\left(
      \cos\phi_{{\bm r}_{l}}-\cos\phi_{{\bm r}_{l'}}
     \right)
     \sin
     \left(
      \theta_{{\bm r}_{l}}-\theta_{{\bm r}_{l'}}
     \right)
    \right]
   \nonumber
   \allowdisplaybreaks \\
   &\quad
   -a_{{\bm r}_{l}}a_{{\bm r}_{l'}}
    \left[
     \left(
      \cos\phi_{{\bm r}_{l}}\cos\phi_{{\bm r}_{l'}}-1
     \right)
     \cos
     \left(
      \theta_{{\bm r}_{l}}-\theta_{{\bm r}_{l'}}
     \right)
    +\sin\phi_{{\bm r}_{l}}\sin\phi_{{\bm r}_{l'}}
     \right.
   \nonumber
   \allowdisplaybreaks \\
   &\quad\quad\quad\quad\quad\!\!
     \left.
   -i\left(
      \cos\phi_{{\bm r}_{l}}-\cos\phi_{{\bm r}_{l'}}
     \right)
     \sin
     \left(
      \theta_{{\bm r}_{l}}-\theta_{{\bm r}_{l'}}
     \right)
    \right]
   \nonumber
   \allowdisplaybreaks \\
   &\quad
   -a_{{\bm r}_{l}}^\dagger a_{{\bm r}_{l'}}
    \left[
     \left(
      \cos\phi_{{\bm r}_{l}}\cos\phi_{{\bm r}_{l'}}+1
     \right)
     \cos
     \left(
      \theta_{{\bm r}_{l}}-\theta_{{\bm r}_{l'}}
     \right)
    +\sin\phi_{{\bm r}_{l}}\sin\phi_{{\bm r}_{l'}}
     \right.
   \nonumber
   \allowdisplaybreaks \\
   &\quad\quad\quad\quad\quad\!\!
     \left.
   -i\left(
      \cos\phi_{{\bm r}_{l}}+\cos\phi_{{\bm r}_{l'}}
     \right)
     \sin
     \left(
      \theta_{{\bm r}_{l}}-\theta_{{\bm r}_{l'}}
     \right)
    \right]
   \nonumber
   \allowdisplaybreaks \\
   &\quad
   -a_{{\bm r}_{l'}}^\dagger a_{{\bm r}_{l}}
    \left[
     \left(
      \cos\phi_{{\bm r}_{l}}\cos\phi_{{\bm r}_{l'}}+1
     \right)
     \cos
     \left(
      \theta_{{\bm r}_{l}}-\theta_{{\bm r}_{l'}}
     \right)
    +\sin\phi_{{\bm r}_{l}}\sin\phi_{{\bm r}_{l'}}
     \right.
   \nonumber
   \allowdisplaybreaks \\
   &\quad\quad\quad\quad\quad
     \left.\!\!
   +i\left(
      \cos\phi_{{\bm r}_{l}}+\cos\phi_{{\bm r}_{l'}}
     \right)
     \sin
     \left(
      \theta_{{\bm r}_{l}}-\theta_{{\bm r}_{l'}}
     \right)
    \right]
   \left.\!\!
    \vphantom{a_{{\bm r}_{l'}}^\dagger}
   \right\},
   \label{E:H(1)3D}
   \allowdisplaybreaks \\
   &
   \mathcal{H}^{\left(\frac{1}{2}\right)}
  =-J\sqrt{\frac{S}{2}}
   \sum_{\langle l,l' \rangle}
   \left\{
    a_{{\bm r}_{l}}^\dagger a_{{\bm r}_{l'}}^\dagger a_{{\bm r}_{l'}}
    \left[
     \cos\phi_{{\bm r}_{l}}\sin\phi_{{\bm r}_{l'}}
     \cos
     \left(
      \theta_{{\bm r}_{l}}-\theta_{{\bm r}_{l'}}
     \right)
   \right.
    -\sin\phi_{{\bm r}_{l}}\cos\phi_{{\bm r}_{l'}}
     \right.
   \nonumber
   \allowdisplaybreaks \\
   &\quad
    \left.
   -i\sin\phi_{{\bm r}_{l'}}
     \sin
     \left(
      \theta_{{\bm r}_{l}}-\theta_{{\bm r}_{l'}}
     \right)
    \right]
   +a_{{\bm r}_{l}}a_{{\bm r}_{l'}}^\dagger a_{{\bm r}_{l'}}
    \left[
     \cos\phi_{{\bm r}_{l}}\sin\phi_{{\bm r}_{l'}}
     \cos
     \left(
      \theta_{{\bm r}_{l}}-\theta_{{\bm r}_{l'}}
     \right)
    -\sin\phi_{{\bm r}_{l}}\cos\phi_{{\bm r}_{l'}}
     \right.
   \nonumber
   \allowdisplaybreaks \\
   &\quad
    \left.
   +i\sin\phi_{{\bm r}_{l'}}
     \sin
     \left(
      \theta_{{\bm r}_{l}}-\theta_{{\bm r}_{l'}}
     \right)
    \right]
   +a_{{\bm r}_{l}}^\dagger a_{{\bm r}_{l}}a_{{\bm r}_{l'}}^\dagger 
    \left[
     \sin\phi_{{\bm r}_{l}}\cos\phi_{{\bm r}_{l'}}
     \cos
     \left(
      \theta_{{\bm r}_{l}}-\theta_{{\bm r}_{l'}}
     \right)
    -\cos\phi_{{\bm r}_{l}}\sin\phi_{{\bm r}_{l'}}
     \right.
   \nonumber
   \allowdisplaybreaks \\
   &\quad
    \left.
   +i\sin\phi_{{\bm r}_{l}}
     \sin
     \left(
      \theta_{{\bm r}_{l}}-\theta_{{\bm r}_{l'}}
     \right)
    \right]
   +a_{{\bm r}_{l}}^\dagger a_{{\bm r}_{l}}a_{{\bm r}_{l'}}
    \left[
     \sin\phi_{{\bm r}_{l}}\cos\phi_{{\bm r}_{l'}}
     \cos
     \left(
      \theta_{{\bm r}_{l}}-\theta_{{\bm r}_{l'}}
     \right)
    -\cos\phi_{{\bm r}_{l}}\sin\phi_{{\bm r}_{l'}}
     \right.
   \nonumber
   \allowdisplaybreaks \\
   &\quad
    \left.
   -i\sin\phi_{{\bm r}_{l}}
     \sin
     \left(
      \theta_{{\bm r}_{l}}-\theta_{{\bm r}_{l'}}
     \right)
    \right]
   \left.\!\!
    \vphantom{a_{{\bm r}_{l'}}^\dagger}
   \right\},
   \label{E:H(1/2)3D}
   \allowdisplaybreaks \\
   &
   \mathcal{H}^{(0)}
  =J
   \sum_{\langle l,l' \rangle}
   \left\{
    a_{{\bm r}_{l }}^\dagger a_{{\bm r}_{l }}
    a_{{\bm r}_{l'}}^\dagger a_{{\bm r}_{l'}}
    \left[
     \sin\phi_{{\bm r}_{l}}\sin\phi_{{\bm r}_{l'}}
     \cos
     \left(
      \theta_{{\bm r}_{l}}-\theta_{{\bm r}_{l'}}
     \right)
    +\cos\phi_{{\bm r}_{l}}\cos\phi_{{\bm r}_{l'}}
    \right]
    \vphantom{\frac{1}{8}}
   \right.
   \nonumber
   \allowdisplaybreaks \\
   &\quad
   -\frac{1}{8}
    a_{{\bm r}_{l}}^\dagger a_{{\bm r}_{l'}}^\dagger 
    \left(
     a_{{\bm r}_{l }}^\dagger a_{{\bm r}_{l }}
    +a_{{\bm r}_{l'}}^\dagger a_{{\bm r}_{l'}}
    \right)
    \left[
     \left(
      \cos\phi_{{\bm r}_{l}}\cos\phi_{{\bm r}_{l'}}-1
     \right)
     \cos
     \left(
      \theta_{{\bm r}_{l}}-\theta_{{\bm r}_{l'}}
     \right)
    +\sin\phi_{{\bm r}_{l}}\sin\phi_{{\bm r}_{l'}}
    \right.
   \nonumber
   \allowdisplaybreaks \\
   &\quad\qquad\qquad\qquad\qquad\qquad\quad\quad\!\!
    \left.
   +i\left(
      \cos\phi_{{\bm r}_{l}}-\cos\phi_{{\bm r}_{l'}}
     \right)
     \sin
     \left(
      \theta_{{\bm r}_{l}}-\theta_{{\bm r}_{l'}}
     \right)
    \right]
   \nonumber
   \allowdisplaybreaks \\
   &\quad
   -\frac{1}{8}
    \left(
     a_{{\bm r}_{l }}^\dagger a_{{\bm r}_{l }}
    +a_{{\bm r}_{l'}}^\dagger a_{{\bm r}_{l'}}
    \right)
    a_{{\bm r}_{l}}a_{{\bm r}_{l'}}
    \left[
     \left(
      \cos\phi_{{\bm r}_{l}}\cos\phi_{{\bm r}_{l'}}-1
     \right)
     \cos
     \left(
      \theta_{{\bm r}_{l}}-\theta_{{\bm r}_{l'}}
     \right)
    +\sin\phi_{{\bm r}_{l}}\sin\phi_{{\bm r}_{l'}}
    \right.
   \nonumber
   \allowdisplaybreaks \\
   &\quad\qquad\qquad\qquad\qquad\qquad\quad\quad\!\!
    \left.
   -i\left(
      \cos\phi_{{\bm r}_{l}}-\cos\phi_{{\bm r}_{l'}}
     \right)
     \sin
     \left(
      \theta_{{\bm r}_{l}}-\theta_{{\bm r}_{l'}}
     \right)
   \right]
   \nonumber
   \allowdisplaybreaks \\
   &\quad
   -\frac{1}{8}
    a_{{\bm r}_{l }}^\dagger
    \left(
     a_{{\bm r}_{l }}^\dagger a_{{\bm r}_{l }}
    +a_{{\bm r}_{l'}}^\dagger a_{{\bm r}_{l'}}
    \right)
    a_{{\bm r}_{l'}}
    \left[
     \left(
      \cos\phi_{{\bm r}_{l}}\cos\phi_{{\bm r}_{l'}}+1
     \right)
     \cos
     \left(
      \theta_{{\bm r}_{l}}-\theta_{{\bm r}_{l'}}
     \right)
    +\sin\phi_{{\bm r}_{l}}\sin\phi_{{\bm r}_{l'}}
    \right.
   \nonumber
   \allowdisplaybreaks \\
   &\quad\qquad\qquad\qquad\qquad\qquad\quad\quad\!\!
    \left.
   -i\left(
      \cos\phi_{{\bm r}_{l}}+\cos\phi_{{\bm r}_{l'}}
     \right)
     \sin
     \left(
      \theta_{{\bm r}_{l}}-\theta_{{\bm r}_{l'}}
     \right)
    \right]
   \nonumber
   \allowdisplaybreaks \\
   &\quad
   -\frac{1}{8}
    a_{{\bm r}_{l'}}^\dagger 
    \left(
     a_{{\bm r}_{l }}^\dagger a_{{\bm r}_{l }}
    +a_{{\bm r}_{l'}}^\dagger a_{{\bm r}_{l'}}
    \right)
    a_{{\bm r}_{l }}
    \left[
     \left(
      \cos\phi_{{\bm r}_{l}}\cos\phi_{{\bm r}_{l'}}+1
     \right)
     \cos
     \left(
      \theta_{{\bm r}_{l}}-\theta_{{\bm r}_{l'}}
     \right)
    +\sin\phi_{{\bm r}_{l}}\sin\phi_{{\bm r}_{l'}}
    \right.
   \nonumber
   \allowdisplaybreaks \\
   &\quad\qquad\qquad\qquad\qquad\qquad\quad\quad\!\!
    \left.
   +i\left(
      \cos\phi_{{\bm r}_{l}}+\cos\phi_{{\bm r}_{l'}}
     \right)
     \sin
     \left(
      \theta_{{\bm r}_{l}}-\theta_{{\bm r}_{l'}}
     \right)
    \right]
   \left.\!\!
    \vphantom{\frac{1}{8}}
   \right\}
   \label{E:H(0)3D},
   \allowdisplaybreaks \\
   &
   \mathcal{H}^{\left(-\frac{1}{2}\right)}
  =\frac{J}{\sqrt{32S}}
   \sum_{\langle l,l' \rangle}
   \left\{
    a_{{\bm r}_{l }}^\dagger a_{{\bm r}_{l }}^\dagger a_{{\bm r}_{l}}
    a_{{\bm r}_{l'}}^\dagger a_{{\bm r}_{l'}}
    \left[
     \cos\phi_{{\bm r}_{l}}\sin\phi_{{\bm r}_{l'}}
    \cos
    \left(
     \theta_{{\bm r}_{l}}-\theta_{{\bm r}_{l'}}
    \right)
    -\sin\phi_{{\bm r}_{l}}\cos\phi_{{\bm r}_{l'}}
    \right.
   \right.
   \nonumber
   \allowdisplaybreaks \\
   &\quad\quad\!
   -i\sin\phi_{{\bm r}_{l'}}
    \left.
     \sin
     \left(
      \theta_{{\bm r}_{l}}-\theta_{{\bm r}_{l'}}
     \right)
    \right]
   +a_{{\bm r}_{l }}^\dagger a_{{\bm r}_{l }}a_{{\bm r}_{l}}
    a_{{\bm r}_{l'}}^\dagger a_{{\bm r}_{l'}}
   \nonumber
   \allowdisplaybreaks \\
   &\quad\ \times
    \left[
     \cos\phi_{{\bm r}_{l}}\sin\phi_{{\bm r}_{l'}}
     \cos
     \left(
      \theta_{{\bm r}_{l}}-\theta_{{\bm r}_{l'}}
     \right)
    -\sin\phi_{{\bm r}_{l}}\cos\phi_{{\bm r}_{l'}}
   +i\sin\phi_{{\bm r}_{l'}}
     \sin
     \left(
      \theta_{{\bm r}_{l}}-\theta_{{\bm r}_{l'}}
     \right)
    \right]
   \nonumber
   \allowdisplaybreaks \\
   &\quad
   +a_{{\bm r}_{l }}^\dagger a_{{\bm r}_{l }}
    a_{{\bm r}_{l'}}^\dagger a_{{\bm r}_{l'}}^\dagger a_{{\bm r}_{l'}}
    \left[
     \sin\phi_{{\bm r}_{l}}\cos\phi_{{\bm r}_{l'}}
     \cos
     \left(
      \theta_{{\bm r}_{l}}-\theta_{{\bm r}_{l'}}
     \right)
    -\cos\phi_{{\bm r}_{l}}\sin\phi_{{\bm r}_{l'}}
    \right.
   \nonumber
   \allowdisplaybreaks \\
   &\quad\quad\!
   +i\sin\phi_{{\bm r}_{l}}
    \left.
     \sin
     \left(
      \theta_{{\bm r}_{l}}-\theta_{{\bm r}_{l'}}
     \right)
    \right]
   +a_{{\bm r}_{l }}^\dagger a_{{\bm r}_{l }}
    a_{{\bm r}_{l'}}^\dagger a_{{\bm r}_{l'}}a_{{\bm r}_{l'}}
   \nonumber
   \allowdisplaybreaks \\
   &\quad\ \times
    \left[
     \sin\phi_{{\bm r}_{l}}\cos\phi_{{\bm r}_{l'}}
     \cos
     \left(
      \theta_{{\bm r}_{l}}-\theta_{{\bm r}_{l'}}
     \right)
    -\cos\phi_{{\bm r}_{l}}\sin\phi_{{\bm r}_{l'}}
   -i\sin\phi_{{\bm r}_{l}}
     \sin
     \left(
      \theta_{{\bm r}_{l}}-\theta_{{\bm r}_{l'}}
     \right)
    \right]
   \left.\!\!
    \vphantom{a_{{\bm r}_{l'}}^\dagger}
   \right\},
   \label{E:H(-1/2)3D}
   \allowdisplaybreaks \\
   &
   \mathcal{H}^{(-1)}
  =-\frac{J}{64S}
   \sum_{\langle l,l' \rangle}
   \left\{
    \left(
     a_{{\bm r}_{l }}^\dagger
     a_{{\bm r}_{l'}}^\dagger
     a_{{\bm r}_{l'}}^\dagger a_{{\bm r}_{l'}}
     a_{{\bm r}_{l'}}^\dagger a_{{\bm r}_{l'}}
    +a_{{\bm r}_{l }}^\dagger
     a_{{\bm r}_{l }}^\dagger a_{{\bm r}_{l }}
     a_{{\bm r}_{l }}^\dagger a_{{\bm r}_{l }}
     a_{{\bm r}_{l'}}^\dagger
   -2a_{{\bm r}_{l }}^\dagger 
     a_{{\bm r}_{l }}^\dagger a_{{\bm r}_{l }}
     a_{{\bm r}_{l'}}^\dagger 
     a_{{\bm r}_{l'}}^\dagger a_{{\bm r}_{l'}}
    \right)
   \right.
   \nonumber
   \allowdisplaybreaks \\
   &\quad\ \times
    \left[
     \left(
      \cos\phi_{{\bm r}_{l}}\cos\phi_{{\bm r}_{l'}}-1
     \right)
     \cos
     \left(
      \theta_{{\bm r}_{l}}-\theta_{{\bm r}_{l'}}
     \right)
    +\sin\phi_{{\bm r}_{l}}\sin\phi_{{\bm r}_{l'}}
    \right.
   \nonumber
   \allowdisplaybreaks \\
   &\quad\quad\!
   +i\left(
      \cos\phi_{{\bm r}_{l}}-\cos\phi_{{\bm r}_{l'}}
     \right)
    \left.
     \sin
     \left(
      \theta_{{\bm r}_{l}}-\theta_{{\bm r}_{l'}}
     \right)
    \right]
   \nonumber
   \allowdisplaybreaks \\
   &\quad
   +\left(
     a_{{\bm r}_{l }}
     a_{{\bm r}_{l'}}^\dagger a_{{\bm r}_{l'}}
     a_{{\bm r}_{l'}}^\dagger a_{{\bm r}_{l'}}
     a_{{\bm r}_{l'}}
    +a_{{\bm r}_{l }}^\dagger a_{{\bm r}_{l }}
     a_{{\bm r}_{l }}^\dagger a_{{\bm r}_{l }}
     a_{{\bm r}_{l }}
     a_{{\bm r}_{l'}}
   -2a_{{\bm r}_{l }}^\dagger a_{{\bm r}_{l }}
     a_{{\bm r}_{l }}
     a_{{\bm r}_{l'}}^\dagger a_{{\bm r}_{l'}}
     a_{{\bm r}_{l'}}
    \right)
   \nonumber
   \allowdisplaybreaks \\
   &\quad\ \times
    \left[
     \left(
      \cos\phi_{{\bm r}_{l}}\cos\phi_{{\bm r}_{l'}}-1
     \right)
     \cos
     \left(
      \theta_{{\bm r}_{l}}-\theta_{{\bm r}_{l'}}
     \right)
    +\sin\phi_{{\bm r}_{l}}\sin\phi_{{\bm r}_{l'}}
    \right.
   \nonumber
   \allowdisplaybreaks \\
   &\quad\quad\!
   -i\left(
      \cos\phi_{{\bm r}_{l}}-\cos\phi_{{\bm r}_{l'}}
     \right)
    \left.
     \sin
     \left(
      \theta_{{\bm r}_{l}}-\theta_{{\bm r}_{l'}}
     \right)
    \right]
   \nonumber
   \allowdisplaybreaks \\
   &\quad
   +\left(
     a_{{\bm r}_{l }}
     a_{{\bm r}_{l'}}^\dagger 
     a_{{\bm r}_{l'}}^\dagger a_{{\bm r}_{l'}}
     a_{{\bm r}_{l'}}^\dagger a_{{\bm r}_{l'}}
    +a_{{\bm r}_{l }}^\dagger a_{{\bm r}_{l }}
     a_{{\bm r}_{l }}^\dagger a_{{\bm r}_{l }}
     a_{{\bm r}_{l }}
     a_{{\bm r}_{l'}}^\dagger 
   -2a_{{\bm r}_{l }}^\dagger a_{{\bm r}_{l }}
     a_{{\bm r}_{l }}
     a_{{\bm r}_{l'}}^\dagger 
     a_{{\bm r}_{l'}}^\dagger a_{{\bm r}_{l'}}
    \right)
   \nonumber
   \allowdisplaybreaks \\
   &\quad\ \times
    \left[
     \left(
      \cos\phi_{{\bm r}_{l}}\cos\phi_{{\bm r}_{l'}}+1
     \right)
     \cos
     \left(
      \theta_{{\bm r}_{l}}-\theta_{{\bm r}_{l'}}
     \right)
    +\sin\phi_{{\bm r}_{l}}\sin\phi_{{\bm r}_{l'}}
    \right.
   \nonumber
   \allowdisplaybreaks \\
   &\quad\quad\!
   +i\left(
      \cos\phi_{{\bm r}_{l}}+\cos\phi_{{\bm r}_{l'}}
     \right)
    \left.
     \sin
     \left(
      \theta_{{\bm r}_{l}}-\theta_{{\bm r}_{l'}}
     \right)
    \right]
   \nonumber
   \allowdisplaybreaks \\
   &\quad
   +\left(
     a_{{\bm r}_{l }}^\dagger 
     a_{{\bm r}_{l'}}^\dagger a_{{\bm r}_{l'}}
     a_{{\bm r}_{l'}}^\dagger a_{{\bm r}_{l'}}
     a_{{\bm r}_{l'}}
    +a_{{\bm r}_{l }}^\dagger 
     a_{{\bm r}_{l }}^\dagger a_{{\bm r}_{l }}
     a_{{\bm r}_{l }}^\dagger a_{{\bm r}_{l }}
     a_{{\bm r}_{l'}}
   -2a_{{\bm r}_{l }}^\dagger 
     a_{{\bm r}_{l }}^\dagger a_{{\bm r}_{l }}
     a_{{\bm r}_{l'}}^\dagger a_{{\bm r}_{l'}}
     a_{{\bm r}_{l'}}
    \right)
   \nonumber
   \allowdisplaybreaks \\
   &\quad\ \times
    \left[
     \left(
      \cos\phi_{{\bm r}_{l}}\cos\phi_{{\bm r}_{l'}}+1
     \right)
     \cos
     \left(
      \theta_{{\bm r}_{l}}-\theta_{{\bm r}_{l'}}
     \right)
    +\sin\phi_{{\bm r}_{l}}\sin\phi_{{\bm r}_{l'}}
    \right.
   \nonumber
   \allowdisplaybreaks \\
   &\quad\quad\!
   -i\left(
      \cos\phi_{{\bm r}_{l}}+\cos\phi_{{\bm r}_{l'}}
     \right)
    \left.
     \sin
     \left(
      \theta_{{\bm r}_{l}}-\theta_{{\bm r}_{l'}}
     \right)
     \right]
   \left.\!\!
    \vphantom{a_{{\bm r}_{l'}}^\dagger}
   \right\}
   \label{E:H(-1)3D}.
\end{align}

\section{Spectral functions and their renormalization}
\label{A:Akw}
In order to investigate effects of the interactions $\mathcal{V}$ on the DC-MLSW dispersion
relations $\varepsilon_{\bm{k}_\nu:\sigma}$, we consider their perturbative corrections to
the unperturbed Green functions
\begin{align}
   &
   G_0(\bm{k}_\nu:\sigma\sigma';i\omega_n)
 =-\int_0^{\beta\hbar}\!d\tau\,
   e^{i\omega_n\tau}
   \left\langle
    \alpha_{\bm{k}_\nu:\sigma }        (\tau)
    \alpha_{\bm{k}_\nu:\sigma'}^\dagger(0   )
   \right\rangle_T
   \nonumber
   \allowdisplaybreaks \\
   &\qquad
 =-\int_0^{\beta\hbar}\!d\tau\,
   e^{i\omega_n\tau}
   e^{-\tau\varepsilon_{\bm{k}:\sigma}/\hbar}
   \left\langle
    \alpha_{\bm{k}_\nu:\sigma }
    \alpha_{\bm{k}_\nu:\sigma'}^\dagger
   \right\rangle_T
   \nonumber
   \allowdisplaybreaks \\
   &\qquad
  =\frac{1-e^{(i\hbar\omega_n-\varepsilon_{\bm{k}:\sigma})\beta}}
        {i\hbar\omega_n-\varepsilon_{\bm{k}:\sigma}}
   (\bar{n}_{\bm{k}:\sigma}+1)\hbar\delta_{\sigma\sigma'}
  =\frac{\hbar\delta_{\sigma\sigma'}}
        {i\hbar\omega_n-\varepsilon_{\bm{k}_\nu:\sigma}}.
   \label{E:G0explicit}
\end{align}
The exact temperature Green function consists of matrix elements of Heisenberg operators in
the exact interacting eigenstates.
Denoting the correlated DC-MSW Hamiltonian by
$\widetilde{\mathcal{H}}
\equiv\widetilde{\mathcal{H}}_{\mathrm{harm}}+\mathcal{V}$,
we introduce the correlated temperature Green functions
\begin{align}
   &
   G(\bm{k}_\nu:\sigma\sigma';\tau)
  \equiv
  -\left\langle
    \mathcal{T}
    \left[
     \alpha_{\bm{k}_\nu:\sigma }        (\tau)^{\mathrm{corr}}
     \alpha_{\bm{k}_\nu:\sigma'}^\dagger(0   )^{\mathrm{corr}}
    \right]
   \right\rangle_T^{\mathrm{corr}}
   \nonumber
   \allowdisplaybreaks \\
   &\qquad
 =-\frac{\mathrm{Tr}
         \left\{
          e^{-\widetilde{\mathcal{H}}/k_{\mathrm{B}}T}
          \mathcal{T}
          \left[
           e^{ \tau\widetilde{\mathcal{H}}/\hbar}
           \alpha_{\bm{k}_\nu:\sigma }
           e^{-\tau\widetilde{\mathcal{H}}/\hbar}
           \alpha_{\bm{k}_\nu:\sigma'}^\dagger
          \right]
         \right\}}
         {\mathrm{Tr}
          \left\{
           e^{-\widetilde{\mathcal{H}}/k_{\mathrm{B}}T}
          \right\}},
   \nonumber
   \allowdisplaybreaks \\
   &
   G(\bm{k}_\nu:\sigma\sigma';i\omega_n)
  \equiv
   \int_0^{\beta\hbar}\!d\tau\,
   e^{i\omega_n\tau}
   G(\bm{k}_\nu:\sigma\sigma';\tau).
   \label{E:Gdef}
\end{align}
In the interaction picture, \eqref{E:Gdef} reads
\begin{align}
   &
   G  (\bm{k}_\nu:\sigma\sigma';i\omega_n)
  =G_0(\bm{k}_\nu:\sigma\sigma';i\omega_n)
  -\sum_{l=1}^\infty
   \frac{(-1)^l}{l!\hbar^l}
   \nonumber
   \allowdisplaybreaks \\
   &\times
   \int_0^{\beta\hbar}\!d\tau\,
   e^{i\omega_n\tau}
   \int\!\cdots\!\int_0^{\beta\hbar}\!d\tau_1 \cdots d\tau_l
   \left\langle
    \mathcal{T}
    \left[
     \mathcal{V}(\tau_1)\cdots\mathcal{V}(\tau_l)
     \alpha_{\bm{k}_\nu:\sigma }        (\tau)
     \alpha_{\bm{k}_\nu:\sigma'}^\dagger(0   )
    \right]
   \right\rangle_T.
   \label{E:perturbedG}
\end{align}
By virtue of the Bloch-De Dominicis theorem \cite{B459}, i.e., the finite-temperature version of
Wick's theorem \cite{W268}, we can evaluate the correlated Green function \eqref{E:perturbedG} as
a series in the unperturbed Green function \eqref{E:G0explicit} and vertex functions \cite{C7127}
(or interparticle potentials \cite{McGrraw-Hill1971}) in $\mathcal{V}$
[cf. $\varLambda_{+-},\varLambda_{++},\varLambda_{++--},\varLambda_{+++-},\varLambda_{++++}$
in \eqref{E:H(0)alpha} and $\varLambda_{++-},\varLambda_{+++}$ in \eqref{E:H(1/2)alpha}].
In the perturbation expansion \eqref{E:perturbedG},
the leading $O(S^0)$ corrections arise at $l=1$, to first order in $\mathcal{H}^{(0)}$, and
at $l=2$, to second order in $\mathcal{H}^{\left(\frac{1}{2}\right)}$.
Denoting their corresponding self-energies by
$\varSigma_1^{(0)}(\bm{k}_\nu:\sigma\sigma')$ [Fig. \ref{F:PSE}(a)] and
$\varSigma_2^{(0)}(\bm{k}_\nu:\sigma\sigma';i\omega_n)$ [Fig. \ref{F:PSE}(b)], respectively,
and the sum of them by $\varSigma^{(0)}(\bm{k}_\nu:\sigma\sigma';i\omega_n)$,
we approximate \eqref{E:perturbedG} by the Dyson equation \cite{McGrraw-Hill1971}
\begin{align}
   &
   G  (\bm{k}_\nu:\sigma\sigma;i\omega_n)
  \simeq
   G_0(\bm{k}_\nu:\sigma\sigma;i\omega_n)
   \sum_{l=0}^\infty
   \left[
    \frac{\varSigma^{(0)}(\bm{k}_\nu:\sigma\sigma;i\omega_n)}{\hbar}
    G_0(\bm{k}_\nu:\sigma\sigma;i\omega_n)
   \right]^l
   \nonumber
   \allowdisplaybreaks \\
   &\quad
  =G_0(\bm{k}_\nu:\sigma\sigma;i\omega_n)
  +G_0(\bm{k}_\nu:\sigma\sigma;i\omega_n)
   \frac{\varSigma^{(0)}(\bm{k}_\nu:\sigma\sigma;i\omega_n)}{\hbar}
   G(\bm{k}_\nu:\sigma\sigma;i\omega_n)
   \label{E:DysonEq}
\end{align}
to obtain the solution
\begin{align}
   G(\bm{k}_\nu:\sigma\sigma;i\omega_{n})
  =\frac{\hbar}{i\hbar\omega_{n}-\varepsilon_{\bm{k}_\nu:\sigma}
  -\varSigma^{(0)}(\bm{k}_\nu:\sigma\sigma;i\omega_{n})}.
   \label{E:GfromDysonEq}
\end{align}

   Let us recall that the MLSW dispersion relations $\varepsilon_{\bm{k}:\sigma}$ are
available from the unperturbed Green functions \eqref{E:G0explicit},
\begin{align}
   &
   A_0(\bm{k}:\sigma\sigma;\omega)
  \equiv
  -\lim_{\eta\rightarrow +0}
   \frac{\mathrm{Im}
         \left[
          G_0(\bm{k}:\sigma\sigma;i\omega_{n})|_{i\omega_{n}=\omega+i\eta}
         \right]}
        {\pi\hbar}
   \nonumber
   \allowdisplaybreaks \\
   &\qquad
 =-\frac{1}{\pi}
   \lim_{\eta\rightarrow +0}
   \mathrm{Im}
   \left[
    \frac{1}{\hbar\omega-\varepsilon_{\bm{k}:\sigma}+i\hbar\eta}
   \right]
  =\delta(\hbar\omega-\varepsilon_{\bm{k}:\sigma}).
   \label{E:A0kw}
\end{align}
Without any correlation, every magnon peak has the form of a delta function, signifying
its infinite lifetime.
Similar to \eqref{E:A0kw}, we can extract the correlated spectral functions from
the renormalized Green functions \eqref{E:GfromDysonEq},
\begin{align}
   &
   A(\bm{k}:\sigma\sigma;\omega)
  \equiv
  -\lim_{\eta\rightarrow +0}
   \frac{\mathrm{Im}
         \left[
          G(\bm{k}:\sigma\sigma;i\omega_{n})|_{i\omega_{n}=\omega+i\eta}
         \right]}
        {\pi\hbar}
   \nonumber
   \allowdisplaybreaks \\
   &\qquad
 =-\frac{1}{\pi}
   \lim_{\eta\rightarrow +0}
   \mathrm{Im}
   \left[
    \frac{1}
         {\hbar\omega
         -\varepsilon_{\bm{k}:\sigma}
         -\varSigma^{(0)}(\bm{k}:\sigma\sigma;\omega+i\eta)
         +i\hbar\eta}
   \right].
   \label{E:Akw}
\end{align}
Unlike bare magnons, correlated magnons may have a finite decay rate.
The primary self-energies
$\varSigma^{(0)}(\bm{k}:\sigma\sigma;\omega+i\eta)
\equiv
 \varSigma_1^{(0)}(\bm{k}:\sigma\sigma)
+\varSigma_2^{(0)}(\bm{k}:\sigma\sigma;\omega+i\eta)$
renormalize the bare magnon excitation energies $\varepsilon_{\bm{k}:\sigma}$ into
\begin{align}
   \varepsilon_{\bm{k}:\sigma}
  +\lim_{\eta\rightarrow +0}\varSigma^{(0)}(\bm{k}:\sigma\sigma;\omega+i\eta)
  \equiv
   \varepsilon_{\bm{k}:\sigma}
  +\Delta\varepsilon_{\bm{k}:\sigma}
  -i\varGamma_{\bm{k}:\sigma}.
   \label{E:RenormEnergy}
\end{align}
When we adopt what they call on-shell approximation \cite{C144416} for the sake of argument
to evaluate the self-energies each at their corresponding bare magnon energy,
the energy corrections and emergent decay rates read
\begin{align}
   &
   \Delta\varepsilon_{\bm{k}:\sigma}
  =\lim_{\eta\rightarrow +0}
   \mathrm{Re}
   \left[
    \varSigma^{(0)}
    \left(
     \bm{k}:\sigma\sigma;\frac{\varepsilon_{\bm{k}:\sigma}}{\hbar}+i\eta
    \right)
   \right],
   \nonumber
   \allowdisplaybreaks \\
   &
   \varGamma_{\bm{k}:\sigma}
 =-\lim_{\eta\rightarrow +0}
   \mathrm{Im}
   \left[
    \varSigma^{(0)}
    \left(
     \bm{k}:\sigma\sigma;\frac{\varepsilon_{\bm{k}:\sigma}}{\hbar}+i\eta
    \right)
   \right].
   \label{E:RenormDispRel&DecayRate}
\end{align}

   The first-order corrections $\varSigma_1^{(0)}(\bm{k}:\sigma\sigma)$ have no dependence on
the Matsubara frequencies $\omega_n$ to begin with and therefore yield no magnon decay.
Suppose the $O(S^0)$ Hamiltonian \eqref{E:H(0)2D} is rewritten in terms of Bogoliubov bosons as
\begin{align}
   &
   {\cal H}^{(0)}
  =E^{(0)}
  +\sum_{\nu=1}^N
   \sum_{\sigma_1,\sigma_2=1}^p
   \left\{
    \varLambda_{+-}
    \left(
     \bm{k}_\nu:\sigma_1,\bm{k}_\nu:\sigma_2
    \right)
    \alpha_{\bm{k}_\nu:\sigma_1}^\dagger
    \alpha_{\bm{k}_\nu:\sigma_2}
   \right.
   \nonumber
   \allowdisplaybreaks \\
   &
   \left.
   +\left[
     \varLambda_{++}
     \left(
      \bm{k}_\nu:\sigma_1,-\bm{k}_\nu:\sigma_2
     \right)
     \alpha_{ \bm{k}_\nu:\sigma_1}^\dagger
     \alpha_{-\bm{k}_\nu:\sigma_2}^\dagger
    +\mathrm{H.c.}
    \right]
   \right\}
  +\sum_{\nu,\mu,\xi=1}^N
   \sum_{\sigma_1,\sigma_2,\sigma_3,\sigma_4=1}^p
   \nonumber
   \allowdisplaybreaks \\
   &\times
   \left\{
         \vphantom{\alpha_{ \bm{k}_\nu            :\sigma_1}^\dagger
                   \alpha_{ \bm{k}_\xi'           :\sigma_2}^\dagger
                   \alpha_{-\bm{k}_\nu -\bm{q}_\mu:\sigma_3}^\dagger
                   \alpha_{-\bm{k}_\xi'+\bm{q}_\mu:\sigma_4}^\dagger
                   }
    \varLambda_{++--}
    \left(
     \bm{k}_\nu            :\sigma_1,
     \bm{k}_\xi'           :\sigma_2,
     \bm{k}_\nu +\bm{q}_\mu:\sigma_3,
     \bm{k}_\xi'-\bm{q}_\mu:\sigma_4
    \right)
    \alpha_{\bm{k}_\nu            :\sigma_1}^\dagger
    \alpha_{\bm{k}_\xi'           :\sigma_2}^\dagger
    \alpha_{\bm{k}_\nu +\bm{q}_\mu:\sigma_3}
    \alpha_{\bm{k}_\xi'-\bm{q}_\mu:\sigma_4}
   \right.
   \nonumber
   \allowdisplaybreaks \\
   &
   +\left[
     \varLambda_{+++-}
     \left(
      \bm{k}_\nu            :\sigma_1,
      \bm{k}_\xi'           :\sigma_2,
     -\bm{k}_\nu +\bm{q}_\mu:\sigma_3,
      \bm{k}_\xi'+\bm{q}_\mu:\sigma_4
     \right)
     \alpha_{ \bm{k}_\nu            :\sigma_1}^\dagger
     \alpha_{ \bm{k}_\xi'           :\sigma_2}^\dagger
     \alpha_{-\bm{k}_\nu +\bm{q}_\mu:\sigma_3}^\dagger
     \alpha_{ \bm{k}_\xi'+\bm{q}_\mu:\sigma_4}
    \right.
   \nonumber
   \allowdisplaybreaks \\
   &
    \left.\!
    +\varLambda_{++++}
     \left(
      \bm{k}_\nu            :\sigma_1,
      \bm{k}_\xi'           :\sigma_2,
     -\bm{k}_\nu -\bm{q}_\mu:\sigma_3,
     -\bm{k}_\xi'+\bm{q}_\mu:\sigma_4
     \right)
     \alpha_{ \bm{k}_\nu            :\sigma_1}^\dagger
     \alpha_{ \bm{k}_\xi'           :\sigma_2}^\dagger
     \alpha_{-\bm{k}_\nu -\bm{q}_\mu:\sigma_3}^\dagger
     \alpha_{-\bm{k}_\xi'+\bm{q}_\mu:\sigma_4}^\dagger
    \right.
   \nonumber
   \allowdisplaybreaks \\
   &
    \left.\!\!
         \vphantom{\alpha_{ \bm{k}_\nu            :\sigma_1}^\dagger
                   \alpha_{ \bm{k}_\xi'           :\sigma_2}^\dagger
                   \alpha_{-\bm{k}_\nu +\bm{q}_\mu:\sigma_3}^\dagger
                   \alpha_{ \bm{k}_\xi'+\bm{q}_\mu:\sigma_4}}
    +\mathrm{H.c.}
    \right]
   \left.\!\!
         \vphantom{\alpha_{ \bm{k}_\nu            :\sigma_1}^\dagger
                   \alpha_{ \bm{k}_\xi'           :\sigma_2}^\dagger
                   \alpha_{-\bm{k}_\nu +\bm{q}_\mu:\sigma_3}^\dagger
                   \alpha_{ \bm{k}_\xi'+\bm{q}_\mu:\sigma_4}}
   \right\},
   \label{E:H(0)alpha}
\end{align}
where
$
    \varLambda_{+-}
    \left(
     \bm{k}_\nu:\sigma_2,\bm{k}_\nu:\sigma_1
    \right)
   =\varLambda_{+-}
    \left(
     \bm{k}_\nu:\sigma_1,\bm{k}_\nu:\sigma_2
    \right)^*
$
and
$
    \varLambda_{++--}
    (
     \bm{k}_\xi'-\bm{q}_\mu:\sigma_4,
     \bm{k}_\nu +\bm{q}_\mu:\sigma_3,
     \bm{k}_\xi'           :\sigma_2,
     \bm{k}_\nu            :\sigma_1
    )
   =\varLambda_{++--}
    (
     \bm{k}_\nu            :\sigma_1,
     \bm{k}_\xi'           :\sigma_2,
     \bm{k}_\nu +\bm{q}_\mu:\sigma_3,
     \bm{k}_\xi'-\bm{q}_\mu:\sigma_4
    )^*
$.
The first-order primary self-energies read
\begin{align}
   &
   \varSigma_1^{(0)}(\bm{k}_\nu:\sigma\sigma)
  =\varLambda_{+-}(\bm{k}_\nu:\sigma,\bm{k}_\nu:\sigma)
   \nonumber
   \allowdisplaybreaks \\
   &
  -\frac{2!2!}{\beta 1!\hbar}
   \sum_{\mu=1}^N
   \sum_{\rho=1}^p
   \varLambda_{++--}
   \left(
    \bm{k}_\nu           :\sigma,
    \bm{k}_\nu+\bm{q}_\mu:\rho,
    \bm{k}_\nu           :\sigma,
    \bm{k}_\nu+\bm{q}_\mu:\rho
   \right)
   \sum_{l=-\infty}^\infty
   G_0(\bm{k}_\nu+\bm{q}_\mu:\rho\rho;i\omega_l)
   \nonumber
   \allowdisplaybreaks \\
   &
  =\varLambda_{+-}(\bm{k}_\nu:\sigma,\bm{k}_\nu:\sigma)
 +4\sum_{\mu=1}^N
   \sum_{\rho=1}^p
   \varLambda_{++--}
   \left(
    \bm{k}_\nu           :\sigma,
    \bm{k}_\nu+\bm{q}_\mu:\rho,
    \bm{k}_\nu           :\sigma,
    \bm{k}_\nu+\bm{q}_\mu:\rho
   \right)
   \bar{n}_{\bm{k}_\nu+\bm{q}_\mu:\rho}
   \label{E:Sigma_1^(0)}
\end{align}
and therefore contribute only to $\Delta\varepsilon_{\bm{k}:\sigma}$.
On the other hand, the analytic continuations of the second-order corrections
$\varSigma_2^{(0)}(\bm{k}:\sigma\sigma;\omega+i\eta)$ each may have a finite imaginary part.
Suppose the $O(S^{\frac{1}{2}})$ Hamiltonian \eqref{E:H(1/2)2D} is rewritten in terms of
Bogoliubov bosons as
\begin{align}
   &
   \mathcal{H}^{\left(\frac{1}{2}\right)}
  =\sum_{\nu,\mu=1}^N
   \sum_{\sigma_1,\sigma_2,\sigma_3=1}^p
   \left[
    \varLambda_{++-}
    \left(
     \bm{k}_\nu:\sigma_1,\bm{k}_\mu'-\bm{k}_\nu:\sigma_2,\bm{k}_\mu':\sigma_3
    \right)
    \alpha_{\bm{k}_\nu:\sigma_1}^\dagger
    \alpha_{\bm{k}_\mu'-\bm{k}_\nu:\sigma_2}^\dagger
    \alpha_{\bm{k}_\mu':\sigma_3}
   \right.
    \nonumber
    \allowdisplaybreaks \\
    &\qquad
  +\left.\!\!
    \varLambda_{+++}
    \left(
     \bm{k}_\nu:\sigma_1,-\bm{k}_\nu-\bm{k}_\mu':\sigma_2,\bm{k}_\mu':\sigma_3
    \right)
    \alpha_{ \bm{k}_\nu:\sigma_1}^\dagger
    \alpha_{-\bm{k}_\nu-\bm{k}_\mu':\sigma_2}^\dagger
    \alpha_{ \bm{k}_\mu':\sigma_3}^\dagger
   +\mathrm{H.c.}
   \right].
   \label{E:H(1/2)alpha}
\end{align}
The second-order primary self-energies read
\begin{align}
   &
   \varSigma_2^{(0)}({{\bm k}_{\nu}:\sigma\sigma;i\omega_n})
 =-\frac{1}{\beta\hbar^{2}}
   \sum_{\mu=1}^N
   \sum_{\rho,\rho'=1}^p
   \sum_{s=\mp}
   f(s)
   |\varLambda_{++s}( \bm{q}_\mu:\rho,\bar{s}\bm{k}_\nu-\bm{q}_\mu:\rho',\bm{k}_\nu:\sigma)|^{2}
   \nonumber
   \allowdisplaybreaks \\
   &\qquad\times
   \sum_{l=-\infty}^{\infty}
   G_0(                  \bm{q}_\mu:\rho \rho ;i       \omega_l          )
   G_0(\bar{s}\bm{k}_\nu-\bm{q}_\mu:\rho'\rho';i\bar{s}\omega_n-i\omega_l)
   \nonumber
   \allowdisplaybreaks \\
   &
  =\sum_{\mu=1}^N
   \sum_{\rho,\rho'=1}^p
   \sum_{s=\mp}
   \frac{f(s)
         \left|
          \varLambda_{++s}(\bm{q}_\mu:\rho,\bar{s}\bm{k}_\nu-\bm{q}_\mu:\rho',\bm{k}_\nu:\sigma)
         \right|^{2}
         \left(
          \bar{n}_{                  \bm{q}_\mu:\rho }
         +\bar{n}_{\bar{s}\bm{k}_\nu-\bm{q}_\mu:\rho'}+1
         \right)}
        { i\bar{s}\hbar\omega_n
         -\varepsilon_{                  \bm{q}_\mu:\rho }
         -\varepsilon_{\bar{s}\bm{k}_\nu-\bm{q}_\mu:\rho'}}
   \label{E:Sigma_2^(0)}
\end{align}
with $f(s)\equiv(\frac{5+s}{2}!)^2/2!$ and $\bar{s}\equiv -s$,
and therefore they contribute to both $\Delta\varepsilon_{\bm{k}:\sigma}$ and
$\varGamma_{\bm{k}:\sigma}$,
\begin{align}
   &
   \lim_{\eta\rightarrow +0}
   \mathrm{Re}
   \left[
    \varSigma_2^{(0)}
    \left(
     {{\bm k}:\sigma\sigma;\frac{\varepsilon_{\bm{k}:\sigma}}{\hbar}+i\eta}
    \right)
   \right]
   \nonumber
   \allowdisplaybreaks \\
   &\qquad
  =\sum_{\mu=1}^N
   \sum_{\rho,\rho'=1}^p
   \sum_{s=\mp}
   \frac{f(s)
         \left|
          \varLambda_{++s}(\bm{q}_\mu:\rho,\bar{s}\bm{k}-\bm{q}_\mu:\rho',\bm{k}:\sigma)
         \right|^{2}
         \left(
          \bar{n}_{              \bm{q}_\mu:\rho }
         +\bar{n}_{\bar{s}\bm{k}-\bm{q}_\mu:\rho'}+1
         \right)}
        {\bar{s}\varepsilon_{\bm{k}:\sigma}
         -\varepsilon_{              \bm{q}_\mu:\rho }
         -\varepsilon_{\bar{s}\bm{k}-\bm{q}_\mu:\rho'}},
   \nonumber
   \allowdisplaybreaks \\
   &
   \lim_{\eta\rightarrow +0}
   \mathrm{Im}
   \left[
    \varSigma_2^{(0)}
    \left(
     {{\bm k}:\sigma\sigma;\frac{\varepsilon_{\bm{k}:\sigma}}{\hbar}+i\eta}
    \right)
   \right]
   \nonumber
   \allowdisplaybreaks \\
   &\qquad
  =\sum_{\mu=1}^N
   \sum_{\rho,\rho'=1}^p
   \sum_{s=\mp}
   sf(s)
   \left|
    \varLambda_{++s}(\bm{q}_\mu:\rho,\bar{s}\bm{k}-\bm{q}_\mu:\rho',\bm{k}:\sigma)
   \right|^{2}
   \left(
    \bar{n}_{              \bm{q}_\mu:\rho }
   +\bar{n}_{\bar{s}\bm{k}-\bm{q}_\mu:\rho'}+1
   \right)
   \nonumber
   \allowdisplaybreaks \\
   &\qquad\qquad\qquad\quad\times
   \delta
   \left(
    \varepsilon_{\bm{k}:\sigma}
   +s\varepsilon_{              \bm{q}_\mu:\rho }
   +s\varepsilon_{\bar{s}\bm{k}-\bm{q}_\mu:\rho'}
   \right).
   \label{E:Sigma_2^(0)Re&Im}
\end{align}
Considering that $\varepsilon_{\bm{k}:\sigma}\geq 0$,
the finite magnon decay rates $\varGamma_{\bm{k}:\sigma}$ come only from vertex functions of
the $\varLambda_{++-}$ type.

\end{document}